\documentclass[12pt]{article}
\usepackage{amssymb}

\usepackage{comment}

\usepackage{amsmath}
\usepackage{graphicx}
\usepackage{theorem}
\usepackage{here}
\usepackage{booktabs}

\usepackage{natbib}

\usepackage{bm}
\usepackage{color}

\theorembodyfont{\rmfamily}

\usepackage[doublespacing]{setspace}

\def\GE{\mathrm{GE}}

\begin{document}
\title{Multilevel Decomposition of Generalized Entropy Measures Using Constrained Bayes Estimation: An Application to Japanese Regional Data}

\author{
Yuki Kawakubo\footnote{Graduate School of Social Sciences, Chiba University, (E-mail: \texttt{kawakubo@chiba-u.jp})} \ and
Kazuhiko Kakamu\footnote{Graduate School of Data Science, Nagoya City University, (E-mail: \texttt{kakamu@ds.nagoya-cu.ac.jp})}
}

\maketitle
\begin{abstract}
We propose a method for multilevel decomposition of generalized entropy (GE) measures that explicitly accounts for nested population structures such as national, regional, and subregional levels.
Standard approaches that estimate GE separately at each level do not guarantee compatibility with multilevel decomposition.
Our method constrains lower-level GE estimates to match higher-level benchmarks while preserving hierarchical relationships across layers.
We apply the method to Japanese income data to estimate GE at the national, prefectural, and municipal levels, decomposing national inequality into between-prefecture and within-prefecture inequality, and further decomposing prefectural GE into between-municipality and within-municipality inequality.

\par\vspace{4mm}

\end{abstract}

\section{Introduction}
\label{sec:int}

The measurement of income inequality has been extensively studied in the literature and remains a crucial issue in economics and other social sciences.
One powerful tool in income inequality analysis is decomposition analysis, which breaks down total income inequality into between-group and within-group inequality \citep{B79, C00}.
Groups can be defined on the basis of various characteristics such as gender, occupation, and educational background.
When groups are defined by regions, decomposition analysis can be extended to a multilevel framework.
This approach provides a more detailed examination of income inequality by decomposing the overall inequality into between-region and within-region inequality, further subdividing within-region inequality into between-subregion and within-subregion inequality, and so forth \citep{C85,A03,KF09}.
However, since the seminal works of \citet{P03,MS08}, it has become evident that decomposition alone is insufficient to analyze income inequality.
It is also essential to examine inequality between different income levels, particularly focusing on the income share held by the top percentiles.
In \citet{AS15}, these two objectives---decomposition analysis and top income analysis---are treated separately in the context of global income inequality.
Performing both simultaneously remains one of the most challenging issues in this field.

Several inequality measures have been proposed in the literature, including the coefficient of variation, mean log deviation (MLD), Theil index \citep{T67}, Atkinson index \citep{A70}, and the widely used Gini coefficient.
Among these, this study focuses on the class of generalized entropy (GE) measures discussed by \citet{C80,S80,CK81} among others.
The GE measure encompasses the MLD, Theil and Atkinson indices as special cases and provides a flexible framework for evaluating income inequality by incorporating a sensitivity parameter. 
Although previous studies, such as \citet{P03,MS08}, have focused mainly on top incomes, it is equally crucial to examine income inequality at lower income levels, particularly in the context of Japan’s aging population and the decreasing size of households \citep{T05,O08}.
As will be discussed later, the GE measure facilitates the analysis of both top and bottom income inequalities, offering a more comprehensive assessment of income distribution.
Furthermore, its additive property enables the decomposition of income inequality into between-group and within-group components, allowing for a more detailed examination of inequality dynamics.

The first key property of the GE measure is its sensitivity parameter, which determines the extent to which the measure responds to changes in different parts of the income distribution.
A higher value of the sensitivity parameter makes the GE measure more responsive to variations in the upper tail of the income distribution.
For instance, as demonstrated in Section \ref{subsec:empirical}, in Japan, the GE measure for Tokyo ranks among the highest when the sensitivity parameter is large but is relatively lower when the parameter is small.
This suggests that income inequality in Tokyo is more pronounced among high-income groups, implying that redistributive policies targeting high-income individuals could be effective in reducing overall inequality.
By allowing for varying degrees of sensitivity, the GE measure provides a nuanced analysis of the effects of redistributive policies on different income groups.

Another significant property of the GE measure is its decomposability.
When a population, such as an entire country, is divided into subpopulations, such as geographical regions, the overall GE measure can be decomposed into within-region inequality (the weighted sum of the GE measures of the individual regions) and between-region inequality (the GE measure of the mean incomes across regions).
\citet{CK81} showed that some axioms, including decomposability, lead to additive inequality measures, but the Gini coefficient does not belong to this class.
The Gini coefficient, defined intuitively as twice the area between the Lorenz curve and the 45-degree line, can also be interpreted as half of the relative mean absolute difference.
Due to this construction, decomposing the Gini coefficient into within-region and between-region components is problematic, as income differences between members of different regions create an interaction effect.
In contrast, additive measures such as the GE measure facilitate precise attribution analysis of inequality and have been widely utilized in empirical research.

\begin{figure}[th]
    \centering
    \includegraphics[width=\textwidth]{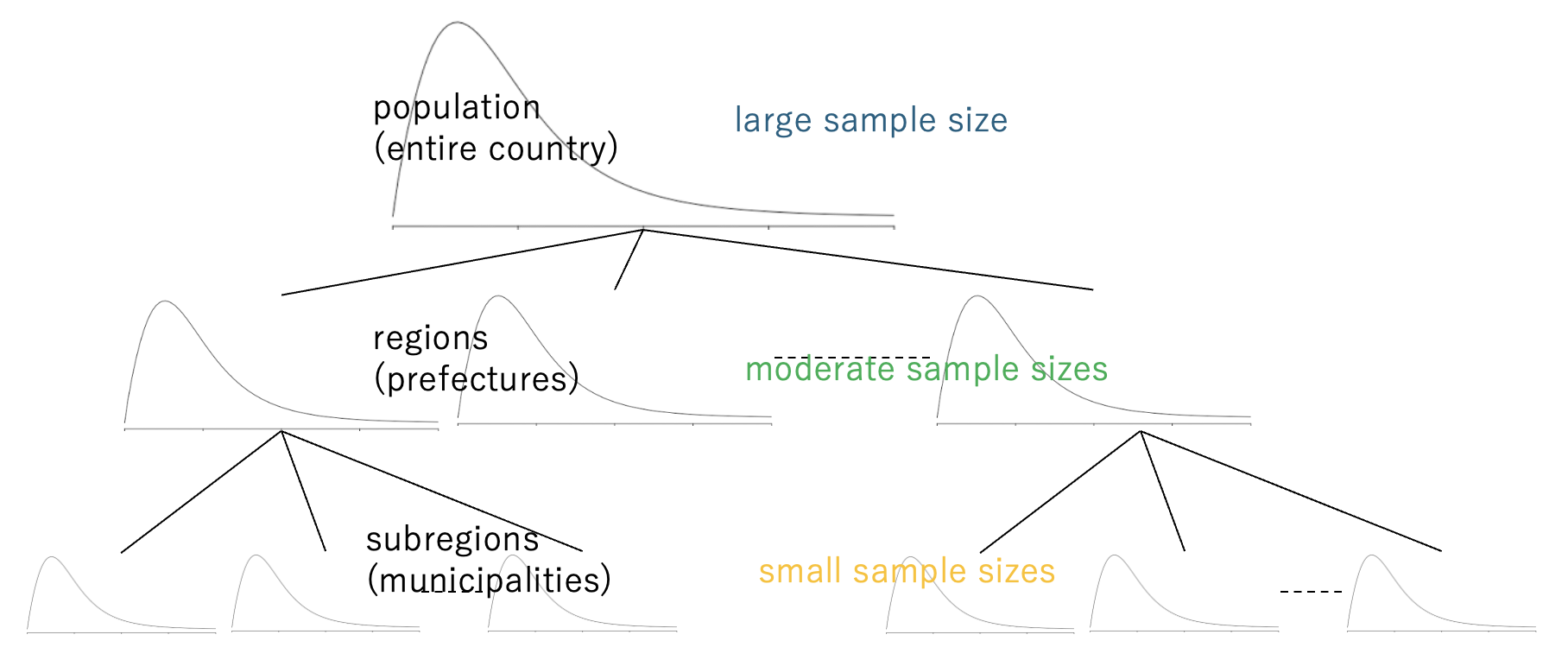}
    \caption{Nested structure of subpopulations}
    \label{fig:punch}
\end{figure}

In nested population structures as shown in Figure \ref{fig:punch}, such as when each region is further divided into subregions, the GE measure supports multilevel decomposition.
Specifically, the GE measure for an entire country can be decomposed into between-region inequality, the weighted sum of between-subregion inequality, and the weighted sum of within-subregion inequality, a process known as multilevel decomposition \citep{KF09}.
In this study, we consider a regional structure wherein Japan is divided into 47 prefectures (regions), each of which is further divided into municipalities (subregions), totaling approximately 1,900 municipalities.
Our goal is to estimate a multilevel decomposition of the national GE measure.

While the GE measure exhibits decomposability by definition in a descriptive statistical approach for a finite population, estimating it in a manner compatible with decomposition for statistical inference presents challenges.
Specifically, for income data, individual-level data is typically unavailable due to privacy concerns, and researchers often rely on grouped data, such as frequency distributions.
In such cases, assuming a parametric income distribution can facilitate statistical inference based on probability distributions.
This approach also enables estimation of the upper tail of the income distribution from limited data, even though prior studies, such as \citet{P03,MS08}, necessitate detailed top-income data.
However, efficiently estimating the GE measure in a manner that maintains compatibility with decomposition remains a challenging statistical problem that requires further methodological advancements.

Previous studies have used a model that assumes some parametric distribution as the income distribution for each subpopulation and that the income distribution of the population is a finite mixture of these parametric distributions \citep{CGRV12,LN16}.
Estimating the GE based on this model provides estimates consistent with decomposition.
However, there may be two drawbacks to this approach.
First, the number of parameters of the income distribution for the population becomes very large when the number of subpopulations is large, so that the estimation of the GE for the overall population would become inefficient.
In the dataset we consider in this paper, the entire country is divided into 47 prefectures, each of which is divided into municipalities, the number of which is about 1900 in all.
While this method might work in situations where the number of regions is around 10, as supposed by \cite{CGRV12} and \cite{LN16}, it would clearly be inefficient for our dataset.
Second, if the sample sizes of subpopulations are not very large, the parametric distributions assumed for the subpopulations need to be simple models such as the family of lognormal distributions, which is known to be a poor fit to actual income data, causing concern that the GE estimation for each subpopulation might be inaccurate.
In our dataset, because there are many municipalities with very small sample sizes, it is difficult to estimate the distribution by fitting a distribution with a large number of parameters, such as the Singh--Maddala distribution, to such municipalities, while a distribution with a small number of parameters, such as the lognormal distribution, does not fit well.

Several parametric distributions that are fit to the actual income data have been proposed and empirically investigated in a large amount of literature.
Therefore, we first assume appropriate families of parametric distributions for the entire country, regions (prefectures), and subregions (municipalities), respectively, depending on their sample sizes.
For example, in our empirical analysis, we assume a generalized beta distribution of the second kind (GB2) with 4 parameters proposed by \cite{McDonald84}, Singh--Maddala distributions with 3 parameters \citep{SM76} for prefectures with moderate sample sizes, and lognormal distributions for municipalities with very small sample sizes in some places.
Estimating national, prefectural, and municipal GEs based on their respective assumed distributions does not, of course, yield results compatible with the decomposition.
Hence, we estimate the GEs for prefectures and municipalities by employing a method called constrained Bayes estimation \citep{Ghosh92, DGSM11}, with the constraint that equality of the decomposition holds.
The original estimates of GE for prefectures are modified using the highly reliable national GE estimate as the benchmark based on data with a sufficiently large sample size and a rich parametric distribution, and the estimates of GE for municipalities are modified using their prefectural GE estimates as benchmarks.
As a result, it is expected that the GEs of municipalities can be estimated with reasonable accuracy even under the situation that only poorly fitting lognormal distributions can be assumed due to the small sample sizes.

From the empirical results, we were able to analyze the structure of income inequality in detail by multilevel decomposition of the GE index.
First, we examine income inequality in Japan by analyzing changes in the parameters of the income distribution and variations in intensity.
Second, by decomposing inequality into between-prefecture and within-prefecture components, we identified a negative correlation between the parameters of the Singh--Maddala distribution, except for Tokyo, which exhibited a distinct trend.
Finally, a multilevel decomposition of inequality revealed that municipalities with higher average income tend to exhibit greater inequality in the upper tail of the distribution.


The rest of the paper is organized as follows.
In Section \ref{sec:GE}, we define the GE in a finite population and explain its decomposability in detail.
We also describe the usefulness of the GE compared to the Gini coefficient using a toy example.
In Section \ref{sec:Estimation}, we propose an estimation method for decomposition of the GE based on grouped data.
Its extension to a multilevel decomposition is explained in Section \ref{sec:multilevel}.
In Section \ref{sec:realdata}, we conduct an empirical analysis using Japanese income data, compare the proposed method with existing methods, and examine the estimation results.
Finally, Section \ref{sec:discussion} concludes the paper with a discussion.

\section{The class of the generalized entropy measures}
\label{sec:GE}

\subsection{Definition of the GE in finite population}
Firstly, we explain the definition of the GE in the finite population.
Let $x_i \ (i=1,\dots,N)$ be the income of the $i$th household in the population, where $N$ is the population size.
The index set of households is denoted by $I = \{ 1,\dots,N \}$, which is taken as the finite population.
We are interested in the distribution of income $\{ x_1,\dots,x_N \}$ in the finite population $I$, typically some inequality measure of the income distribution.

The class of generalized entropy (GE) measures of income inequality 
is defined as
\begin{equation}
    \label{eqn:GE_def}
    \GE_\theta = \frac{1}{\theta(\theta - 1)} \frac{1}{N} \sum_{i \in I} \left\{ \left( \frac{x_i}{\mu} \right)^\theta - 1 \right\}, \quad \theta \not= 0, 1,
\end{equation}
where $\mu = N^{-1} \sum_{i=1}^N x_i$ denotes the mean income in the population $I$.
$\GE_\theta$ has two limiting cases for $\theta \to 0$ and $\theta \to 1$.
For the case of $\theta \to 0$,
\begin{equation*}
    \GE_\theta \to -\frac{1}{N} \sum_{i \in I}\log \left( \frac{x_i}{\mu} \right) =: T_0,
\end{equation*}
which is called the mean log deviation (MLD).
Another limiting case is for $\theta \to 1$:
\begin{equation*}
    \GE_\theta \to \frac{1}{N} \sum_{i \in I} \frac{x_i}{\mu} \log \left( \frac{x_i}{\mu} \right) =: T_1,
\end{equation*}
which is known as the Theil entropy index.
The parameter $\theta$ in GE represents the sensitivity to the importance given to changes in the upper tail of the distribution.
$\GE_\theta$ becomes more sensitive to changes at the top of the distribution as $\theta$ increases; on the other hand, it gives more importance to differences at the bottom of the distribution as $\theta$ decreases.
In the next subsection, we explain the roll of $\theta$ with a toy example.

\subsection{The roll of the sensitivity parameter in the GE}
\begin{figure}[h]
    \centering
    \includegraphics[width=\textwidth]{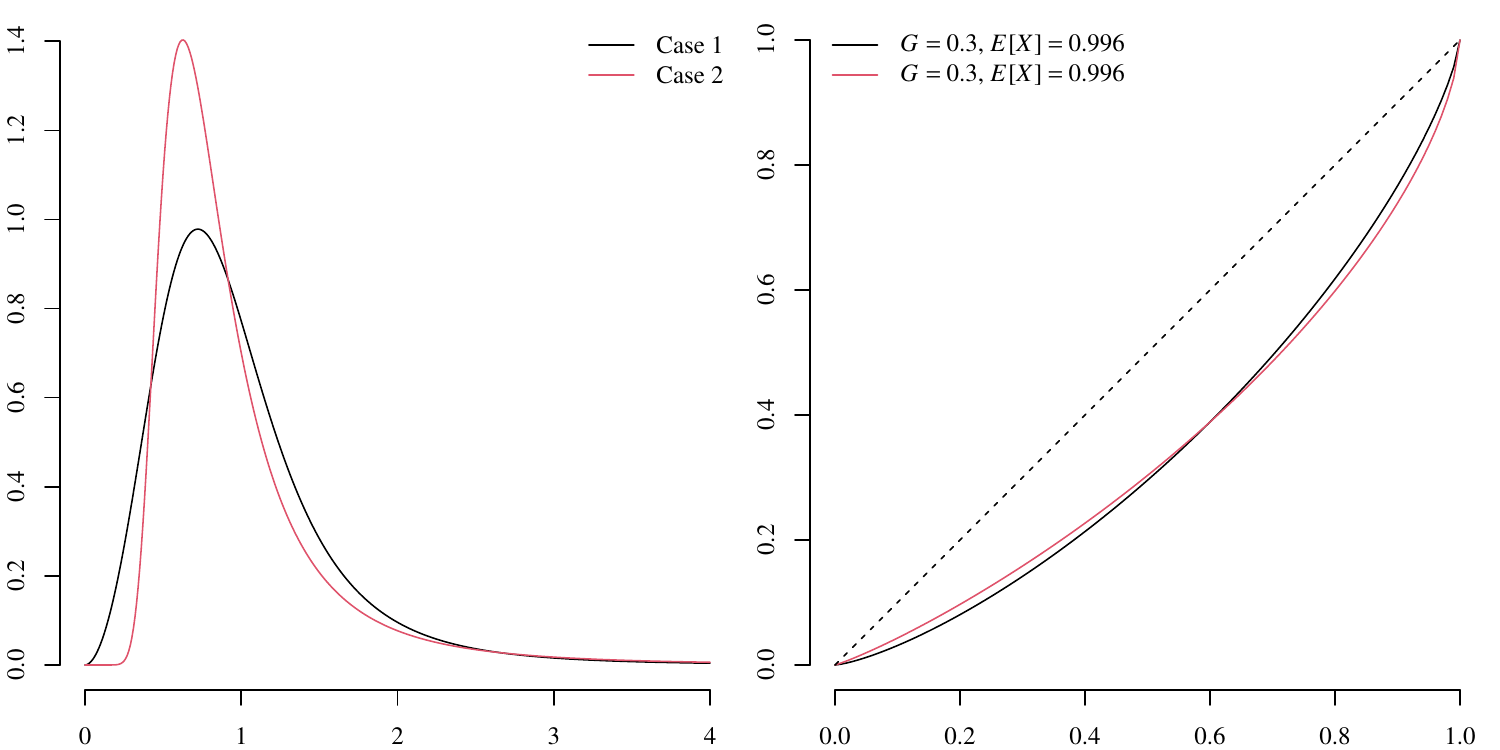}
    \caption{Two income distributions with same Gini coefficients and means}
    \label{fig:two-dist}
\end{figure}

We would like to use a toy example to show how the GE can be useful in comparison with the Gini coefficient.
Figure \ref{fig:two-dist} (left and right) displays a toy example: income distributions with different shapes and their corresponding Lorenz curves, respectively.
From the figure, we can observe that the shapes of the Lorenz curves are different since the shapes of the income distributions are different.
However, if we calculate the Gini coefficients and means, they show the same values in both cases.
Therefore, it is difficult to distinguish these two distributions from the Gini coefficients and means.
On the other hand, the generalized entropy (GE) can distinguish the difference assuming different $\theta$.
Table \ref{tab:two-dist} shows the GE for different $\theta$ for two cases.
From the table, we can confirm that the income inequality in Case 1 is greater when $\theta = -2$, while the income inequality of Case 2 is greater when $\theta = 2$.
Moreover, the order of the income inequality changes between $\theta = 0$ and 1.
This result is consistent with the shapes of the Lorenz curves.
Therefore, the GE is useful when we are interested in the details of the income distributions.

\begin{table}[h]
    \centering
    \caption{The generalized entropy for different $\theta$}
    \label{tab:two-dist}
    \begin{tabular}{lrrrrr}
        \hline
        $\theta$ &    $-2$ &    $-1$ &     $0$ &    $1$ &     $2$\\
        \hline
        Case 1   & $0.371$ & $0.196$ & $0.156$ & $0.155$ & $0.195$\\
        Case 2   & $0.160$ & $0.144$ & $0.147$ & $0.183$ & $0.357$\\
        \hline
    \end{tabular}
\end{table}

\subsection{Decomposability of the GE}
Next, we consider the situation that the population $I$ is divided into $J$ non-overlapping subpopulations $I_1,\dots,I_J$, that is, it follows that
$$
I = \bigcup_{j=1}^J I_j \quad \mathrm{and} \quad I_j \cap I_{j'} = \emptyset \ (j \not= j').
$$
In this paper, the entire country is considered the population and the regions are treated as subpopulations.
Consequently, we will use the terms country and region interchangeably with population and subpopulation, respectively.
However, subpopulations are not necessarily those classified by regions, but can be constituted as groups classified by demographic variables such as sex and age groups.
The size of the subpopulation $I_j$, the number of households in the $j$th region, is denoted by $N_j$, so that $\#(I_j) = N_j$ and $N = \sum_{j=1}^J N_j$.
We define the population share of $I_j$ as $\lambda_j = N_j / N$, so we can see that $\sum_{j=1}^J \lambda_j = 1$.
Furthermore, we define the mean income of $I_j$ as $\mu_j = N_j^{-1} \sum_{i \in I_j} x_i$.
Then, the GE measures of the $j$th region are
$$
\GE_{\theta,j} = \frac{1}{\theta(\theta - 1)} \frac{1}{N_j} \sum_{i \in I_j} \left\{ \left( \frac{x_i}{\mu_j} \right)^\theta - 1 \right\}, \quad \theta \not= 0, 1.
$$
As the two limiting cases of $\GE_{\theta,j}$, the MLD and the Theil entropy index of $I_j$ are
$$
T_{0j} := -\frac{1}{N_j} \sum_{i \in I_j}\log \left( \frac{x_i}{\mu_j} \right), \quad T_{1j} := \frac{1}{N_j} \sum_{i \in I_j} \frac{x_i}{\mu_j} \log \left( \frac{x_i}{\mu_j} \right),
$$
respectively.

An attractive feature of the GE measures is the following additive decomposability:
\begin{equation}
    \label{eqn:decomp}
    \GE_\theta = \underbrace{ \sum_{j=1}^J \lambda_j^{1-\theta} s_j^\theta \mathrm{GE}_{\theta,j} }_{\text{within-region inequality}} + \underbrace{ \frac{1}{\theta(\theta - 1)} \left\{ \sum_{j=1}^J \lambda_j \left( \frac{\mu_j}{\mu} \right)^\theta - 1 \right\} }_{\text{between-region inequality}},
\end{equation}
where $s_j = \lambda_j\mu_j / \mu$ stands for the income share of $I_j$.
The first term on the right hand side of \eqref{eqn:decomp} is called the within-region inequality, and the second term of that is called the between-region inequality.
As the limiting cases, MLD and the Theil entropy index can be also decomposed as
$$
T_0 =  \sum_{j=1}^J \lambda_j T_{0j} + \sum_{j=1}^J \lambda_j \log \left( \frac{\mu}{\mu_j} \right), \quad T_1 = \sum_{j=1}^J s_j T_{1j} + \sum_{j=1}^J s_j \log \left( \frac{\mu_j}{\mu} \right).
$$
The two terms of the decomposition of the MLD and the Theil entropy index are also called the within-region inequality and the between-region inequality.
The within-region inequality can be interpreted as the weighted average of the inequality of $J$ regions with the weights $\lambda_j^{1-\theta} s_j^\theta$, though the weights do not necessarily sum up to one unless $\theta \to 0$ or $\theta \to 1$, which are the cases for the MLD and the Theil entropy index.
On the other hand, the between-region inequality is understood to represent the GE measure if all households in each region had the average income $\mu_j$ for that region.

\section{Estimation strategy}
\label{sec:Estimation}
In this section, we discuss the strategy to simultaneously estimate GE measures both for the entire country $I$ and the regions $I_j$, thereby obtaining the income inequality decomposition.
We consider the problem in the situation that household income is observed in the form of grouped data for the entire country and for each of the regions.

\subsection{Estimating GEs based on grouped data}
\label{subsec:groupeddata}
Let $\mathcal{S} \subseteq I$ be the index set of $i$ representing the households sampled in some sample survey.
The sample size of the survey is denoted by $n = \#(\mathcal{S})$.
We cannot directly observe household income $x_i$ for $i \in \mathcal{S}$, but we can observe the frequency distribution $\bm{y} = (y_1,\dots,y_G)^\top$, where
$$
y_g = \sum_{i \in \mathcal{S}} I(c_{g-1} \leq x_i < c_g), \quad (g=1,\dots,G),
$$
for the pre-specified boundary values $0 = c_0 < c_1 < \dots < c_G = +\infty$.

In order to estimate the GEs based on grouped data with the number of groups around $G=10$, we assume some parametric income distribution.
Let $F(\cdot \mid \bm{\omega})$ and $f(\cdot \mid \bm{\omega})$ be the cumulative distribution function and the probability density function of the assumed parametric income distribution with the parameter vector $\bm{\omega}$, respectively.
Then, the likelihood function based on the grouped data $\bm{y}$ is
$$
L(\bm{\omega}) \propto \prod_{g=1}^G \left\{ F(c_g \mid \bm{\omega}) - F(c_{g-1} \mid \bm{\omega}) \right\}^{y_g}.
$$

The GE measures of a parametric distribution with density function $f(\cdot \mid \bm{\omega})$ are defined as the expression for the GE measures of a finite population given by \eqref{eqn:GE_def}, with the finite population mean (arithmetic mean) replaced by the mean of the probability distribution (expectation) with respect to the density $f(\cdot \mid \bm{\omega})$ as follows:
\begin{equation}
    \label{eqn:GE_para}
        \GE_\theta(\bm{\omega}) = \frac{1}{\theta(\theta-1)} \left\{ \int_0^\infty \left( \frac{x}{\mu} \right)^\theta f(x \mid \bm{\omega}) \mathrm{d}x - 1 \right\},
\end{equation}
where $\mu = \int_0^\infty x f(x \mid \bm{\omega})\mathrm{d}x$ is the mean of the distribution.
In the same way, the MLD and the Theil entropy index of a parametric distribution with density $f(\cdot \mid \bm{\omega})$ are
\begin{equation}
    \label{eqn:Theil_para}
        T_0(\bm{\omega}) = -\int_0^\infty \log\left( \frac{x}{\mu} \right) f(x \mid \bm{\omega}) \mathrm{d}x, \quad T_1(\bm{\omega}) = \int_0^\infty \frac{x}{\mu} \log\left( \frac{x}{\mu} \right) f(x \mid \bm{\omega}) \mathrm{d}x,
\end{equation}
respectively.
For various parametric probability distributions used as income distributions, $\GE_\theta(\bm{\omega}), T_0(\bm{\omega})$ and $T_1(\bm{\omega})$ are explicitly expressed as a function of the parameter vector $\bm{\omega}$ \citep{SJR17}.
Thus, once the parameter vector $\bm{\omega}$ of the income distribution is estimated based on the likelihood function $L(\bm{\omega})$, its GEs can also easily be estimated.

\subsection{Estimating the decomposition of GEs}
\label{subsec:GEsub}
We next consider simultaneously estimating the GEs for both the entire country and the regions.
Define the sample index set for each region as $\mathcal{S}_j = \mathcal{S} \cap I_j$ and its sample size as $n_j = \#(\mathcal{S}_j)$.
We observe the frequency distribution of household income in each region $\bm{y}_{\mathrm{R},j} = (y_{j1},\dots,y_{jG})^\top$, where
$$
y_{jg} = \sum_{i \in \mathcal{S}_j} I(c_{g-1} \leq x_i < c_g), \quad (g=1,\dots,G).
$$
The boundary values $c_g$'s and the number of groups $G$ do not necessarily have to be the same for the entire country and the regions.
The collection of grouped data at the region level is denoted by $\bm{y}_{\mathrm{R}} = (\bm{y}_{\mathrm{R},1}^\top,\dots,\bm{y}_{\mathrm{R},J}^\top)^\top$.
We also assume that the population size $N$ and the size of each region $N_j$ are known.

In order to estimate not only the GEs of the entire country but also those of the regions, we may assume parametric distributions for the country and each region separately.
Then, the parametric distribution assumed for the country $I$ is estimated based on $\bm{y}$, and the parametric distribution assumed for the region $I_j$ is estimated based on $\bm{y}_j$.
However, this method does not yield estimates that are compatible with the decomposition of the GEs.

One way to obtain estimates compatible with the decomposition is to model the parametric distribution of the entire country as a finite mixture of parametric distributions of regions.
Let $f(x \mid \bm{\omega})$ and $f_j(x \mid \bm{\omega}_j)$ be the density functions of the parametric distribution assumed for the entire country $I$ and that for the region $I_j$, respectively.
If $f$ is the finite mixture of $f_j$'s with known weights $\lambda_j = N_j/N$, that is, $f(x \mid \bm{\omega}) = \sum_{j=1}^J \lambda_j f_j(x \mid \bm{\omega}_j)$ where $\bm{\omega} = ( \bm{\omega}_1^\top,\dots,\bm{\omega}_J^\top )^\top$, $\GE_\theta (\bm{\omega})$ for $f$ is decomposed as
\begin{equation}
    \label{eqn:decomp_para}
    \GE_\theta (\bm{\omega}) = \sum_{j=1}^J \lambda_j^{1-\theta} s_j^\theta \GE_{\theta,j}(\bm{\omega}_j) + \frac{1}{\theta(\theta - 1)} \left\{ \sum_{j=1}^J \lambda_j \left( \frac{\mu_j}{\mu} \right)^\theta - 1 \right\},
\end{equation}
where $\mu_j = \int_0^\infty x f_j(x \mid \bm{\omega}) \mathrm{d}x$ is the mean of $f_j$ and
$$
\GE_{\theta,j}(\bm{\omega}_j) = \frac{1}{\theta(\theta-1)} \left\{ \int_0^\infty \left( \frac{x}{\mu_j} \right)^\theta f_j(x \mid \bm{\omega}_j) \mathrm{d}x - 1 \right\}.
$$
Then, we can estimate $\GE_{\theta,j}(\bm{\omega}_j)$ and $\mu_j$ based on the observation $\bm{y}_j$.
Furthermore, we obtain estimates of the country mean $\mu = \sum_{j=1}^J \lambda_j \mu_j$ and the income share $s_j = \lambda_j \mu_j / \mu$ from the estimates of $\mu_j$, so that we obtain the decomposition in \eqref{eqn:decomp_para}.
At this point, the estimate of $\GE_\theta(\bm{\omega})$ for the entire country is already obtained as the sum of the two terms on the right-hand side of \eqref{eqn:decomp_para}.
This approach, assuming a finite mixture distribution, is used in \cite{CGRV12}, \cite{LN16} and others.

However, there are two possible drawbacks to this approach as well.
One is that if the number of regions $J$ is large, the number of parameters for the entire country modeled by the finite mixture will be very large, which would make the estimation of the GEs for the country inefficient.
Another drawback is that if the sample size of the region is not very large, the parametric distribution assumed for the region needs to be a simple model (with the small number of parameters), and the accuracy of the estimation of GEs for the region is likely to be reduced.

We therefore propose the following estimation method that addresses these problems.
First, we assume a parametric distribution $f(\cdot \mid \bm{\omega})$ for the entire country $I$ and a prior distribution $\bm{\omega}$, and then estimate GEs for the country, denoted by $\widehat{\GE}_\theta$, based on the model $f$, the $\pi(\bm{\omega})$ and the grouped data at the country level $\bm{y}$.
Let $\widehat{\GE}_0$ denote the estimate of the MLD and $\widehat{\GE}_1$ denote the estimate of the Theil entropy index.
By choosing the appropriate parametric distribution depending on the sample size $n$ and the number of groups $G$, the GEs for the entire country can be efficiently estimated.
Next, we assume a parametric distribution $f_j(\cdot \mid \bm{\omega}_j)$ for each region $I_j$ and a prior distribution $\pi(\bm{\omega}_j)$, and then obtain the mean income estimate $\hat{\mu}_j$ for $I_j$ based on the model $f_j$, the prior $\pi(\bm{\omega}_j)$ and the grouped data $\bm{y}_j$.
We also obtain the mean income estimate for the entire country $I$ as $\hat{\mu} = \sum_{j=1}^J \lambda_j \hat{\mu}_j$.
It is noted that the population share $\lambda_j$ is assumed to be known.
Based on the estimates $\hat{\mu}_j$ and $\hat{\mu}$, we estimate the between-region inequality in \eqref{eqn:decomp} as
\begin{equation}
    \label{eqn:between_est}
    \widehat{B} = \frac{1}{\theta(\theta - 1)} \left\{ \sum_{j=1}^J \lambda_j \left( \frac{\hat{\mu}_j}{\hat{\mu}} \right)^\theta - 1 \right\}.
\end{equation}
If we want to estimate the MLD ($\theta = 0$ case), we have $\widehat{B} = \sum_{j=1}^J \lambda_j \log(\hat{\mu} / \hat{\mu}_j)$, and if we focus on the Theil entropy index ($\theta = 1$ case), we have $\widehat{B} = \sum_{j=1}^J \hat{s}_j \log(\hat{\mu}_j / \hat{\mu})$ for $\hat{s}_j = \lambda_j \hat{\mu}_j / \hat{\mu}$.
Lastly, under the constraint of the decomposition \eqref{eqn:decomp}, we estimate $\GE_{\theta,j}$ using a constrained Bayes estimator, yielding an estimate of within-region inequality which is compatible with the decomposition.
In the next subsection, we explain the constrained Bayes estimation in detail.

\subsection{Constrained Bayes estimation of GEs for the regions}
\label{subsec:CB}

To derive the constrained Bayes estimators of GEs for the regions, we assume a parametric distribution $f_j(\cdot \mid \bm{\omega}_j)$ for each region and prior distributions for the parameter vectors $\bm{\omega}_j$, independently.
Basically, the distribution $f_j$ and the prior of $\bm{\omega}_j$ assumed here are the same as those used to estimate the between-region inequality.
It is noted that once the parametric distribution $f_j(\cdot \mid \bm{\omega}_j)$ is assumed, the GEs are expressed as a function of the parameter vector $\bm{\omega}_j$, denoted by $\GE_{\theta,j}(\bm{\omega}_j)$.
To discuss the estimation of the MLD and the Theil entropy index in a unified manner, let $\GE_{0,j}(\bm{\omega}_j)$ and $\GE_{1,j}(\bm{\omega}_j)$ be the MLD and the Theil entropy index of the density $f_j$, respectively.

Under the quadratic loss, the Bayes estimator of $\GE_{\theta,j}(\bm{\omega}_j)$ is the posterior mean, that is,
$$
\widehat{\GE}^\mathrm{B}_{\theta,j} := E \left[\GE_{\theta,j}(\bm{\omega}_j) \mid \bm{y}_{\mathrm{R},j} \right] = \underset{d_j} {\operatorname{argmin}} \ E \left[ \left\{ d_j - \GE_{\theta,j}(\bm{\omega}_j) \right\}^2 \mid \bm{y}_{\mathrm{R},j} \right],
$$
where the expectation is taken with respect to the posterior distribution of $\bm{\omega}_j$, namely $\pi(\bm{\omega}_j \mid \bm{y}_{\mathrm{R}}) = \pi(\bm{\omega}_j \mid \bm{y}_{\mathrm{R},j})$.
However, of course, this estimator is not compatible with the decomposition.
Therefore, we consider the problem of minimizing the posterior risk under the constraint of the compatibility of the decomposition, given by
\begin{equation}
    \label{eqn:minimization}
    \min_{d_1,\dots,d_J} \sum_{j=1}^J \phi_j E\left[ \left\{ d_j - \GE_{\theta,j}(\bm{\omega}_j) \right\}^2 \mid \bm{y}_{\mathrm{R}} \right] \quad \mathrm{s.t.} \ \ \widehat{\GE}_\theta = \sum_{j=1}^J w_jd_j + \hat{B},
\end{equation}
where $\phi_j \geq 0$ is some weight for the loss function, $w_j = \lambda_j^{1-\theta} \hat{s}_j^\theta$ for $\hat{s}_j = \lambda_j \hat{\mu}_j / \hat{\mu}$, and $\hat{\mu}_j$, $\hat{\mu}$, $\hat{B}$ and $\widehat{\GE}_\theta$ are the estimates derived in Section \ref{subsec:GEsub}.
The weight $\phi_j$ may depend on the observation $\bm{y}_{\mathrm{R}}$ but not on $\GE_{\theta,j}(\bm{\omega}_j)$.
We call the minimizer of \eqref{eqn:minimization} the constrained Bayes estimator, given by
\begin{equation}
    \label{eqn:CB}
    \widehat{\GE}^\mathrm{CB}_{\theta,j} = \widehat{\GE}^\mathrm{B}_{\theta,j} + \frac{r_j}{q} \left( \widehat{\GE}_\theta - \widehat{B} - \overline{\GE}^w_\theta \right), \quad j=1,\dots,J,
\end{equation}
where $\overline{\GE}^w_\theta = \sum_{j=1}^J w_j \widehat{\GE}^\mathrm{B}_{j,\theta}$, $r_j = w_j / \phi_j$ and $q = \sum_{j=1}^J w_j^2 / \phi_j$.
Then, we obtain the estimate of the within-region inequality as
\begin{equation}
    \label{eqn:within_est}
    \widehat{W} = \sum_{j=1}^J w_j \widehat{\GE}_{\theta,j}^\mathrm{CB} = \sum_{j=1}^J \lambda_j^{1-\theta} \hat{s}_j^\theta \widehat{\GE}_{\theta,j}^\mathrm{CB}.
\end{equation}
We can see from \eqref{eqn:CB} that the constrained Bayes estimates can be easily obtained once the Bayes estimates are computed by a numerical method such as MCMC.
A factor $\widehat{\GE}_\theta - \widehat{B} - \overline{\GE}^w_\theta$ in the second term of \eqref{eqn:CB} is the residual when we estimate the GEs of the entire country based on $f$ and those of the regions based on $f_j$, respectively, where we call this estimation procedure `separate method'.
We compare our proposed method with the separate method in Section \ref{subsec:comparison} and show the residual numerically in Tables \ref{tab:multi1} and \ref{tab:multi2}.

The constrained Bayes estimates are compatible with the decomposition, which is obvious from their construction.
Moreover, the constrained Bayes estimators are expected to be robust to model misspecification.
Because the sample size of the region is relatively small, it is appropriate to assume a parametric distribution with a small number of parameters.
However, this is often not a good fit, which is the second drawback of the approach using a finite mixture distribution discussed in Section \ref{subsec:GEsub}.
On the other hand, the sample size of the entire country is relatively large, and the estimates of GEs of the country based on an appropriate parametric distribution assumption are expected to be reliable.
In addition, the estimate of between-region inequality is relatively reliable because it depends only on the mean income of each region.
The constrained Bayes estimators of GEs for the regions can be interpreted as a modified form of the unconstrained Bayes estimator (posterior mean) with estimates of GEs for the entire country and the between-region inequality as benchmarks, which are relatively reliable estimates.
Hence, the constrained Bayes estimator is also called the benchmarked estimator and is often used in the field of small area estimation \citep{DGSM11, KHT14, GKK15}.

\bigskip
\noindent
\textit{Remark 1} The choice of the weight $\phi_j$ in the loss function is discussed in \cite{DGSM11}.
When $\phi_j = w_j$, the constrained Bayes estimator \eqref{eqn:CB} can be written as $\widehat{\GE}^\mathrm{CB}_{\theta,j} = \widehat{\GE}^\mathrm{B}_{\theta,j} + (\widehat{\GE}_\theta - \widehat{B} - \overline{\GE}^w_\theta) / \sum_{j'=1}^J w_{j'}$, which is in the form of the residuals (multiplied by a constant) added to the Bayes estimator for all regions uniformly.
Furthermore, the constrained Bayes estimator includes the so-called raking estimator, $(\widehat{\GE}^\mathrm{B}_{\theta,j} / \overline{\GE}^w_\theta) \times (\widehat{\GE}_\theta - \widehat{B})$, for the case that $\phi_j = w_j / \widehat{\GE}^\mathrm{B}_{\theta,j}$.

\bigskip
To conclude this section, we summarize the procedures of the proposed estimating method:
\begin{enumerate}
    \item Assume a parametric distribution $f(\cdot \mid \bm{\omega})$ for the entire country $I$ and a prior distribution $\pi(\bm{\omega})$. Then, estimate the GEs for the country, denoted by $\widehat{\GE}_\theta$, based on the model $f$, the prior $\pi(\bm{\omega})$ and the country-level grouped data $\bm{y}$.

    \item Assume a parametric distribution $f_j(\cdot \mid \bm{\omega})$ for each region $I_j$ and prior distributions $\pi(\bm{\omega}_j)$. Then, obtain the estimate of the mean income $\hat{\mu}_j$ for $I_j$ and $\hat{\mu} = \sum_{j=1}^J \lambda_j \hat{\mu}_j$ for the entire country $I$, based on the model $f_j$, the prior $\pi(\bm{\omega}_j)$ and the region-level grouped data $\bm{y}_{\mathrm{R},j}$. Using $\hat{\mu}_j$ and $\hat{\mu}$, estimate the between-region inequality as $\widehat{B}$ defined by \eqref{eqn:between_est}.

    \item Specifying the weights $\phi_j$ in the loss function, obtain the constrained Bayes estimates $\widehat{\GE}_{\theta,j}^\mathrm{CB}$ for the regions defined as \eqref{eqn:CB}. Then, obtain the estimate of the within-region inequality as $\widehat{W}$ defined by \eqref{eqn:within_est}.
\end{enumerate}

\section{Multilevel decomposition of the inequality}
\label{sec:multilevel}

We next consider the situation that each region (subpopulation) is further divided into subregions (sub-subpopulations).
For example, Japan is divided into 47 prefectures, and there are on average about 40 municipalities within each prefecture.
Section \ref{sec:realdata} analyzes Japanese household income data in the form of grouped frequency distributions with known income brackets, available at the national, prefectural and municipal levels.

It is assumed that each region $I_j \ (j=1,\dots,J)$ is divided into non-overlapping subregions $I_{j1},\dots,I_{jK_j}$, that is, it follows that
$$
I_j = \bigcup_{k=1}^{K_j} I_{jk} \quad \mathrm{and} \quad I_{jk} \cap I_{jk'} = \emptyset \ (k \not= k').
$$
The size of the subregion $I_{jk}$ is denoted by $N_{jk}$, so $\#(I_{jk}) = N_{jk}$ and $N_j = \sum_{k=1}^{K_j} N_{jk}$.
The population share of $I_{jk}$ in the region $I_j$ is defined by $\lambda_{jk} = N_{jk} / N_j$, therefore, we can see that $\sum_{k=1}^{K_j} \lambda_{jk} = 1$ for $j=1,\dots,J$.
Furthermore, we define the mean income of $I_{jk}$ as $\mu_{jk} = N_{jk}^{-1} \sum_{i \in I_{jk}} x_i$.
The sample index set of each subregion is defined as $\mathcal{S}_{jk} = \mathcal{S} \cap I_{jk}$ and its sample size is $n_{jk} = \#(\mathcal{S}_{jk})$.
We observe the frequency distribution of household income in each subregion $\bm{y}_{\mathrm{SR},jk} = (y_{jk1},\dots,y_{jkG})^\top$, where $y_{jkg} = \sum_{i \in \mathcal{S}_{jk}} I(c_{g-1} \leq x_i \leq c_g)$ for $g=1,\dots,G$.
The collection of the subregion-level grouped data is denoted by $\bm{y}_{\mathrm{SR}}$.

The GE measures of the subregion $I_{jk}$ are
$$
\GE_{\theta, jk} = \frac{1}{\theta(\theta - 1)} \frac{1}{N_{jk}} \sum_{i \in I_{jk}} \left\{ \left( \frac{x_i}{\mu_{jk}} \right)^\theta - 1 \right\}, \quad \theta \not= 0,1.
$$
As the two limiting cases of $\GE_{\theta,jk}$, the MLD and the Theil entropy index are
$$
T_{0jk} := -\frac{1}{N_{jk}} \sum_{i \in I_{jk}} \log \left( \frac{x_i}{\mu_{jk}} \right), \quad \mathrm{and} \quad T_{1jk} := \frac{1}{N_{jk}} \sum_{i \in I_{jk}} \frac{x_i}{\mu_{jk}} \log \left( \frac{x_i}{\mu_{jk}} \right),
$$
respectively.
In this situation, the GE measures in each region can be additively decomposed again as
\begin{equation}
    \label{eqn:decomp_sub}
    \GE_{\theta,j} = \underbrace{ \sum_{k=1}^{K_j} \lambda_{jk}^{1-\theta} s_{jk}^\theta \mathrm{GE}_{\theta,jk} }_{\text{within-subregion inequality}} + \underbrace{ \frac{1}{\theta(\theta - 1)} \left\{ \sum_{k=1}^{K_j} \lambda_{jk} \left( \frac{\mu_{jk}}{\mu_j} \right)^\theta - 1 \right\} }_{\text{between-subregion inequality}},
\end{equation}
where $s_{jk} = \lambda_{jk} \mu_{jk} / \mu_j$ stands for the income share of $I_{jk}$ in $I_j$.
In the same way, the MLD and the Theil entropy index can be also decomposed as
$$
T_{0j} = \sum_{k=1}^{K_j} \lambda_{jk} T_{0jk} + \sum_{k=1}^{K_j} \lambda_{jk} \log \left( \frac{\mu_j}{\mu_{jk}} \right), \quad \mathrm{and} \quad T_{1j} = \sum_{k=1}^{K_j} s_{jk} T_{1jk} + \sum_{k=1}^{K_j} s_{jk} \log \left( \frac{\mu_{jk}}{\mu_j} \right),
$$
respectively \citep{KF09}.

To estimate the decomposition \eqref{eqn:decomp_sub}, we start from the constrained Bayes estimates of each region $I_j$, denoted by $\widehat{\GE}_{\theta,j}^\mathrm{CB}$ in \eqref{eqn:CB}.
Next, we assume a parametric distribution $f_{jk}(\cdot \mid \bm{\omega}_{jk})$ for each subregion $I_{jk}$ and a prior distribution $\pi(\bm{\omega}_{jk})$, and then obtain the mean income estimate $\hat{\mu}_{jk}$ for $I_{jk}$ based on the model $f_{jk}$, the prior $\pi(\bm{\omega}_{jk})$ and the subregion-level grouped data $\bm{y}_{\mathrm{SR},jk}$.
We also obtain the mean income estimate for the region $I_j$ as $\Tilde{\mu}_j = \sum_{k=1}^{K_j} \hat{\mu}_{jk}$, which is not identical to the estimate $\hat{\mu}_j$ based on the model $f_j$, the prior $\pi(\bm{\omega}_j)$ and the observation $\bm{y}_{\mathrm{R},j}$, derived in Section \ref{subsec:GEsub}.
Using $\hat{\mu}_{jk}$ and $\Tilde{\mu}_j$, we estimate the between-subregion inequality in \eqref{eqn:decomp_sub} as
\begin{equation}
    \label{eqn:between_sub_est}
    \widehat{B}_j = \frac{1}{\theta(\theta - 1)} \left\{ \sum_{k=1}^{K_j} \lambda_{jk} \left( \frac{\hat{\mu}_{jk}}{\Tilde{\mu}_j} \right)^\theta - 1 \right\}.
\end{equation}
If we estimate the MLD (case $\theta = 0$), we have $\widehat{B}_j = \sum_{k=1}^{K_j} \lambda_{jk} \log(\Tilde{\mu}_j / \hat{\mu}_{jk})$, and if we focus on the Theil entropy index (case $\theta = 1$), we have $\widehat{B}_j = \sum_{k=1}^{K_j} \hat{s}_{jk} \log(\hat{\mu}_{jk} / \Tilde{\mu}_j)$ for $\hat{s}_{jk} = \lambda_{jk} \hat{\mu}_{jk} / \Tilde{\mu}_j$.
Finally, we obtain the constrained Bayes estimator of the GEs for the subregion, $\GE_{\theta,jk}$, as the solution of the following minimization problem:
\begin{equation*}
    \min_{d_{j1},\dots,d_{jK_j}} \sum_{k=1}^{K_j} \phi_{jk} E\left[ \left\{ d_{jk} - \GE_{\theta,jk}(\bm{\omega}_{jk}) \right\}^2 \mid \bm{y}_{\mathrm{SR}} \right] \quad \mathrm{s.t.} \ \ \widehat{\GE}_{\theta,j}^\mathrm{CB} = \sum_{k=1}^{K_j} w_{jk}d_{jk} + \hat{B}_j,
\end{equation*}
where $\phi_{jk} \geq 0$ is some weight that satisfies $\sum_{k=1}^{K_j} \phi_{jk} = 1$, $w_{jk} = \lambda_{jk}^{1-\theta} \hat{s}_{jk}^\theta$ for $\hat{s}_{jk} = \lambda_{jk} \hat{\mu}_{jk} / \Tilde{\mu}_j$, and $\GE_{\theta,jk}(\bm{\omega}_{jk})$ is the GE of the parametric distribution $f_{jk}(\cdot \mid \bm{\omega}_{jk})$.
The solution is
\begin{equation}
    \label{eqn:CB_sub}
    \widehat{\GE}_{\theta,jk}^\mathrm{CB} = \widehat{\GE}_{\theta,jk}^\mathrm{B} + \frac{r_{jk}}{q_j} \left( \widehat{\GE}_{\theta,j}^\mathrm{CB} - \widehat{B}_j - \overline{\GE}_{\theta,j}^w \right), \quad j=1,\dots,J, \ k=1,\dots,K_j,
\end{equation}
where $\widehat{\GE}_{\theta,jk}^\mathrm{B} = E[\GE_{\theta,jk}(\bm{\omega}_{jk}) \mid \bm{y}_{\mathrm{SR,jk}}]$ is the Bayes estimate (posterior mean) of $\widehat{\GE}_{\theta,jk}(\bm{\omega}_{jk})$, $\overline{\GE}^w_{\theta,j} = \sum_{k=1}^{K_j} w_{jk} \widehat{\GE}^\mathrm{B}_{jk,\theta}$, $r_{jk} = w_{jk} / \phi_{jk}$ and $q_j = \sum_{k=1}^{K_j} w_{jk}^2 / \phi_{jk}$.
Then, we obtain the estimate of the within-subregion inequality for the region $I_j$ as
\begin{equation}
    \label{eqn:within_sub_est}
    \widehat{W}_{j} = \sum_{k=1}^{K_j} w_{jk} \widehat{\GE}_{\theta,jk}^\mathrm{CB} = \sum_{k=1}^{K_j} \lambda_{jk}^{1-\theta} \hat{s}_{jk}^\theta \widehat{\GE}_{\theta,jk}^\mathrm{CB}.
\end{equation}
Using \eqref{eqn:between_est}, \eqref{eqn:within_est}, \eqref{eqn:between_sub_est} and \eqref{eqn:within_sub_est}, we have the estimate of the multilevel decomposition of the GE as follows:
\begin{equation}
    \label{eqn:multi_decomp}
    \begin{split}
            \widehat{\GE}_\theta &= \underbrace{ \sum_{j=1}^J w_j \widehat{\GE}_{\theta,j}^\mathrm{CB} }_{=\widehat{W}} + \widehat{B}, \\
            &= \sum_{j=1}^J w_j \underbrace{ \sum_{k=1}^{K_j} w_{jk} \widehat{\GE}_{\theta,jk}^\mathrm{CB} }_{=\widehat{W}_j} + \sum_{j=1}^J w_j \widehat{B}_j + \widehat{B}.
    \end{split}
\end{equation}

Finally, we summarize the procedures of the proposed estimating method, where steps 1--3 are the same as the procedures summarized in the end of Section \ref{subsec:CB}
\begin{enumerate}
    \item Assume a parametric distribution $f(\cdot \mid \bm{\omega})$ for the entire country $I$ and a prior distribution $\pi(\bm{\omega})$. Then, estimate GEs for the country, denoted by $\widehat{\GE}_\theta$, based on the model $f$, the prior $\pi(\bm{\omega})$ and the grouped data $\bm{y}$.

    \item Assume a parametric distribution $f_j(\cdot \mid \bm{\omega}_j)$ for each region $I_j$ and prior distributions $\pi(\bm{\omega}_j)$. Then, obtain the mean income estimate $\hat{\mu}_j$ for $I_j$ and $\hat{\mu} = \sum_{j=1}^J \lambda_j \hat{\mu}_j$ for the entire country $I$, based on the model $f_j$, the prior $\pi(\bm{\omega}_j)$ and the grouped data $\bm{y}_{\mathrm{R},j}$. Using $\hat{\mu}_j$ and $\hat{\mu}$, estimate the between-region inequality as $\widehat{B}$ defined by \eqref{eqn:between_est}.

    \item Specifying the weights $\phi_j$ in the loss function, obtain the constrained Bayes estimates $\widehat{\GE}_{\theta,j}^\mathrm{CB}$ for the regions defined as \eqref{eqn:CB}. Then, obtain the estimate of the within-region inequality as $\widehat{W}$ defined by \eqref{eqn:within_est}.

    \item Assume a parametric distribution $f_{jk}(\cdot \mid \bm{\omega}_{jk})$ for each subregion $I_{jk}$ and prior distributions $\pi(\bm{\omega}_{jk})$. Then, obtain the mean income estimate $\hat{\mu}_{jk}$ for the subregion $I_{jk}$ and $\Tilde{\mu}_j = \sum_{k=1}^{K_j} \lambda_j \hat{\mu}_{jk}$ for the region $I_j$, based on the model $f_{jk}$, the prior $\pi(\bm{\omega}_{jk})$ and the subregion-level grouped data $\bm{y}_{\mathrm{SR},jk}$. Using $\hat{\mu}_{jk}$ and $\Tilde{\mu}_j$, estimate the between-subregion inequality as $\widehat{B}_j$ defined as \eqref{eqn:between_sub_est}.

    \item Specifying the weights $\phi_{jk}$ in the loss function, obtain the constrained Bayes estimates $\widehat{\GE}_{\theta,jk}^\mathrm{CB}$ for the subregions defined as \eqref{eqn:CB_sub}. Then, obtain the estimate of the within-subregion inequality as $\widehat{W}_j$ defined by \eqref{eqn:within_sub_est}.

    \item Obtain the multilevel decomposition of the GE as \eqref{eqn:multi_decomp}.
\end{enumerate}

\section{Real data analysis of Japanese household income}
\label{sec:realdata}

\subsection{Data summary}
\label{subsec:datasummary}
We apply our proposed method to Japanese income data, which is of the form of grouped data of household income by country, prefecture, and municipality, collected by the Housing and Land Survey (HLS), conducted by the Statistics Bureau of Japan once every five years.
The HLS is a large sample survey, sampling approximately 10 percent of all households.
Since the Statistics Bureau of Japan publishes the estimated frequency distribution, the published estimate of the frequency of the $g$th group multiplied by $0.1$ is treated as the observation $y_g$, $y_{jg}$ and $y_{jkg}$ for the country-level, prefecture-level, and municipality-level data, respectively.
We apply the proposed method to income data for the years 2003, 2008 and 2013.
Although the number of groups $G$ and the boundary values $c_g$ vary slightly from year to year depending on the design of the survey, as a typical example, Table \ref{tab:relative} shows the relative frequency of household income (in millions of yen) for the entire country in 2013 with $G = 10$ and $(c_0,c_1,\dots,c_{G-1},c_G) = (0,1,2,3,4,5,7,10,15,20,+\infty)$.

\begin{table}
    \centering
    \begin{tabular}{lllllllllll} \hline 
         income class&  0--1&  1--2&  2--3&  3--4&  4--5&  5--7&  7--10&  10--15& 15--20& 20 above\\
         relative frequency&  0.068&  0.139&  0.178&  0.157&  0.126&  0.159&  0.110&  0.047& 0.009 & 0.006\\ \hline
    \end{tabular}
    \caption{Relative frequency of household income (in millions of yen) for the entire country in 2013}
    \label{tab:relative}
\end{table}

The assumed sample size across the country for the HLS in 2013 is approximately 5 million households.
The summary statistics of the assumed sample size for the prefecture-level data are presented in Table \ref{tab:samplesize_pref}, and those for the municipality-level data are presented in Table \ref{tab:samplesize_muni}.
Although the sample size for the country-level data is very large, the sample size for the municipality-level data is very small, less than 1072 in a quarter of the municipalities.
We therefore assume a generalized beta distribution of the second kind (GB2) with 4 parameters for the income distribution of the entire country $f$, a Singh--Maddala distribution with 3 parameters for the prefectural income distribution $f_j$, and a lognormal distribution with 2 parameters for the municipal income distribution $f_{jk}$.
In the next subsection, we will review the properties of these parametric income distributions and discuss the methods for their Bayesian estimation.

\begin{table}[h]
    \centering
    \begin{tabular}{llllll} \hline
        Min.   & 1st Qu. & Median & Mean & 3rd Qu. & Max. \\
        21310  & 41445 & 68600 & 110855 & 110555 & 647260 \\ \hline
    \end{tabular}
    \caption{Summary statistics of the assumed sample size for the prefecture-level data in 2013}
    \label{tab:samplesize_pref}
\end{table}

\begin{table}[h]
    \centering
    \begin{tabular}{llllll} \hline
        Min.   & 1st Qu. & Median & Mean & 3rd Qu. & Max. \\
        196  & 1072 & 2109 & 3984 & 5102 & 45145 \\ \hline
    \end{tabular}
    \caption{Summary statistics of the assumed sample size for the municipality-level data in 2013}
    \label{tab:samplesize_muni}
\end{table}

\subsection{Parametric income distributions and their Bayesian estimation}
\label{subsec:para}

A positive random variable $X$ follows the GB2 distribution denoted by $X \sim \mathrm{GB2}(a,b,p,q)$ when $X = bT^{1/a}$ for the power transformation parameter $a>0$ and the scale parameter $b>0$, where $T = U / (1-U)$ for $U \sim \mathrm{Beta}(p,q)$ with shape parameters $p>0$ and $q>0$.
Thus, the cumulative distribution function (cdf) $F_X(\cdot \mid a,b,p,q)$ of $\mathrm{GB2}(a,b,p,q)$ can be expressed using the cdf $F_U(\cdot \mid p,q)$ of $\mathrm{Beta}(p,q)$ as
$$
F_X(x \mid a,b,p,q) = F_U \left( \frac{(x/b)^a}{1 + (x/b)^a} \mid p,q \right).
$$
Therefore, the likelihood function based on grouped data can be easily evaluated because the cdf of the beta distribution is implemented in most statistical software.
GB2 distribution is known to fit the income distribution, so we assume it as the income distribution of the entire country.

To perform a Bayesian estimation, we assume that the four parameters $a,b,p$ and $q$ are independently and identically distributed to $\mathrm{IG}(1,1)$ as the prior distributions, where $\mathrm{IG}(\alpha, \beta)$ denotes the inverse gamma distribution with the density $\pi(x \mid \alpha, \beta) \propto x^{-\alpha - 1} e^{-\beta/x}$.
To draw a sample from the posterior distribution, we use the Metropolis--Hastings (MH) algorithm with a random walk chain \citep{CG00,K16}.
We run the MH algorithm with 10,000 iterations and discard the first 2,000 samples as the burn-in period.

To derive the explicit expression of GEs for the parametric distribution \eqref{eqn:GE_para}, we need to know the $\theta$th moment of $X \sim \mathrm{GB2}(a,b,p,q)$, which is given by
\begin{equation}
    \label{eqn:moment_GB2}
    E[X^\theta] = \frac{ b^\theta \Gamma(p + \theta/a) \Gamma(q - \theta/a) }{ \Gamma(p) \Gamma(q) }, \quad -ap < \theta < aq,
\end{equation}
where $\Gamma(\cdot)$ is the gamma function.
The MLD and the Theil entropy index \eqref{eqn:Theil_para} of $\mathrm{GB2}(a,b,p,q)$ are
\begin{equation}
    \label{eqn:Theil_GB2}
    \begin{split}
            T_0 &= -\frac{\psi(p) - \psi(q)}{a} + \log \frac{\Gamma(p + 1/a) \Gamma(q - 1/a)}{\Gamma(p) \Gamma(q)}, \\
            T_1 &= \frac{\psi(p+1/a) - \psi(q - 1/a)}{a} - \log \frac{\Gamma(p + 1/a)\Gamma(q - 1/a)}{\Gamma(p)\Gamma(q)}, \quad q > 1/a,
    \end{split}
\end{equation}
where $\psi(\cdot)$ is the digamma function.

Although the GB2 distribution fits the income distribution well, the estimation is not stable for the sample size of prefecture-level data due to the large number of parameters.
Therefore, we assume a Singh--Maddala distribution with the parameter $p$ set to 1 in the GB2 distribution, denoted by $\mathrm{SM}(a,b,q) = \mathrm{GB}2(a,b,1,q)$, as the parametric income distribution for each prefecture.
The Singh--Maddala distribution is known to fit Japanese income data.
The $\theta$th moment, MLD, and Theil entropy index of the Singh--Maddala distribution are those of the GB2 distribution given by \eqref{eqn:moment_GB2} and \eqref{eqn:Theil_GB2} with $p=1$.

For the parametric distribution for each municipality, we assume a lognormal distribution $\mathrm{LN}(\xi,\sigma^2)$ with the cdf $\Phi((\log x - \xi) / \sigma)$, where $\Phi(\cdot)$ is the cdf of the standard normal distribution.
Although the lognormal distribution is known to be a poor fit, we assume it to be the parametric distribution for municipalities because the sample size is small for many municipalities, as shown in Table \ref{tab:samplesize_muni}.
The misspecification of the distribution assumption is expected to be corrected when estimating the GEs of municipalities by the constrained Bayes estimators, which use stable prefectural and entire country estimates as benchmarks.
The $\theta$th moment of $\mathrm{LN}(\xi, \sigma^2)$ is $\exp( \xi \theta + \sigma^2 \theta^2 / 2 )$ and the MLD and Theil entropy index are both the MLD and Theil entropy index are $T_0 = T_1 = \sigma^2 / 2$.

\subsection{Comparison with existing methods}
\label{subsec:comparison}
We compare the proposed method with the two existing methods discussed in Section \ref{subsec:GEsub}.
The first method assumes parametric distributions for the entire country, for each prefecture, and for each municipality, separately.
In the same way as in the proposed method, the GB2 distribution is used for the entire country, the Singh--Maddala distribution for each prefecture, and the lognormal distribution for each municipality.
In this method, we estimate the between-prefecture inequality and GEs for each prefecture based on Singh--Maddala distribution and estimate the between-municipality inequality and GEs for each municipality based on lognormal distribution.
A disadvantage of this approach is that it does not yield estimates compatible with the decomposition.
There are two residuals: the residual generated in the process of decomposing the GE of the entire country into the between-prefecture inequality and the within-prefecture inequality, and the residual generated in the process of decomposing the within-prefecture inequality into the between-municipality inequality and the within-municipality inequality.

The second method uses a finite mixture distribution with known weights based on the population shares.
This method assumes a lognormal distribution for the income distribution of each municipality and that the income distribution of each prefecture is a finite mixture of these distributions.
Then, the income distribution of the entire country is assumed to be a mixture of the prefectural income distributions.
In other words, the income distribution of the entire country is a finite mixture of lognormal distributions with known weights.
In this model, the national, prefectural, and municipal GE estimates are compatible with the decomposition.

\begin{table}
    \centering
    \begin{tabular}{lrrrrrr}
        \hline
         & \multicolumn{3}{c}{$\theta = -1$} & \multicolumn{3}{c}{$\theta = 0$} \\
         \cline{2-4} \cline{5-7}
         & Prop & Separate & Mixture & Prop & Separate & Mixture \\
         \hline
        $\widehat{\GE}_\theta$       & 0.50379 & 0.50379 & 0.37288 & 0.27407 & 0.27407 & 0.27618 \\
        $\widehat{B}$         & 0.00713 & 0.00713 & 0.00716 & 0.00692 & 0.00692 & 0.00696 \\
        residual-prefecture   & \multicolumn{1}{c}{--} & 0.02452 & \multicolumn{1}{c}{--} & \multicolumn{1}{c}{--} & 0.00277 & \multicolumn{1}{c}{--} \\
        $\sum_{j=1}^J w_j \widehat{B}_j$  & 0.00480 & 0.00480 & 0.00481 & 0.00477 & 0.00477 & 0.00476 \\
        $\sum_{j=1}^J w_j \widehat{W}_j$  & 0.49186 & 0.36096 & 0.36092 & 0.26238 & 0.26447 & 0.26446 \\
        residual-municipality & \multicolumn{1}{c}{--} & 0.10638 & \multicolumn{1}{c}{--} & \multicolumn{1}{c}{--} & -0.00486 & \multicolumn{1}{c}{--} \\
        \hline
    \end{tabular}
    \caption{Estimates of multilevel decomposition of GEs by three methods for $\theta = -1$ and $\theta = 0$ (MLD) based on the HLS in 2013.}
    \label{tab:multi1}
\end{table}

\begin{table}
    \centering
    \begin{tabular}{lrrrrrr}
        \hline
         & \multicolumn{3}{c}{$\theta = 1$} & \multicolumn{3}{c}{$\theta = 2$} \\
         \cline{2-4} \cline{5-7}
         & Prop & Separate & Mixture & Prop & Separate & Mixture \\
         \hline
        $\widehat{\GE}_\theta$    & 0.24900 & 0.24900 & 0.27574 & 0.31508 & 0.31508 & 0.37078 \\
        $\widehat{B}$         & 0.00676 & 0.00676 & 0.00680 & 0.00664 & 0.00664 & 0.00667 \\
        residual-prefecture   & \multicolumn{1}{c}{--} & 0.00192 & \multicolumn{1}{c}{--} & \multicolumn{1}{c}{--} & 0.00585 & \multicolumn{1}{c}{--} \\
        $\sum_{j=1}^J w_j \widehat{B}_j$  & 0.00487 & 0.00487 & 0.00486 & 0.00512 & 0.00512 & 0.00511 \\
        $\sum_{j=1}^J w_j \widehat{W}_j$  & 0.23737 & 0.26406 & 0.26408 & 0.30332 & 0.35893 & 0.35899 \\
        residual-municipality & \multicolumn{1}{c}{--} & -0.02862 & \multicolumn{1}{c}{--} & \multicolumn{1}{c}{--} & -0.06146 & \multicolumn{1}{c}{--} \\
        \hline
    \end{tabular}
    \caption{Estimates of multilevel decomposition of GEs by three methods for $\theta = 1$ (Theil) and $\theta = 2$ based on the HLS in 2013.}
    \label{tab:multi2}
\end{table}

\begin{figure}
    \centering
    \includegraphics[width=10cm]{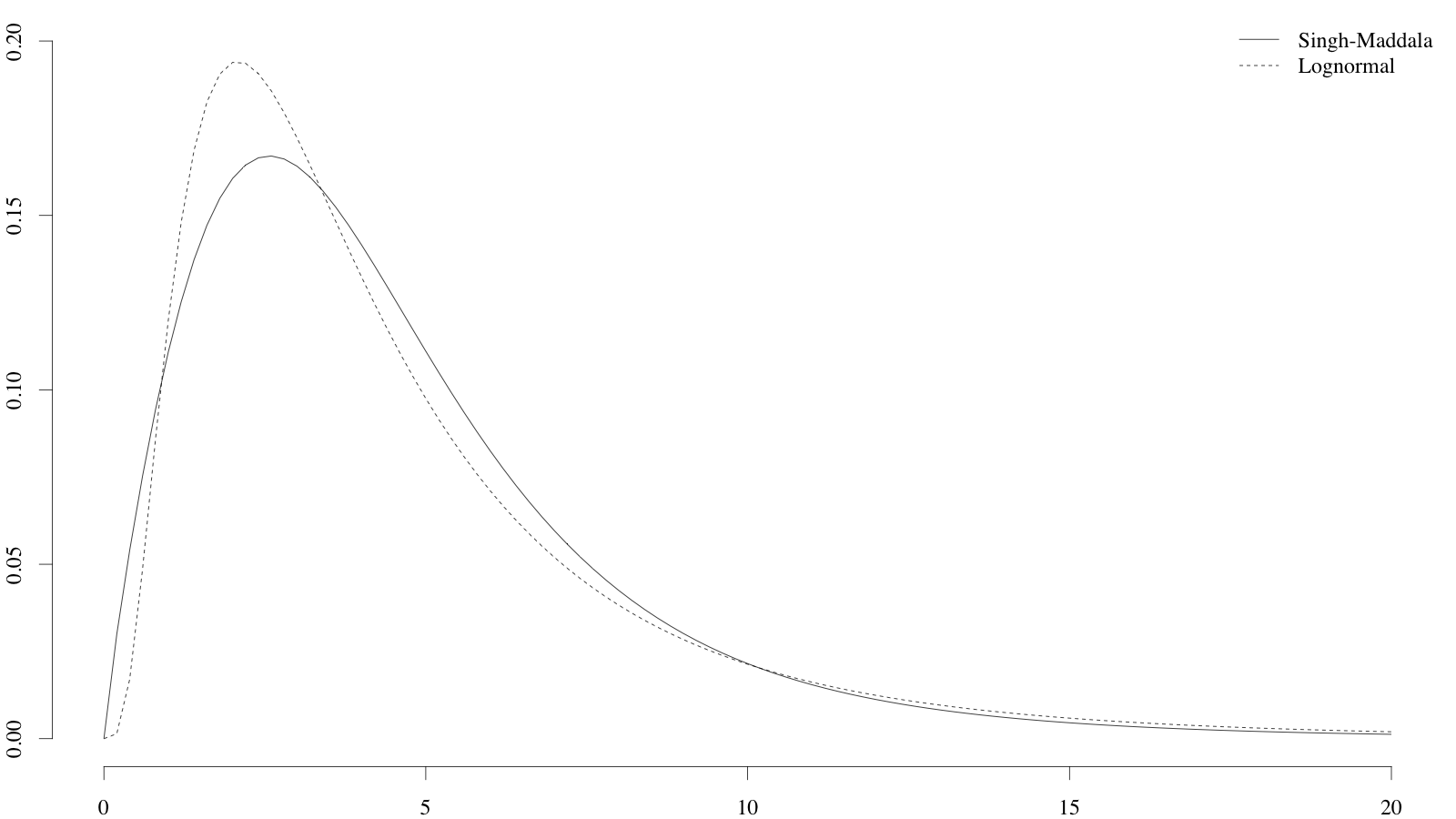}
    \caption{Density estimates of Singh--Maddala and lognormal distributions based on the grouped income data of the entire country.}
    \label{fig:SMvsLN}
\end{figure}

Tables \ref{tab:multi1} and \ref{tab:multi2} show the estimates of multilevel decomposition of GEs \eqref{eqn:multi_decomp} by three methods based on HLS in 2013.
The GE estimates for the entire country using the proposed method and the separate method are exactly the same because both methods assume the GB2 distribution.
For the same reason, the estimates of between-prefecture inequality and of between-municipality inequality by the proposed method and the separate method are also exactly the same, respectively.
However, the estimates of within-municipality inequality differ remarkably between the proposed and the separate methods, especially for $\theta = -1$ and $\theta = 2$.
This is because the GE is more sensitive to changes in the tail of the distribution when the value of $\theta$ is large or small.
For $\theta = -1$, which is the case that the sensitivity to the lower tail of the distribution is high, the separate method tends to underestimate the GEs of municipalities (i.e., within-municipality inequality).
As a result, the residual resulting from the decomposition into the between-municipality and within-municipality inequality is positive.
On the other hand, as $\theta$ increases, the sensitivity to the upper tail of the distribution increases.
In contrast, in this case, the separate method tends to overestimate the GEs of municipalities (i.e., within-municipality inequality).
This result suggests that the lognormal distribution underestimates inequality in the low-income group and overestimates inequality in the high-income group when applied to the Japanese income distribution.
Figure \ref{fig:SMvsLN} shows the results of the estimate of the density functions of the Singh--Maddala and lognormal distributions based on the grouped data for the entire country.
At the lower tail of the distribution, the density of the lognormal distribution is smaller than that of the Singh--Maddala distribution, while the opposite is true at the upper tail of the distribution.
Since the proposed method modifies the GE estimates of municipalities based on the lognormal distribution using the GE estimate of prefectures based on the Singh--Maddala distribution as a benchmark, it is consistent with this result.
\footnote{
The Singh--Maddala distribution seems to fit comparatively well for prefectures because the residual of prefectures is relatively small.
However, it is worthwhile to examine the case that the fit of the hypothetical distribution is poor because it is not well known whether our proposed method works.
Therefore, we examine the case where the lognormal distribution is assumed to be a hypothetical distribution for prefectures.
These results are summarized in Appendix \ref{sec:App_DataAnalysis}.
}

The misspecification of fitting a lognormal distribution to the municipal income distribution also has serious negative consequences for the method that assumes a lognormal mixture distribution.
For the estimation of between-prefecture and between-municipality inequality, the proposed method and the mixture method produce close estimates.
On the other hand, similar to the separate method, the mixture method underestimates within-municipality inequality for small $\theta$ and overestimates it for large $\theta$.
Since Japan consists of a very large number of municipalities, the GE estimates for the entire country based on the mixture method differ substantially from the one based on the proposed method, especially when $\theta = -1$ and $\theta = 2$.
Because the GB2 distribution with four parameters is sufficiently flexible, the fit to the national income distribution of Japan is considered good.
Therefore, the GE estimates for the entire country based on the GB2 distribution are reliable, and we are skeptical of the accuracy of the mixture method, which yields estimates that differ largely from GE estimates based on GB2 distribution.

\subsection{Results of empirical analysis}
\label{subsec:empirical}
\begin{figure}
    \centering
    \includegraphics[width=10cm]{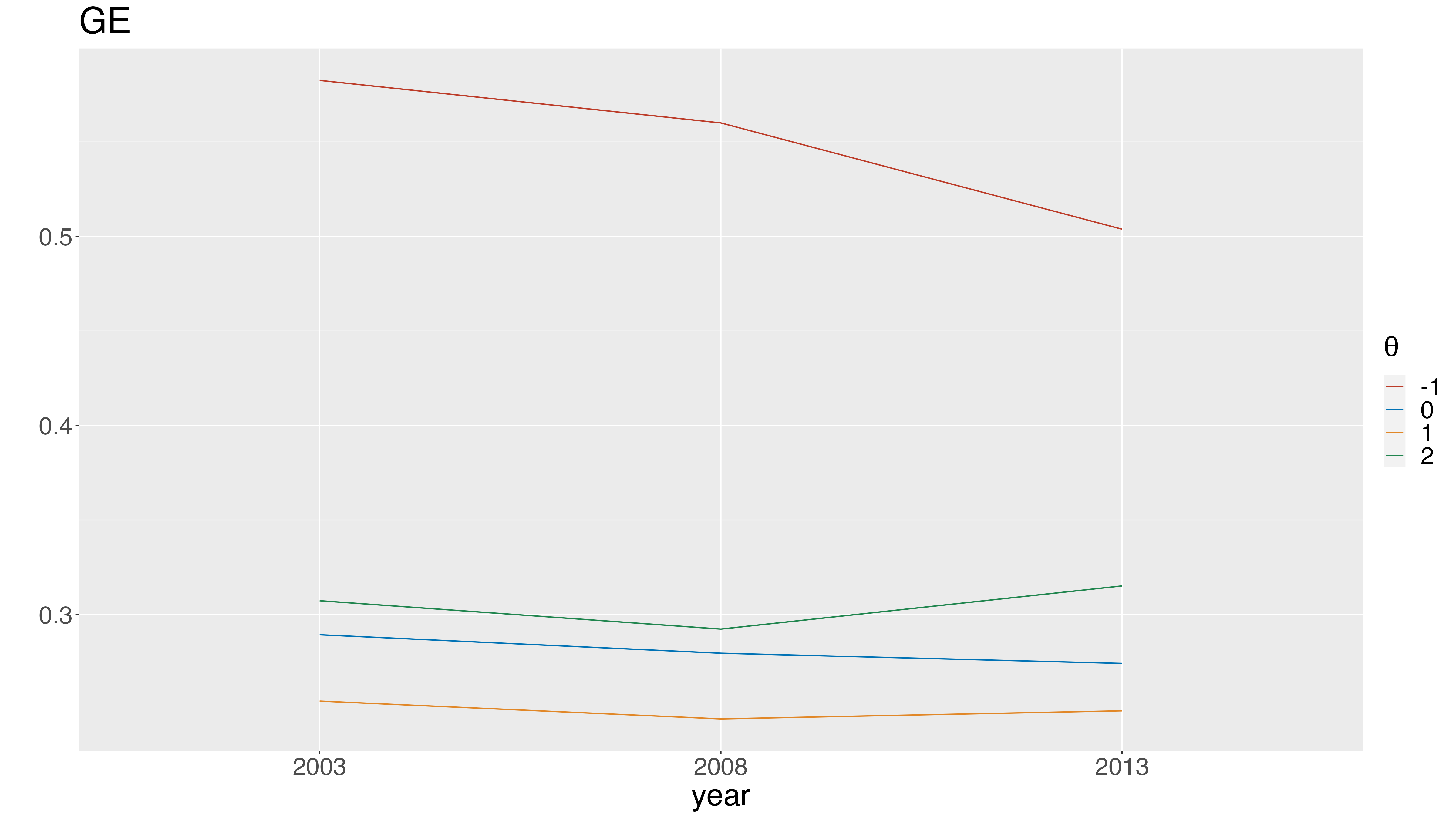}
    \caption{GE for different $\theta$}
    \label{fig:GE_entire_country}
\end{figure}

At the end of this section, we analyze the empirical results based on the multilevel decomposition.
First, we examine the estimates of generalized entropy (GE) indices for the entire country over time, using the GB2 distribution.
Figure \ref{fig:GE_entire_country} presents the estimated GE indices, $\widehat{\GE}_\theta$, for four values of $\theta$ over three years: 2003, 2008, and 2013.
In the figure, a monotonic decrease is observed when $\theta = -1$, where GE is more sensitive to the lower tail of the income distribution.
However, no clear trend is evident for $\theta = 0, 1$, and $2$. Table \ref{tab:GB2_country} reports the posterior estimates for the GB2 distributions.
The results indicate that the income inequality measured by $\theta = -1$ is the highest among all values of $\theta$ each year, corresponding to an increase in the parameter $a$ and a decrease in $q$.
This suggests that the GB2 distribution is not only a flexible tool for modeling income distribution, but also enables a detailed analysis of income inequality through GE indices based on estimated parameters.

\begin{table}
    \centering
    \caption{Posterior estimates for GB2 distribution (Country)}
    \label{tab:GB2_country}
    \begin{tabular}{lrrrrrr}
        \hline
        Year & \multicolumn{2}{c}{2003} & \multicolumn{2}{c}{2008} & \multicolumn{2}{c}{2013}\\
        \hline
        $a$  & 1.916 & (0.011) & 2.024 & (0.008) & 2.119 & (0.009)\\
        $b$  & 8.196 & (0.037) & 7.854 & (0.028) & 6.192 & (0.017)\\
        $p$  & 0.875 & (0.008) & 0.828 & (0.004) & 0.840 & (0.005)\\
        $q$  & 2.459 & (0.028) & 2.341 & (0.019) & 1.904 & (0.014)\\
        \hline
    \end{tabular}
    
    Note: Posterior means and standard deviations (in parentheses) are shown.
\end{table}

Next, we examine the decomposition of the GE indices for the entire country into within-prefecture and between-prefecture inequality, as expressed in \eqref{eqn:decomp}.
As shown in Tables \ref{tab:multi1} and \ref{tab:multi2}, the between-prefecture inequality remains relatively small compared to the within-prefecture inequality across all values of $\theta$.
This finding suggests that the overall income inequality in the country is primarily driven by disparities within individual prefectures rather than differences between them.
To further investigate this, we analyze within-prefecture inequality in greater detail by examining the GE indices at the prefectural level.

\begin{table}
    \centering
    \begin{tabular}{lrrrrrr}
        \hline
         & \multicolumn{3}{c}{$\theta = -1$} & \multicolumn{3}{c}{$\theta = 0$} \\
         \cline{2-4} \cline{5-7}
         & 2003 & 2008 & 2013 & 2003 & 2008 & 2013 \\
         \hline
        $\widehat{\GE}_\theta$       & 0.58255 & 0.56003 & 0.50379 & 0.28923 & 0.27946 & 0.27407 \\
        $\widehat{B}$         & 0.00714 & 0.00889 & 0.00713 & 0.00688 & 0.00855 & 0.00692 \\
        $\sum_{j=1}^J w_j \widehat{B}_j$  & 0.00467 & 0.00462 & 0.00480 & 0.00462 & 0.00445 & 0.00477 \\
        $\sum_{j=1}^J w_j \widehat{W}_j$  & 0.57074 & 0.54651 & 0.49186 & 0.27773 & 0.26647 & 0.26238 \\
        \hline
    \end{tabular}
    \caption{Estimates of multilevel decomposition for $\theta = -1$ and $\theta = 0$ (MLD) based on the HLS in 2003, 2008 and 2013.}
    \label{tab:multi1}
\end{table}

\begin{table}
    \centering
    \begin{tabular}{lrrrrrr}
        \hline
         & \multicolumn{3}{c}{$\theta = 1$} & \multicolumn{3}{c}{$\theta = 2$} \\
         \cline{2-4} \cline{5-7}
         & 2003 & 2008 & 2013 & 2003 & 2008 & 2013 \\
         \hline
        $\widehat{\GE}_\theta$    & 0.25414 & 0.24473 & 0.24900 & 0.30725 & 0.29224 & 0.31508 \\
        $\widehat{B}$         & 0.00667 & 0.00828 & 0.00676 & 0.00651 & 0.00809 & 0.00664 \\
        $\sum_{j=1}^J w_j \widehat{B}_j$  & 0.00473 & 0.00441 & 0.00487 & 0.00501 & 0.00451 & 0.00512 \\
        $\sum_{j=1}^J w_j \widehat{W}_j$  & 0.24274 & 0.23204 & 0.23737 & 0.29573 & 0.27965 & 0.30332 \\
        \hline
    \end{tabular}
    \caption{Estimates of multilevel decomposition of GEs for $\theta = 1$ (Theil) and $\theta = 2$ based on the HLS in 2003, 2008 and 2013.}
    \label{tab:multi2}
\end{table}

\begin{figure}
    \centering
    \includegraphics[width=5cm]{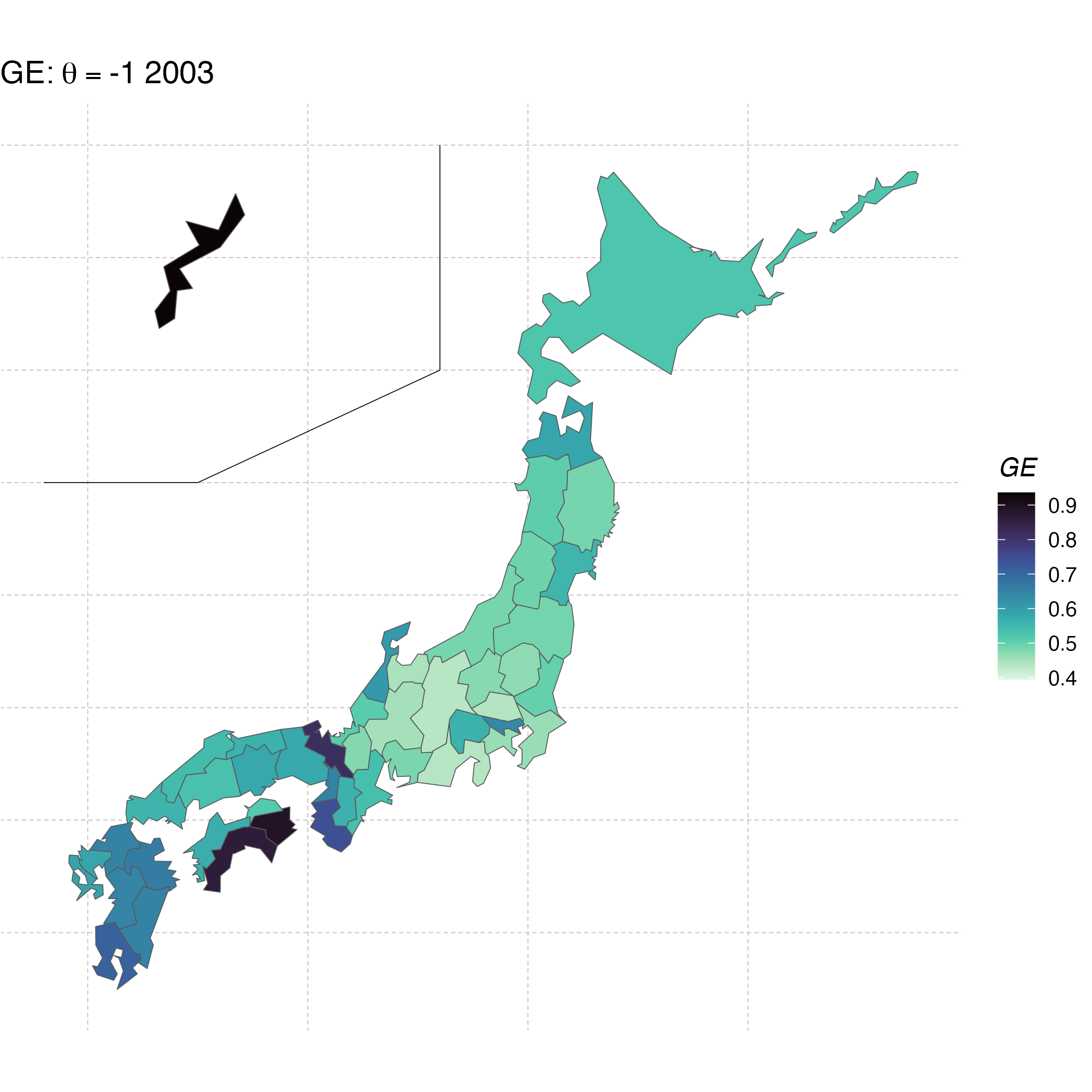}
    \includegraphics[width=5cm]{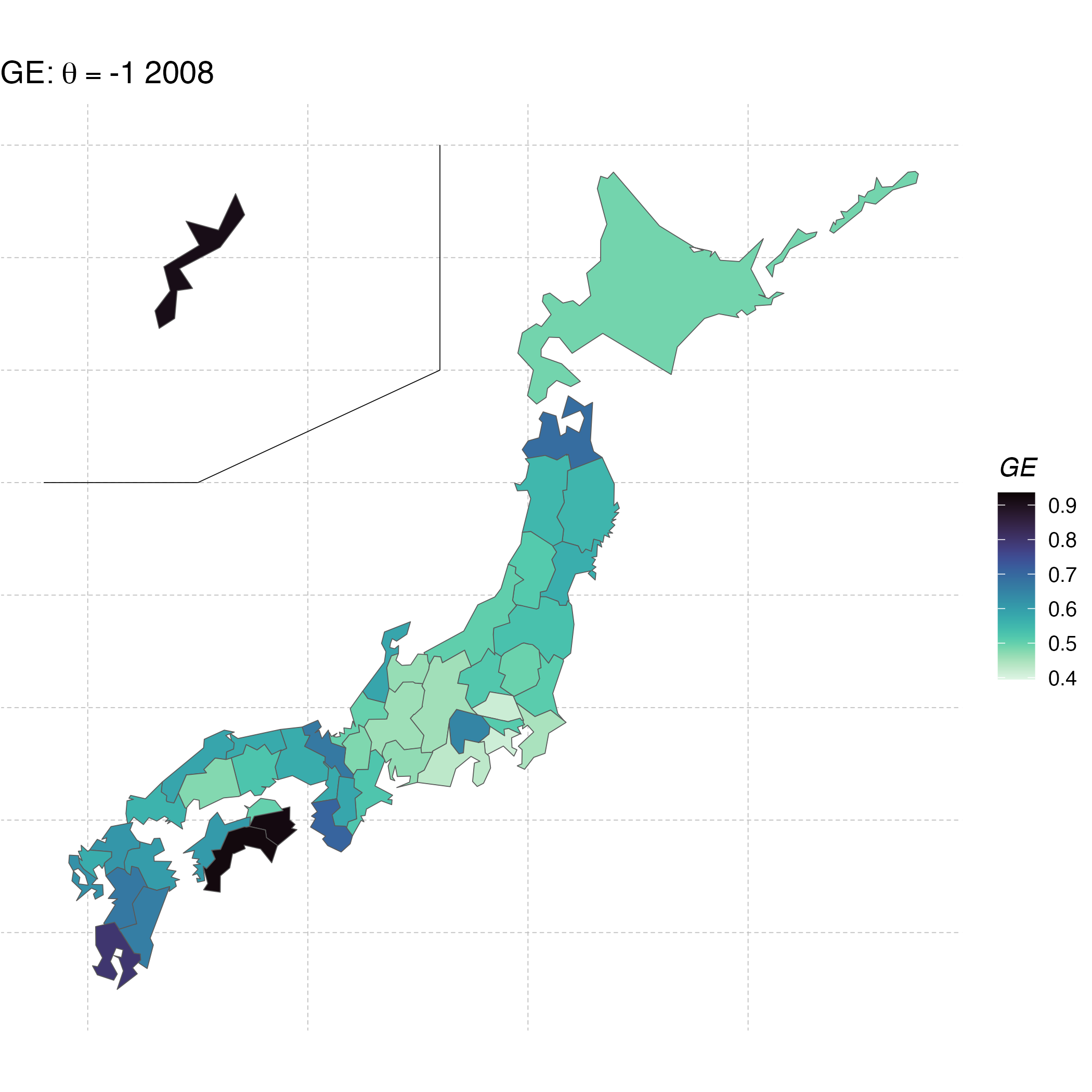}
    \includegraphics[width=5cm]{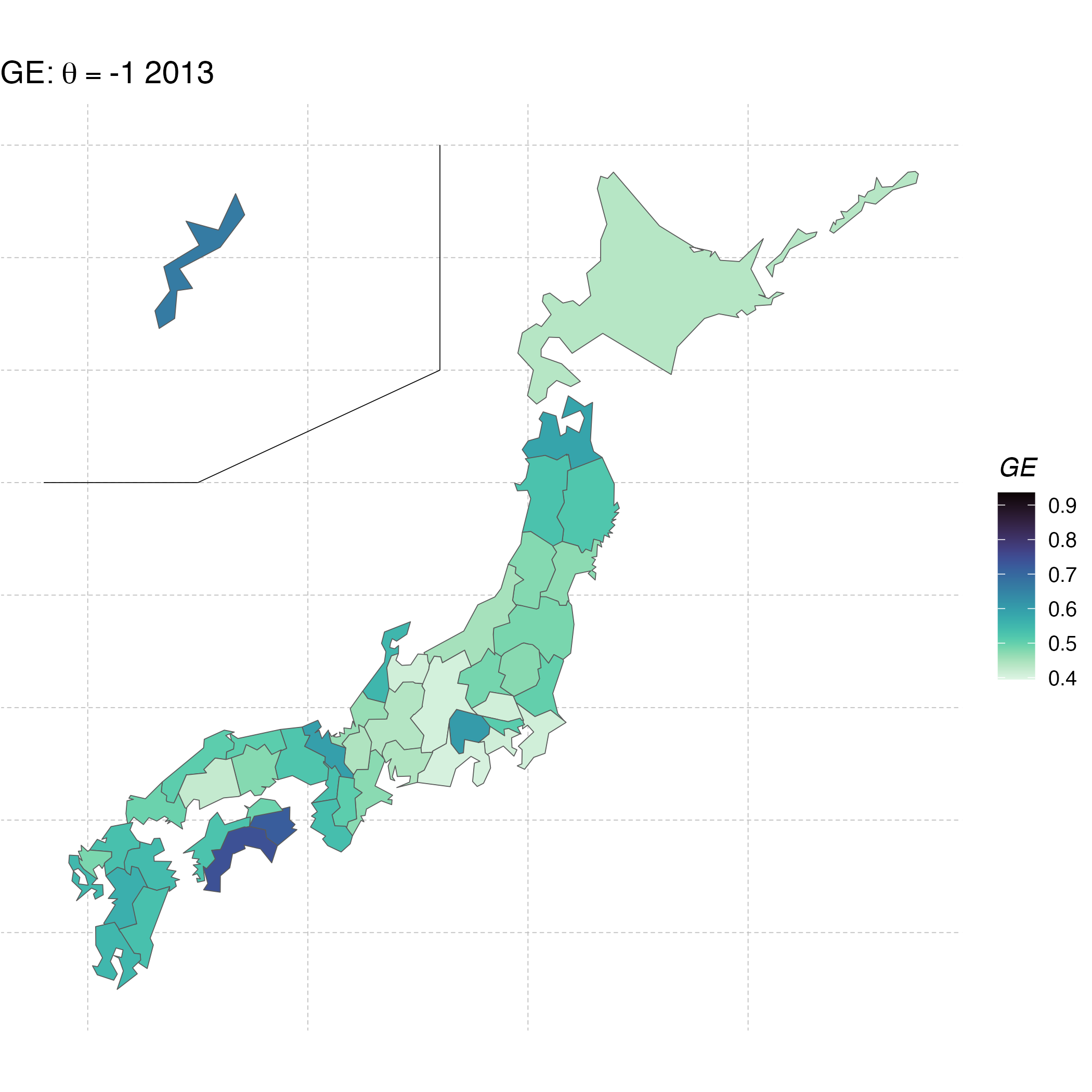}

    \includegraphics[width=5cm]{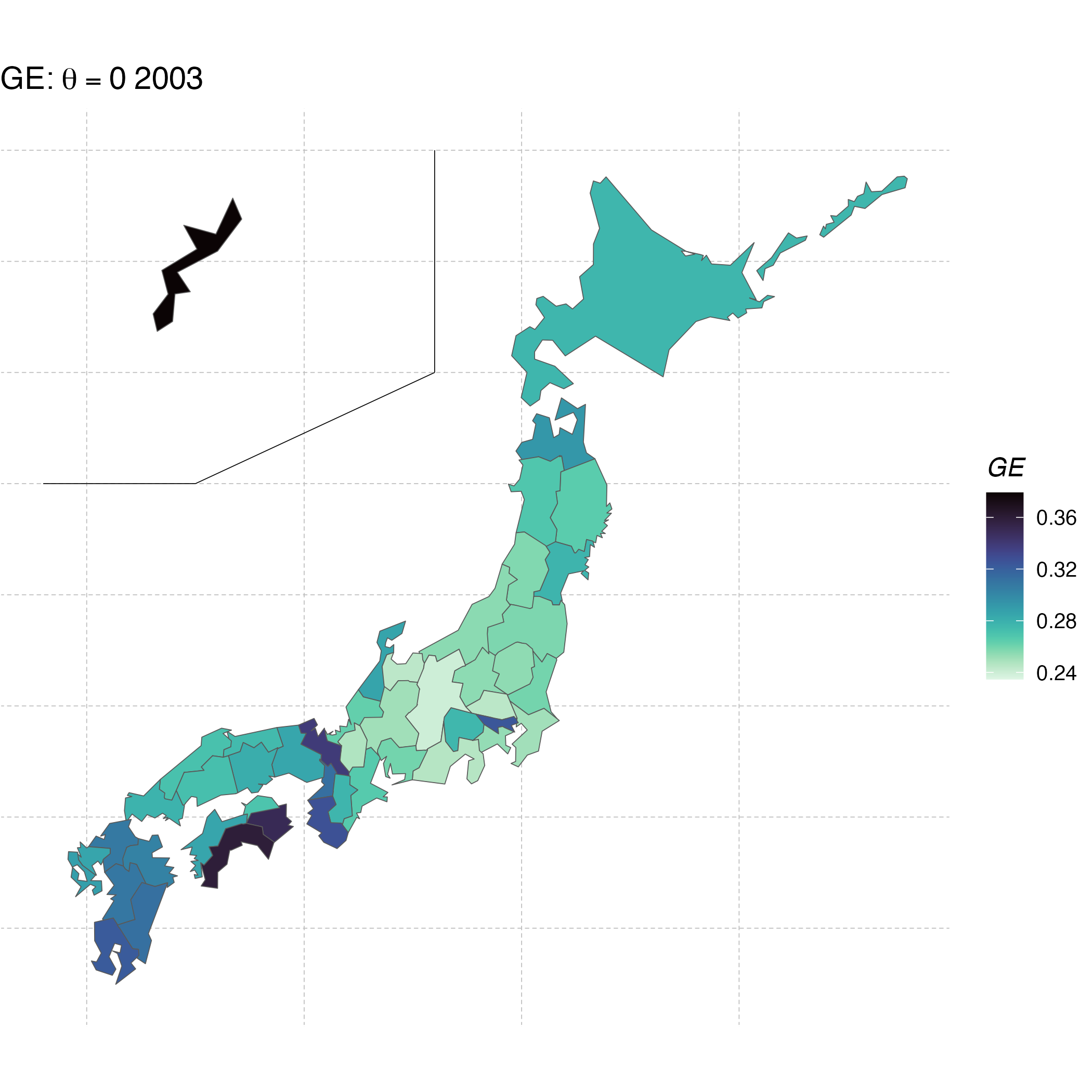}
    \includegraphics[width=5cm]{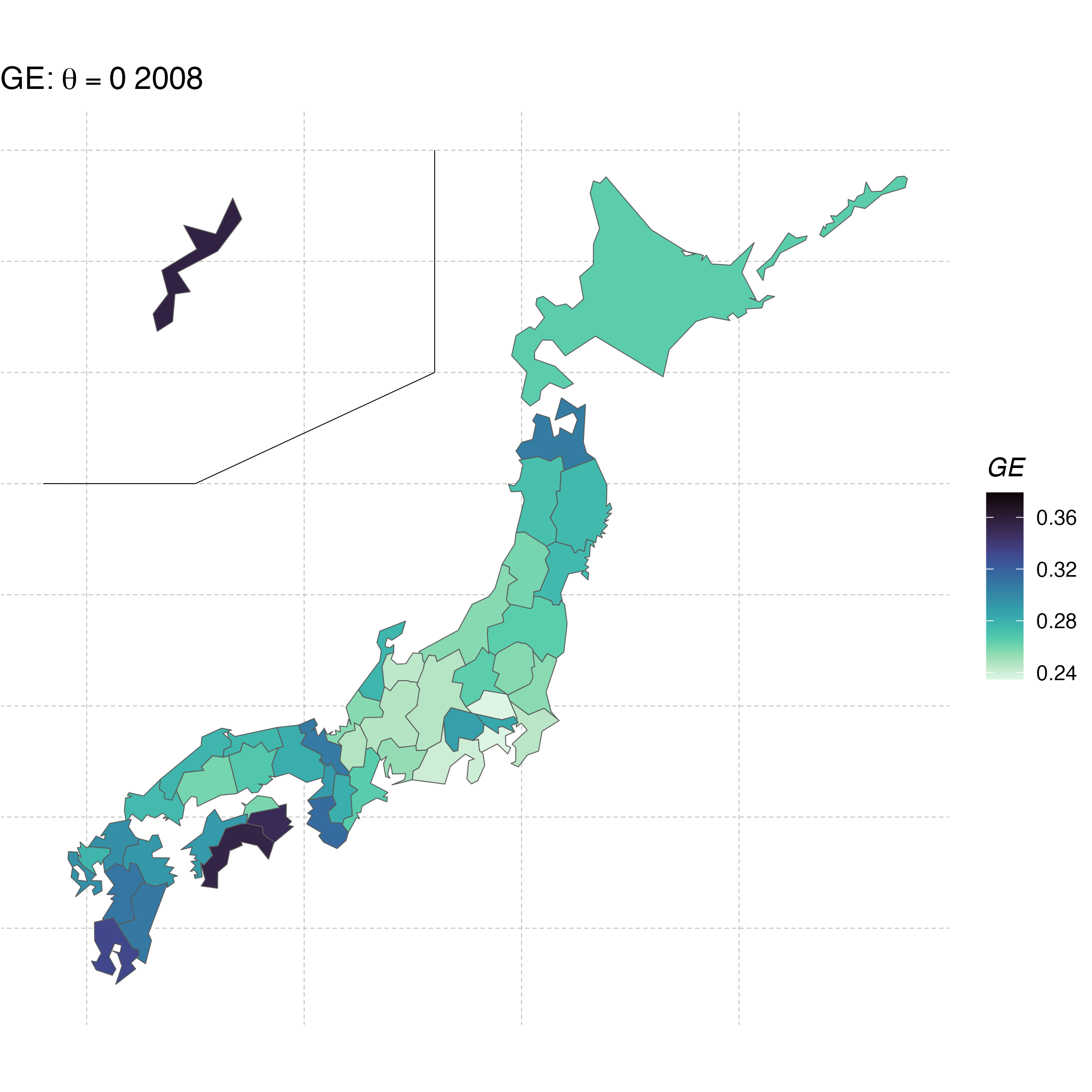}
    \includegraphics[width=5cm]{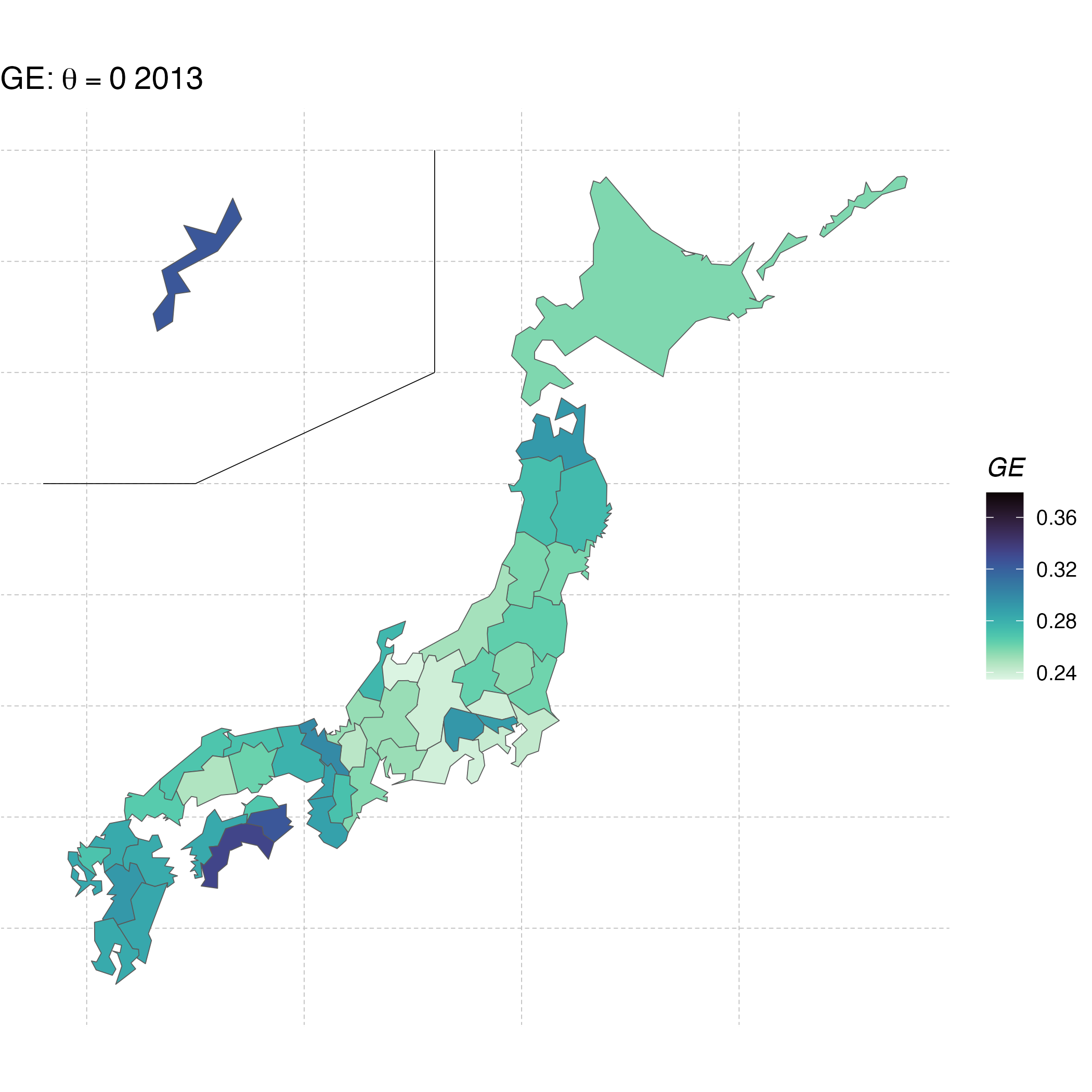}

    \includegraphics[width=5cm]{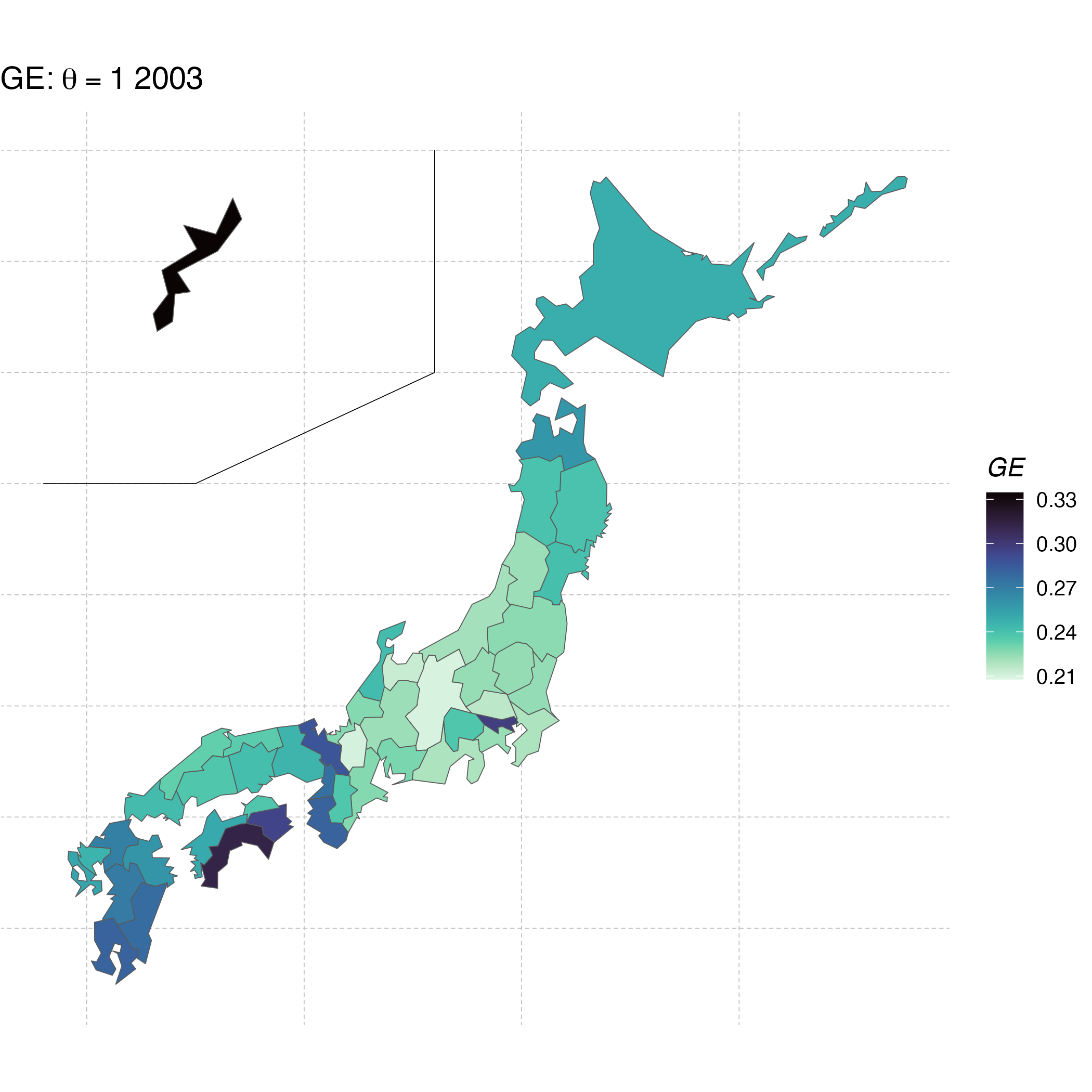}
    \includegraphics[width=5cm]{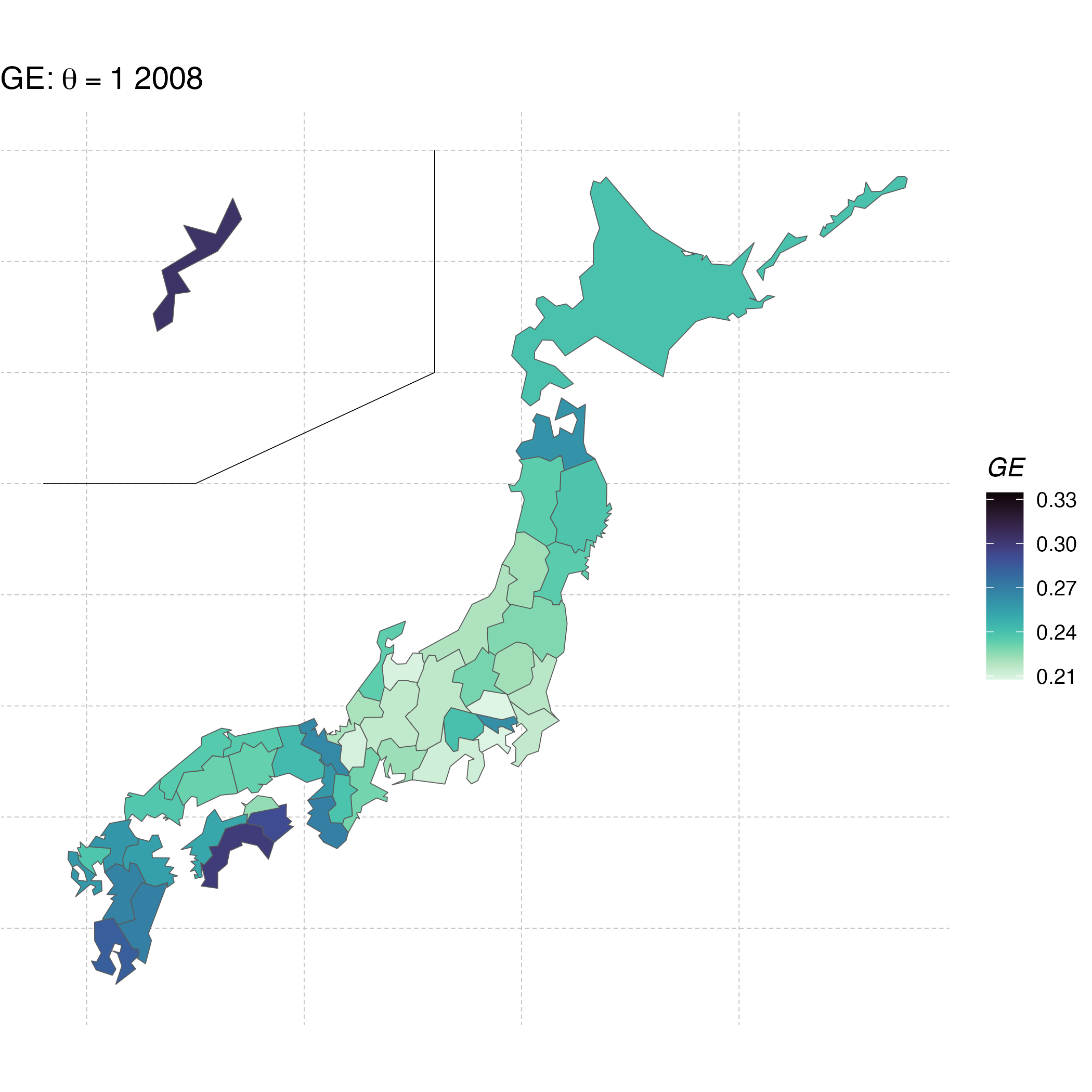}
    \includegraphics[width=5cm]{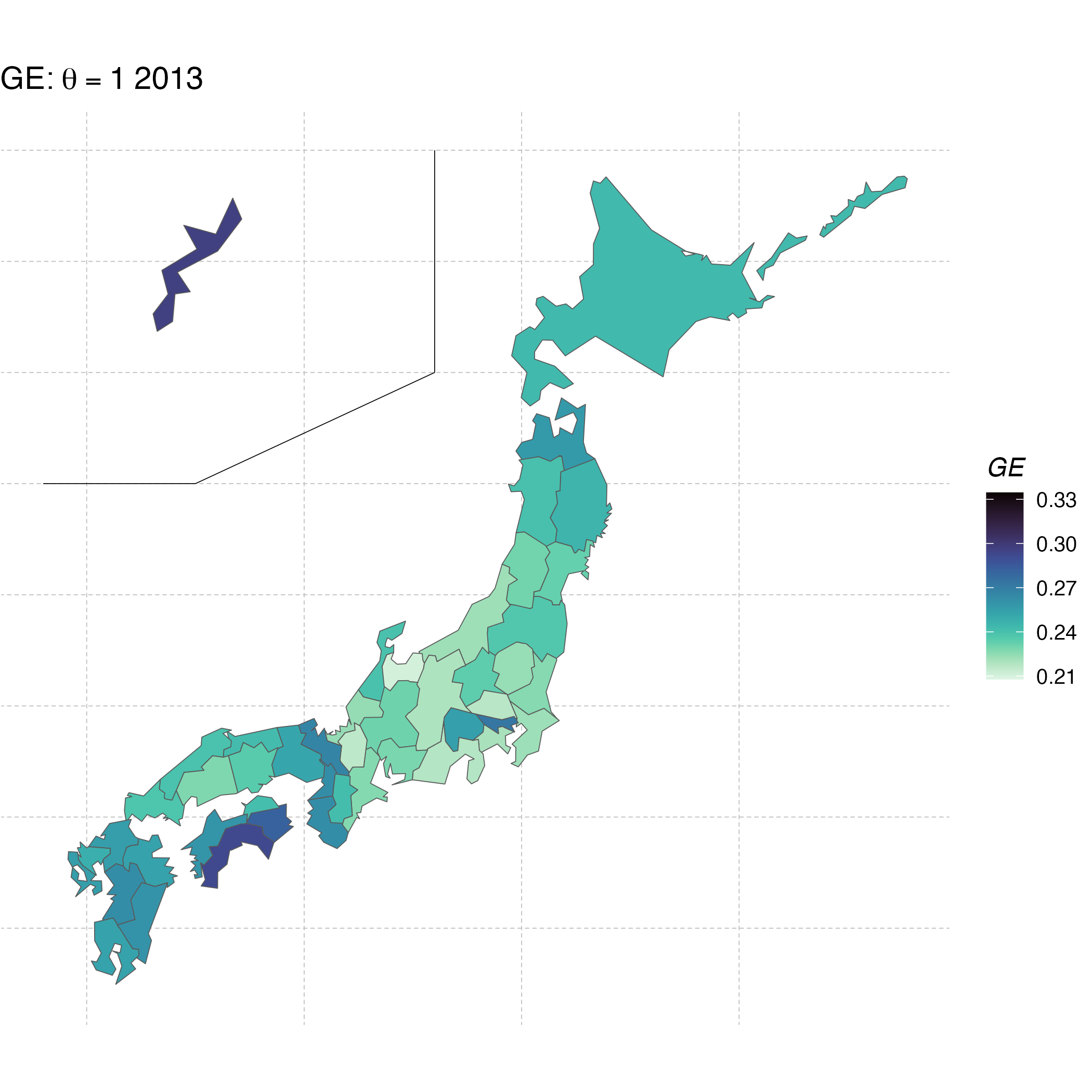}

    \includegraphics[width=5cm]{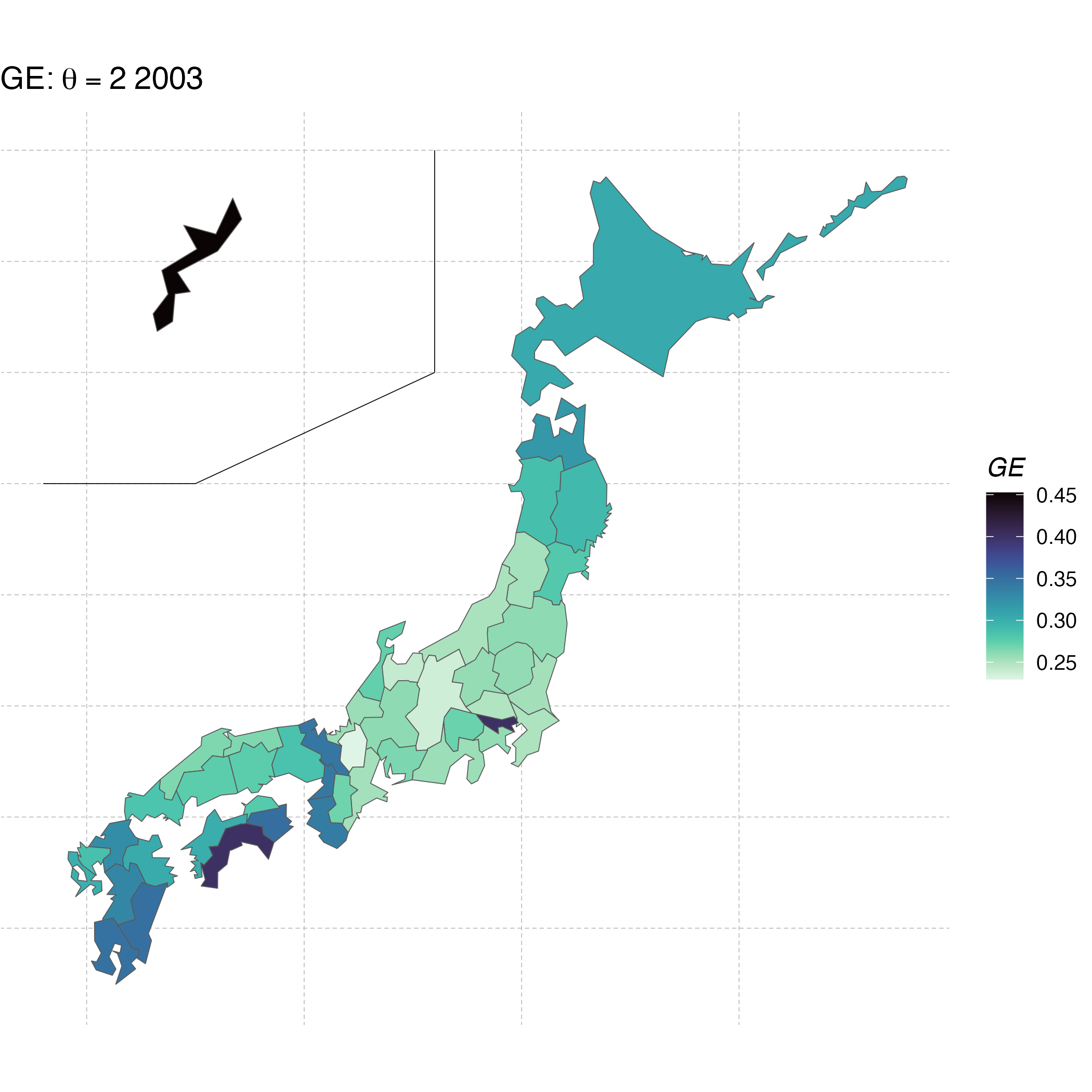}
    \includegraphics[width=5cm]{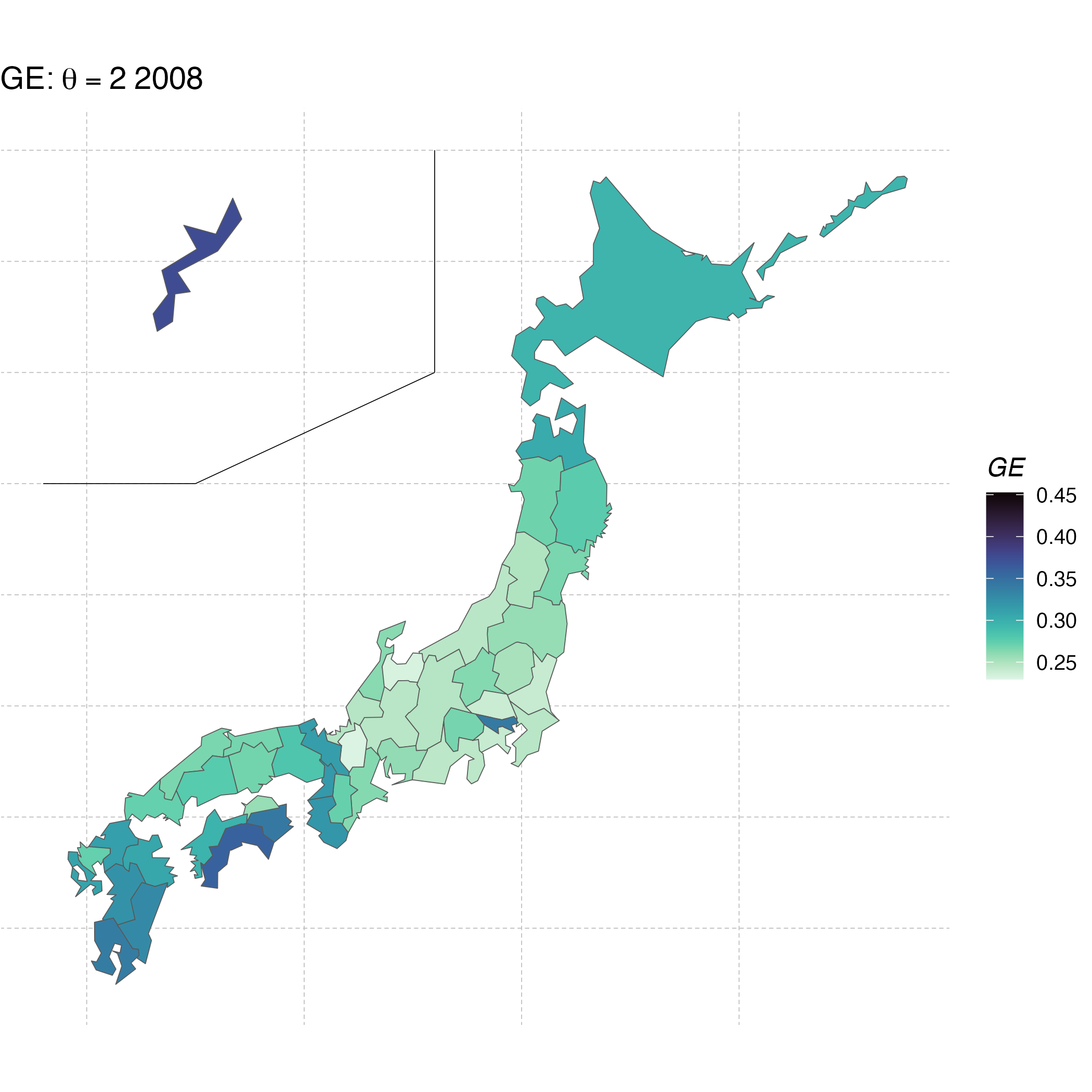}
    \includegraphics[width=5cm]{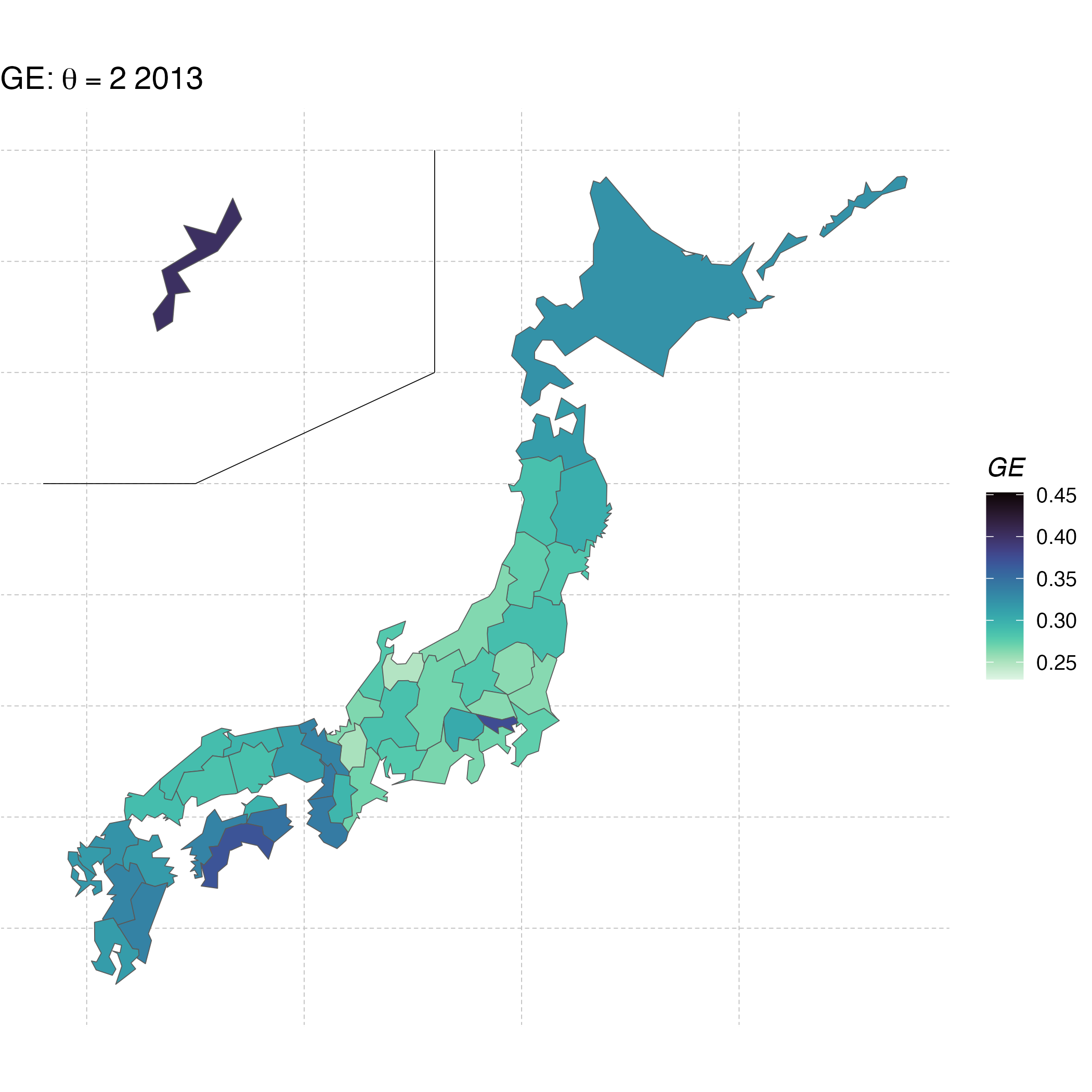}
    \caption{$\widehat{\GE}^{\mathrm{CB}}_{\theta,j}$ for different $\theta$}
    \label{fig:GEmap}
\end{figure}

The constrained Bayes estimates of the GE indices for 47 prefectures, $\widehat{\GE}^\mathrm{CB}_{\theta,j}$, as defined in \eqref{eqn:CB}, are presented in Figure \ref{fig:GEmap} using choropleth maps.
Examining the changes over time, we observe that the shading of the color becomes lighter when $\theta = -1$, whereas for other values of $\theta$, the shading remains relatively stable.
This trend aligns with the overall changes in national inequality shown in Figure \ref{fig:GE_entire_country}.
On the other hand, when comparing prefectures within the same year, we find clear variations in the magnitude of inequality between regions.
Furthermore, as $\theta$ increases from $-1$ to $2$, some prefectures exhibit minimal changes in inequality, while others experience significant fluctuations.
These findings suggest that the overall income inequality in Japan is primarily influenced by a subset of prefectures with high inequality.
This is reflected in the relatively small differences in color shading across most prefectures, with the notable exception of Tokyo, which stands out.

\begin{figure}
     \centering
     \includegraphics[width=5cm]{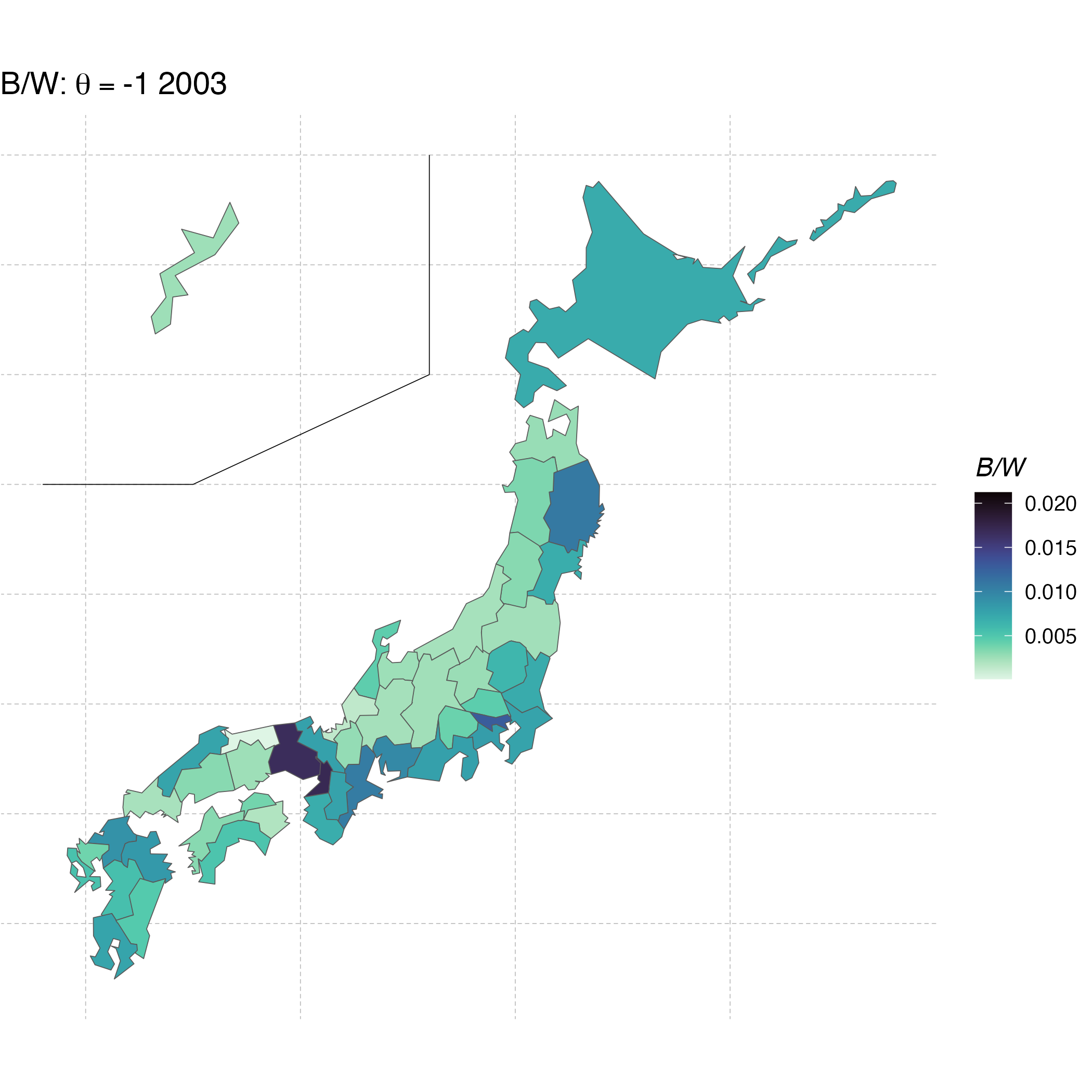}
     \includegraphics[width=5cm]{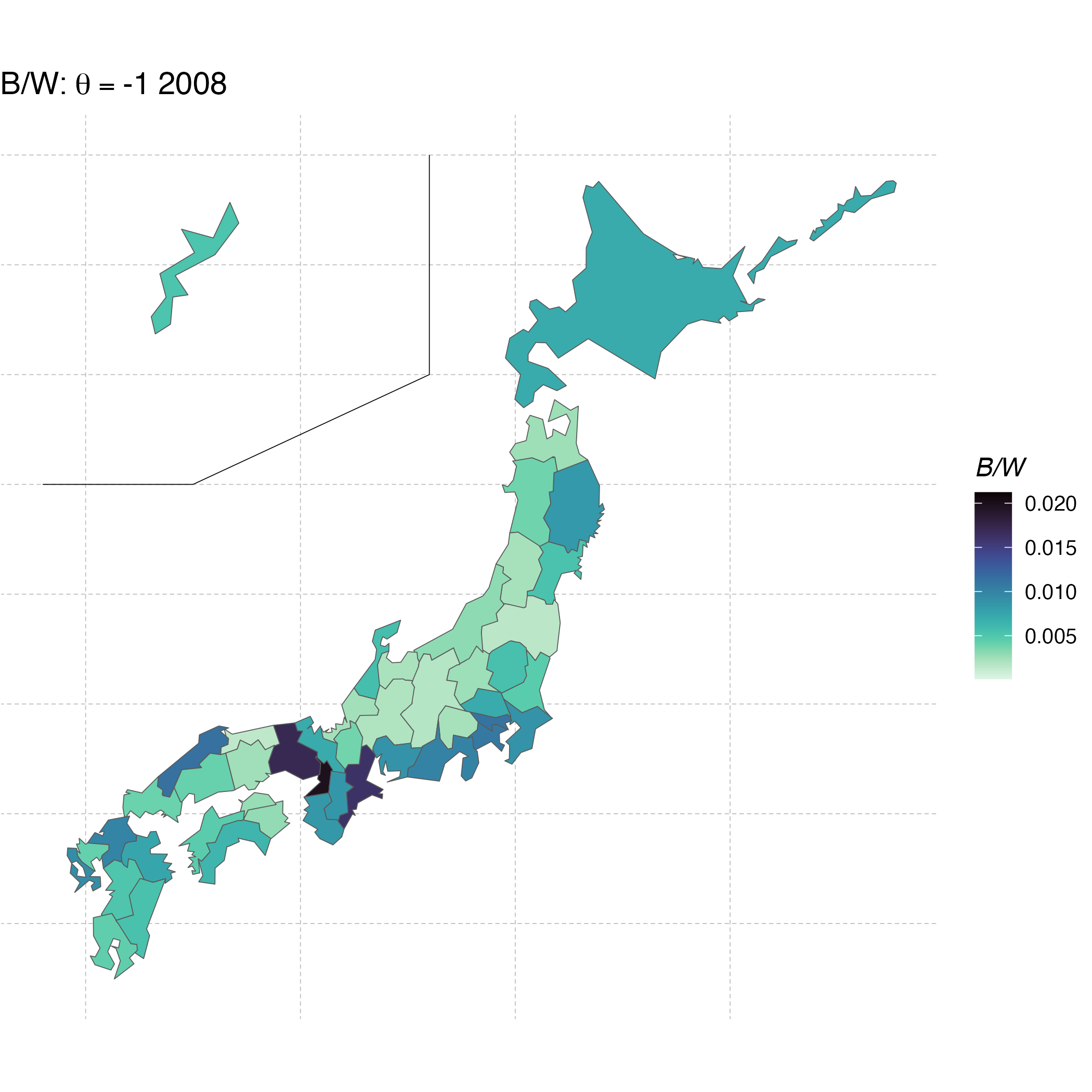}
     \includegraphics[width=5cm]{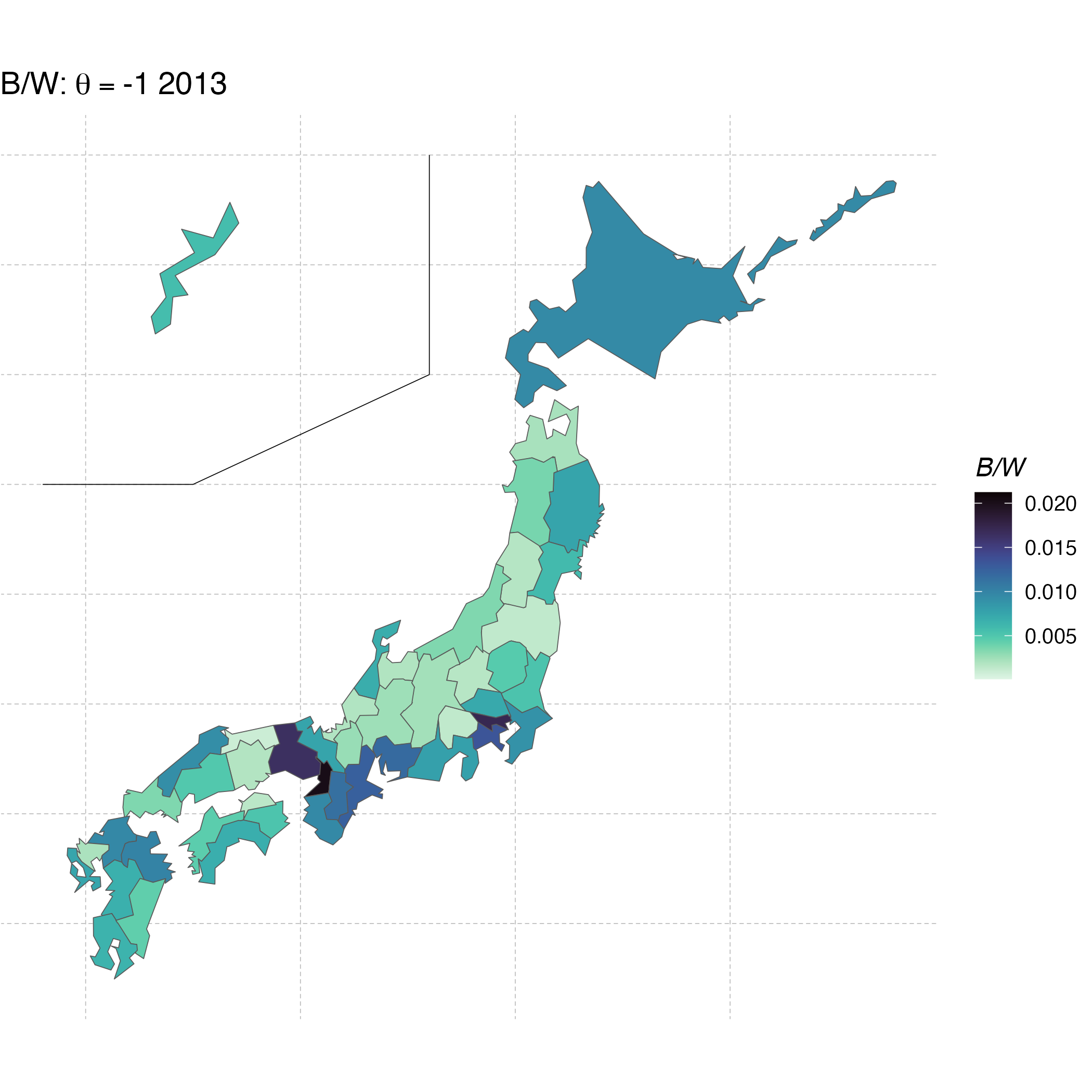}

     \includegraphics[width=5cm]{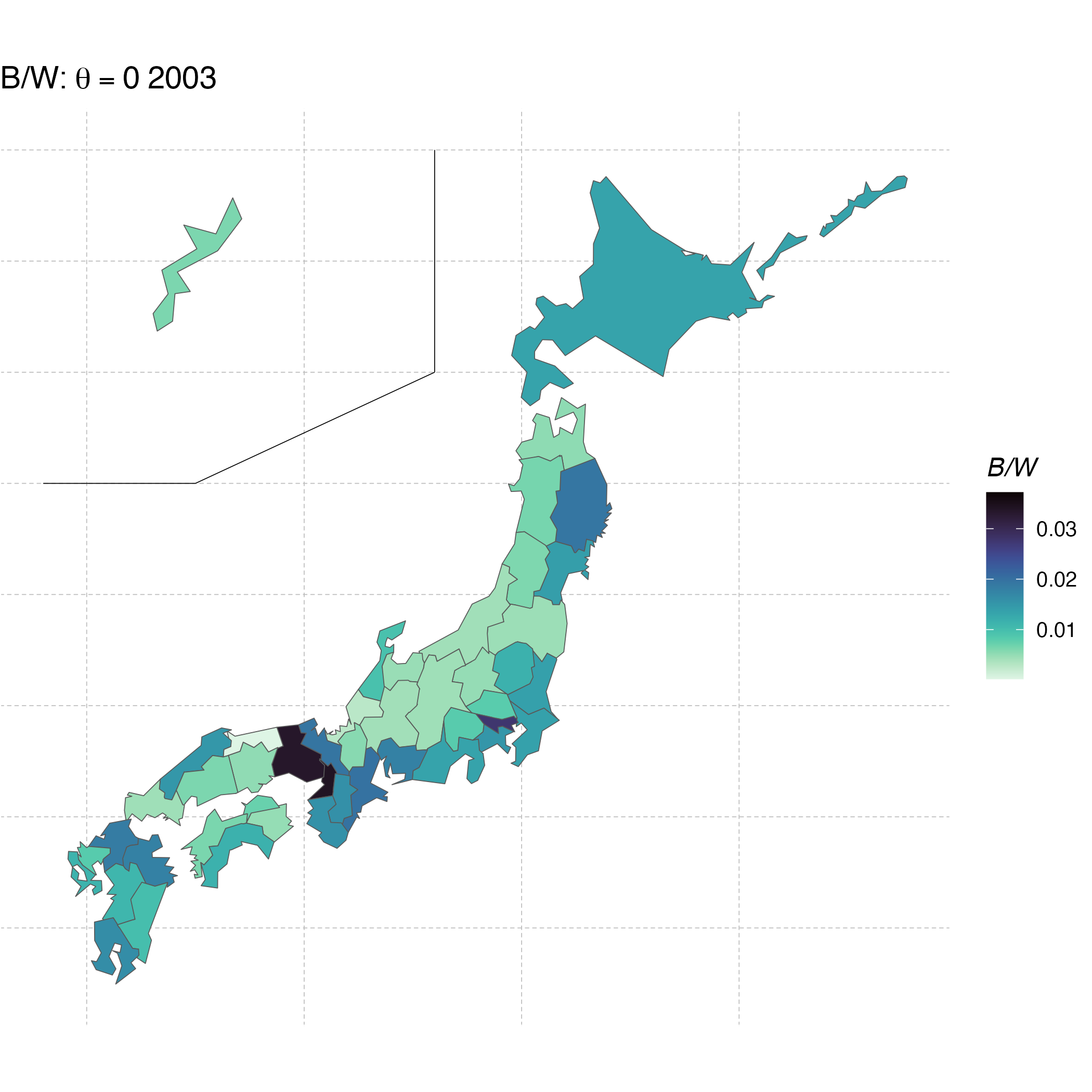}
     \includegraphics[width=5cm]{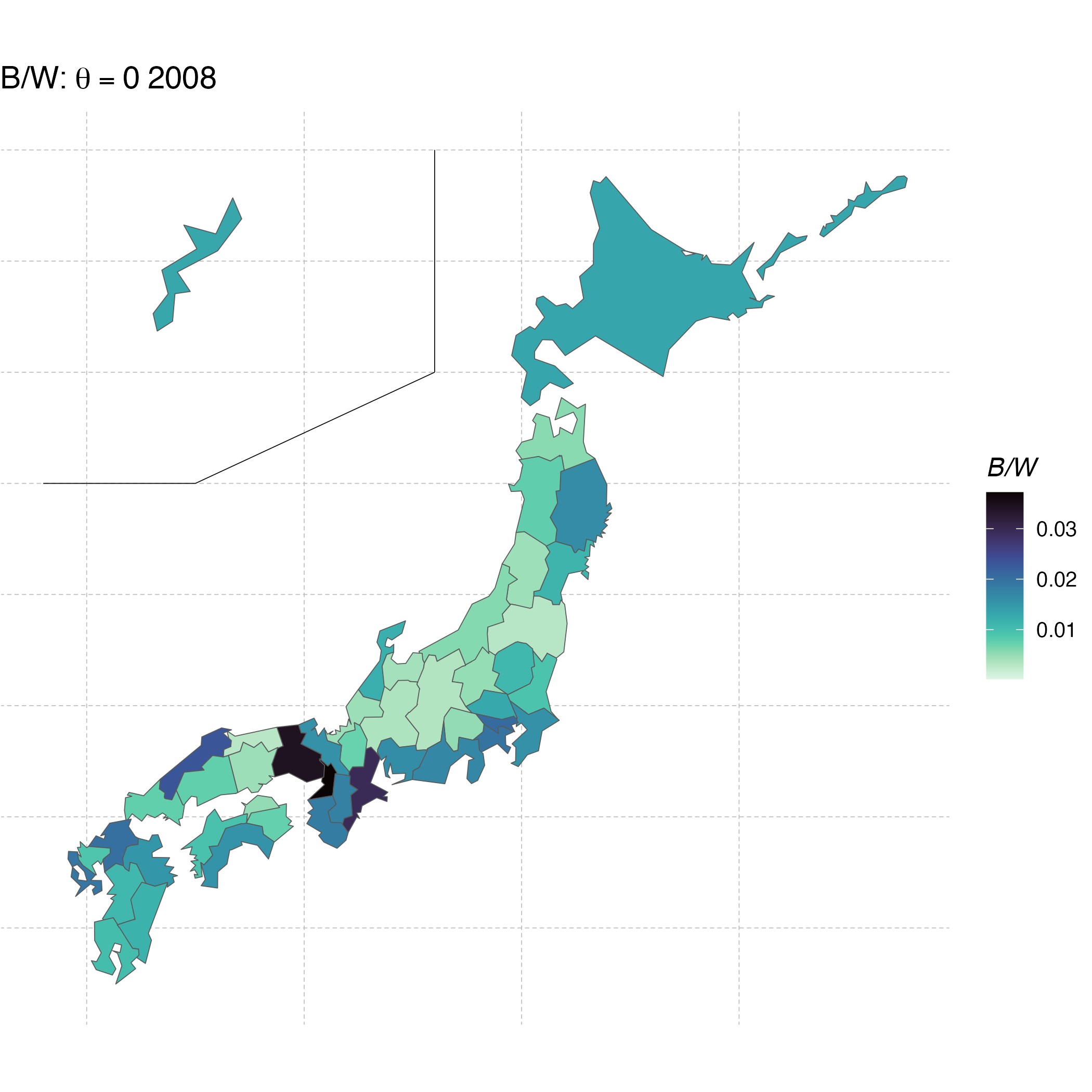}
     \includegraphics[width=5cm]{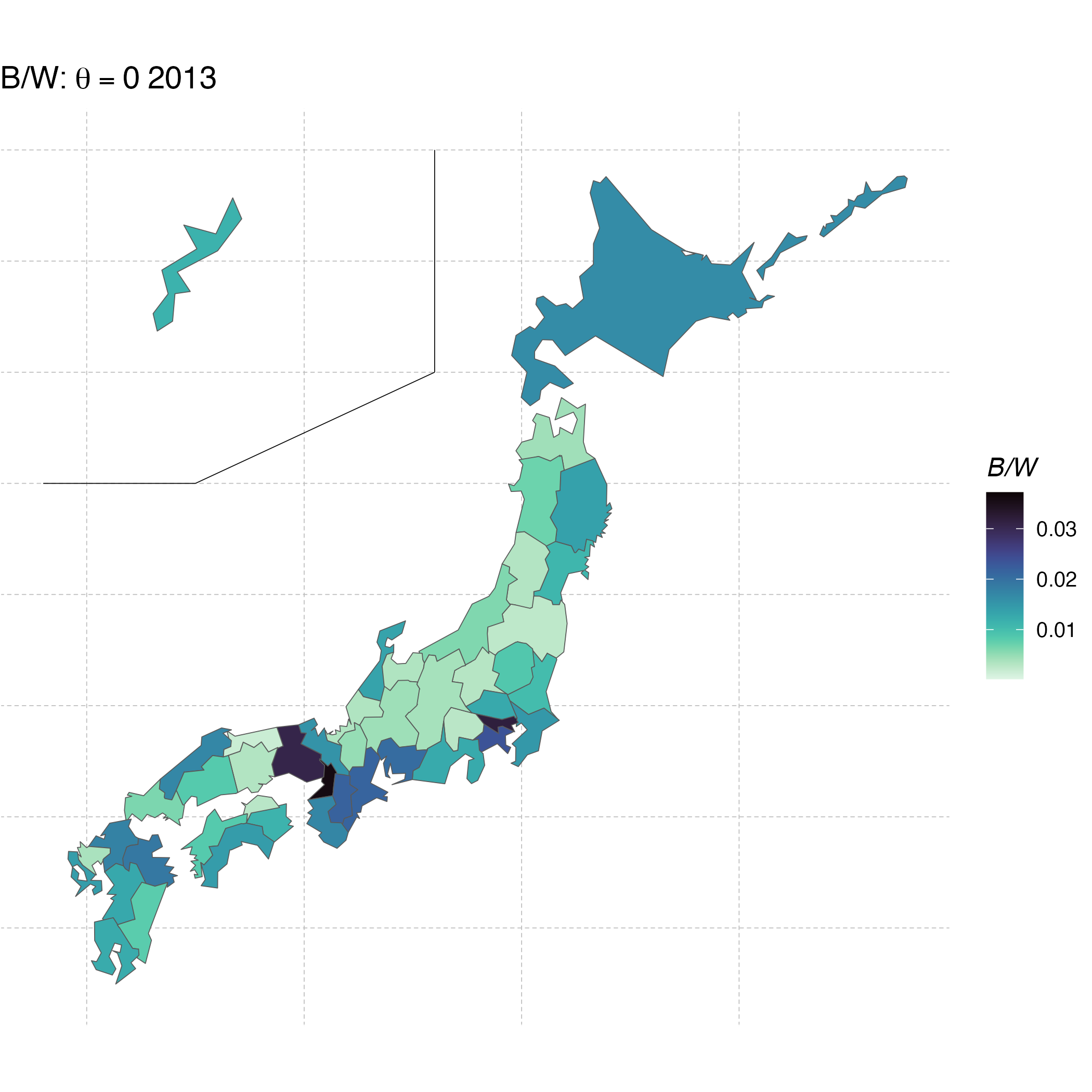}

     \includegraphics[width=5cm]{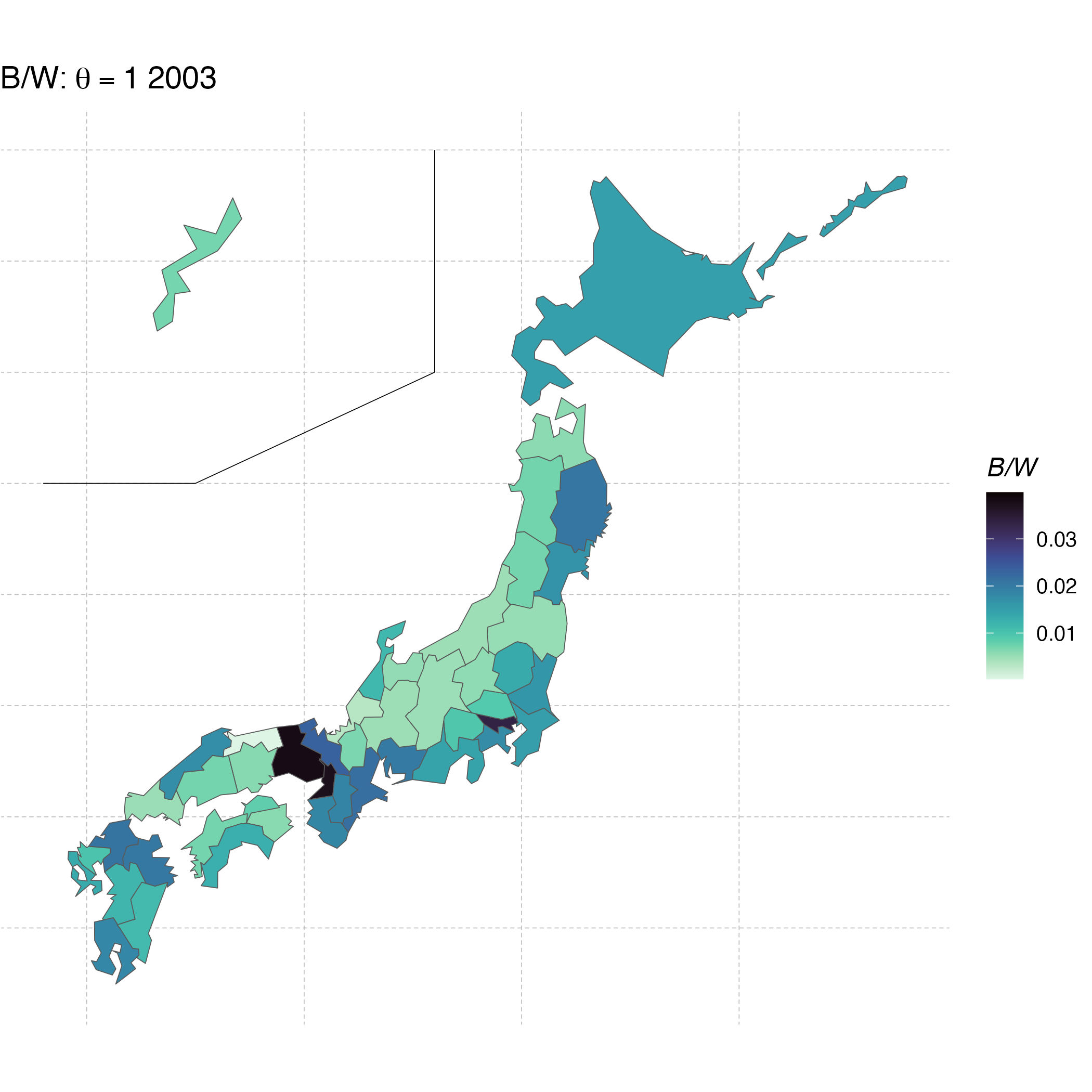}
     \includegraphics[width=5cm]{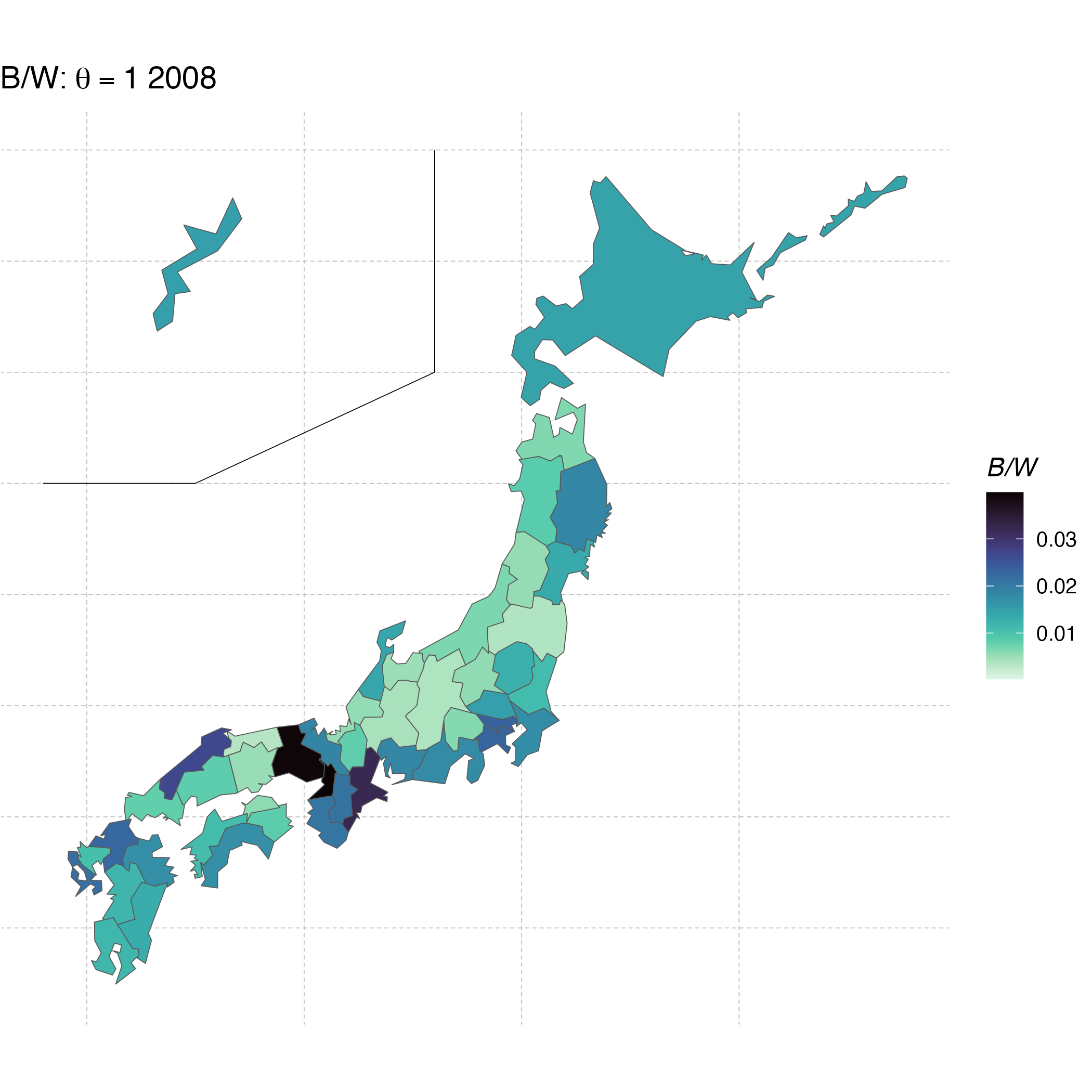}
     \includegraphics[width=5cm]{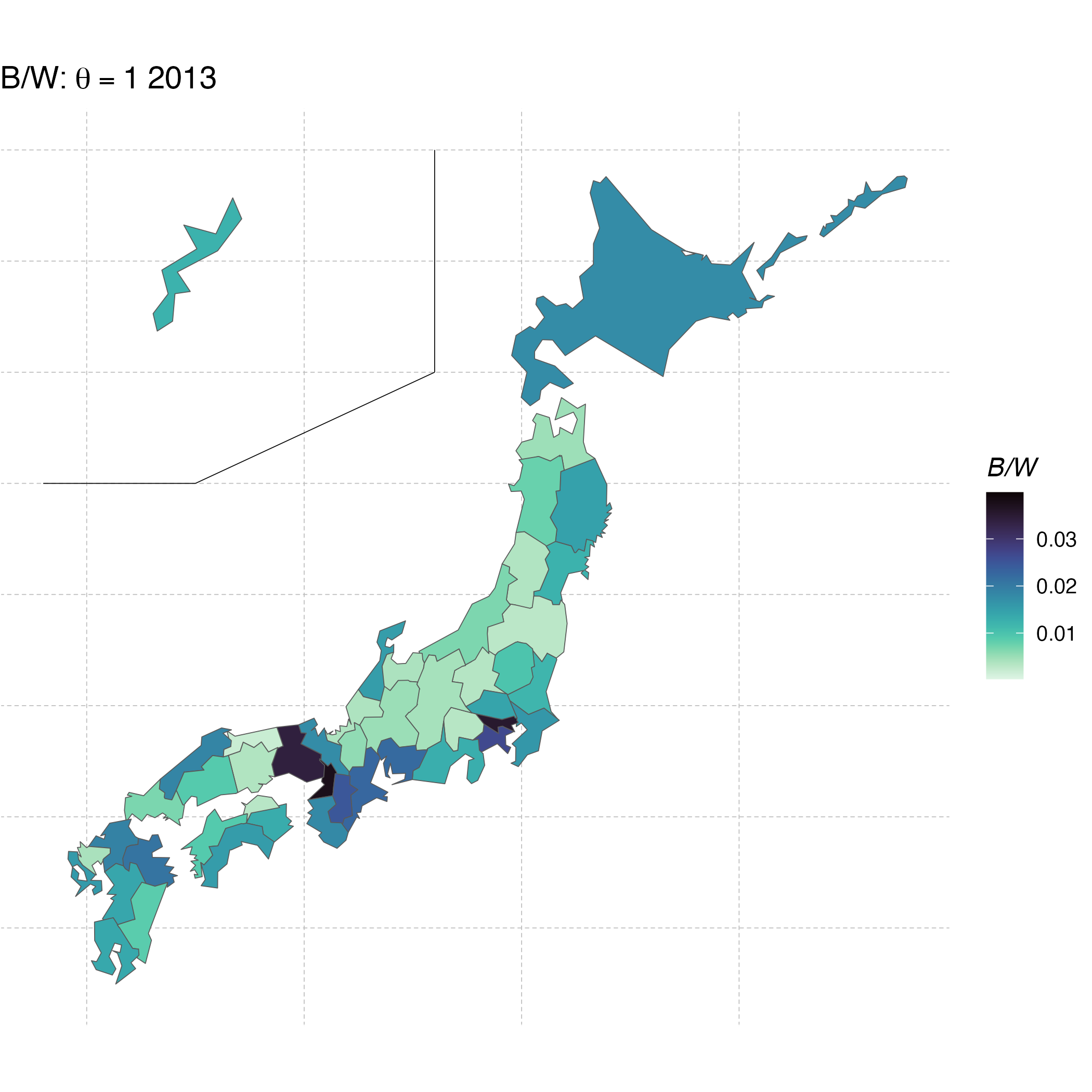}

     \includegraphics[width=5cm]{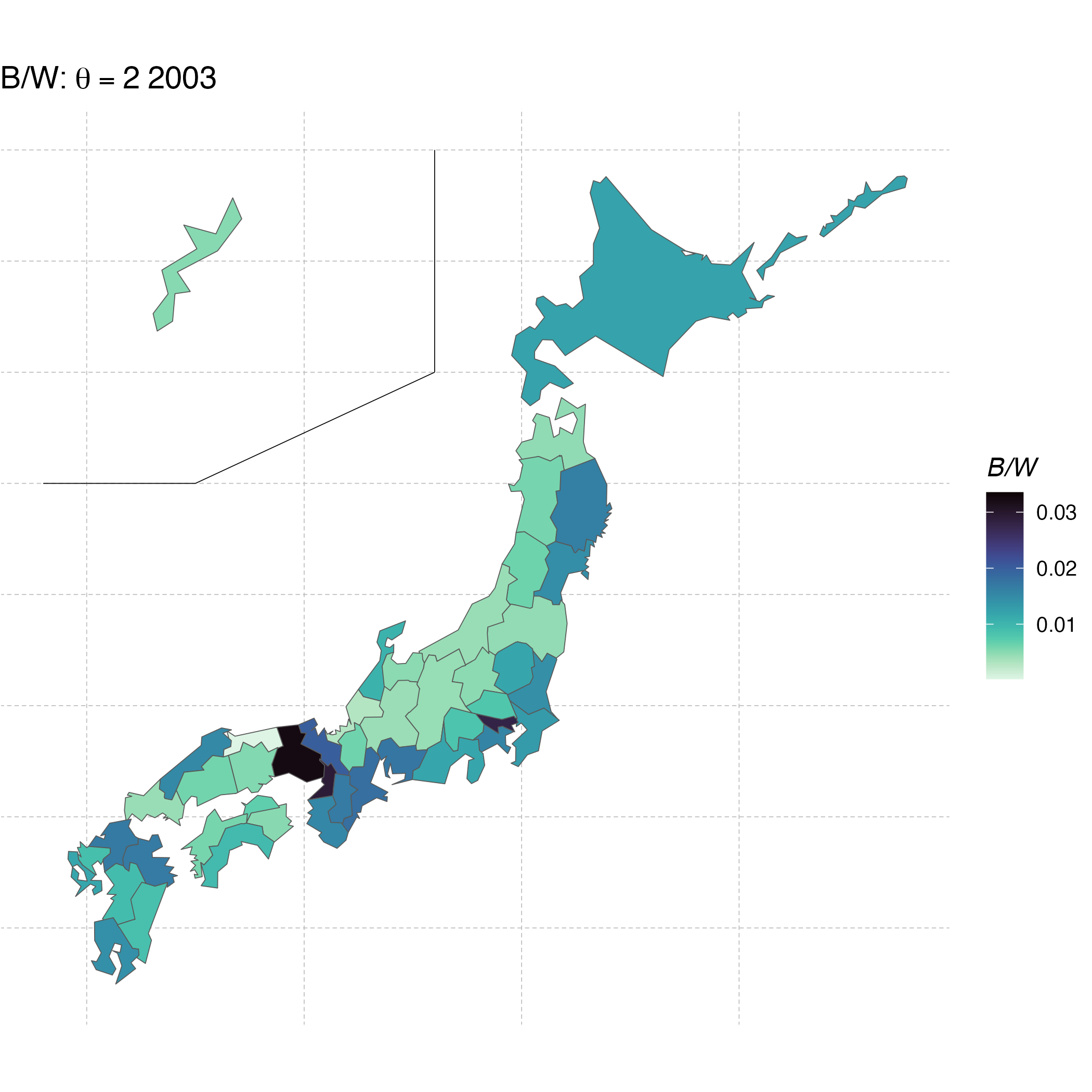}
     \includegraphics[width=5cm]{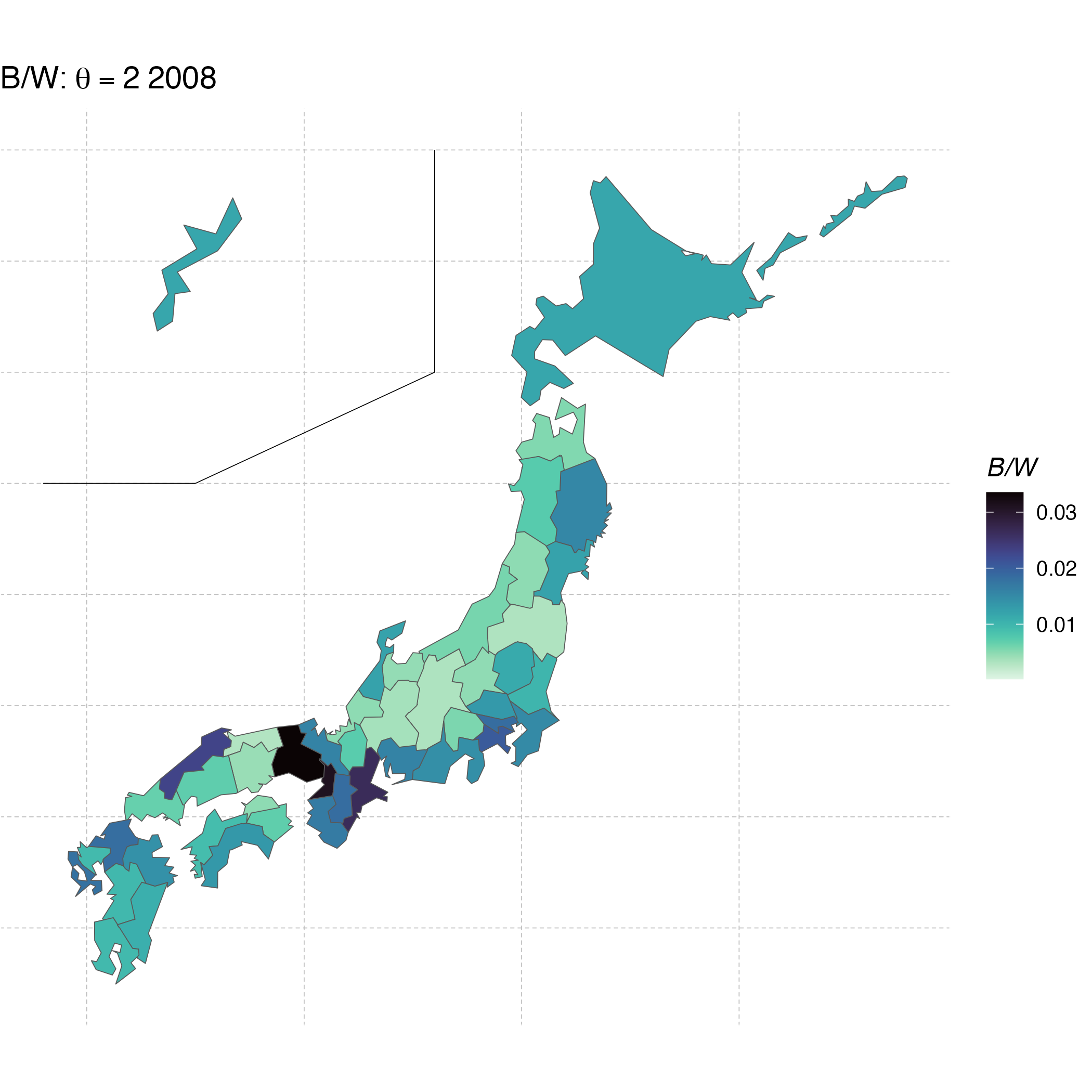}
     \includegraphics[width=5cm]{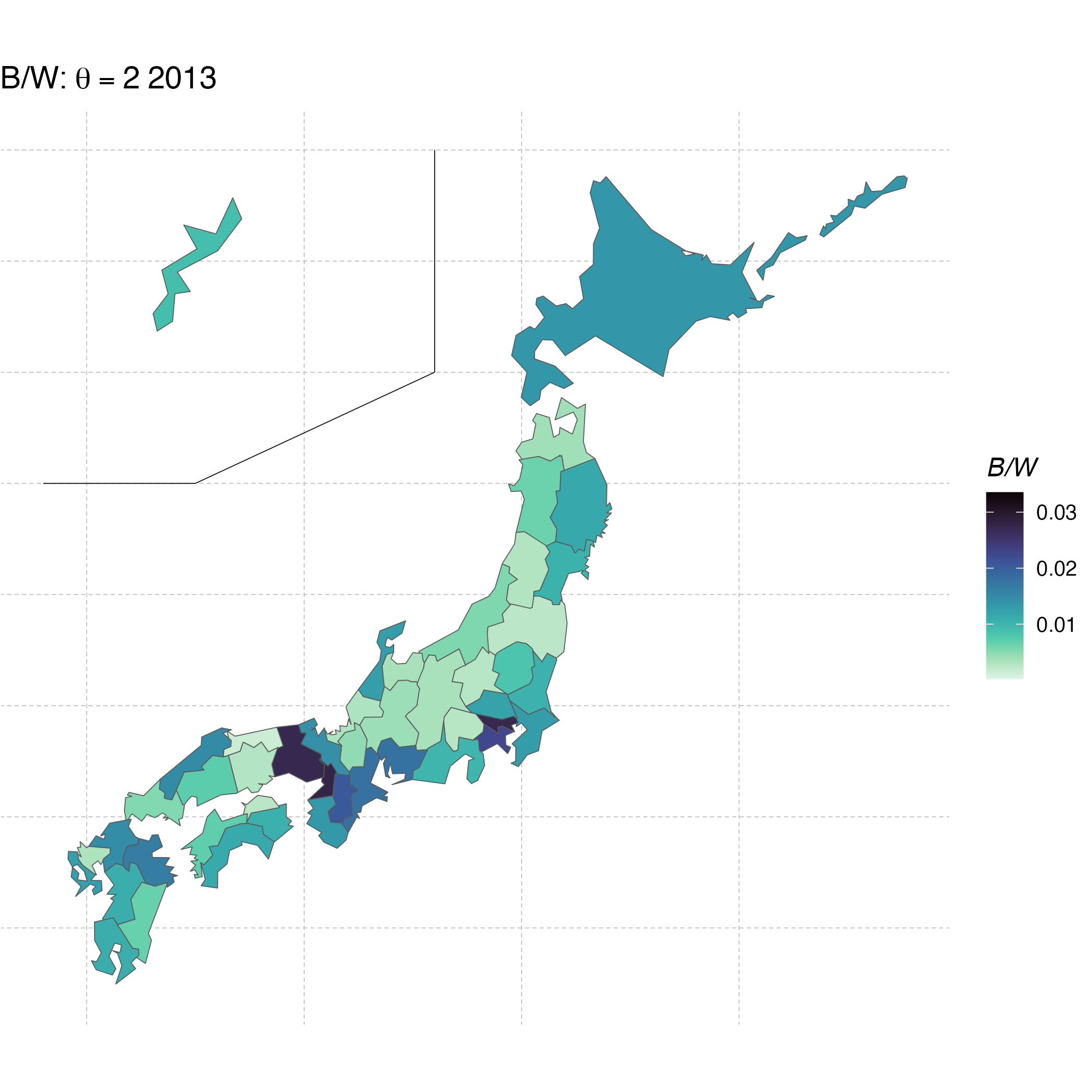}
     \caption{B/W ratio for different $\theta$}
     \label{fig:BWmap}
\end{figure}

To analyze income inequality across prefectures in greater detail, we calculated the ratio of between-municipality inequality to within-municipality inequality for each prefecture, which we refer to as the B/W ratio.
From Tables \ref{tab:multi1} and \ref{tab:multi2}, we confirm that the effect of between-municipality inequality on overall prefectural inequality is as small as that of between-prefecture inequality.
However, the relative importance of within-municipality inequality has not yet been examined.
The B/W ratio captures this effect, with a lower B/W ratio indicating that within-municipality income inequality plays a more significant role compared to between-municipality income inequality within a given prefecture.
Figure \ref{fig:BWmap} presents the choropleth maps of the B/W ratio.
These maps reveal that the relative importance of between-municipality inequality varies across prefectures.
Notably, the impact of between-municipality inequality is pronounced in metropolitan areas such as Tokyo.
Conversely, we do not observe significant variations across different time periods or values of $\theta$, in contrast to the differences observed in Figure \ref{fig:GEmap}.
This suggests that the relative importance of between-municipality income inequality remains fairly consistent over time and across different values of $\theta$ within each prefecture.
Therefore, to gain a more comprehensive understanding of overall income inequality, it is essential to closely examine income inequality at the prefectural level.

\begin{figure}
    \centering
    \includegraphics[width=5.2cm]{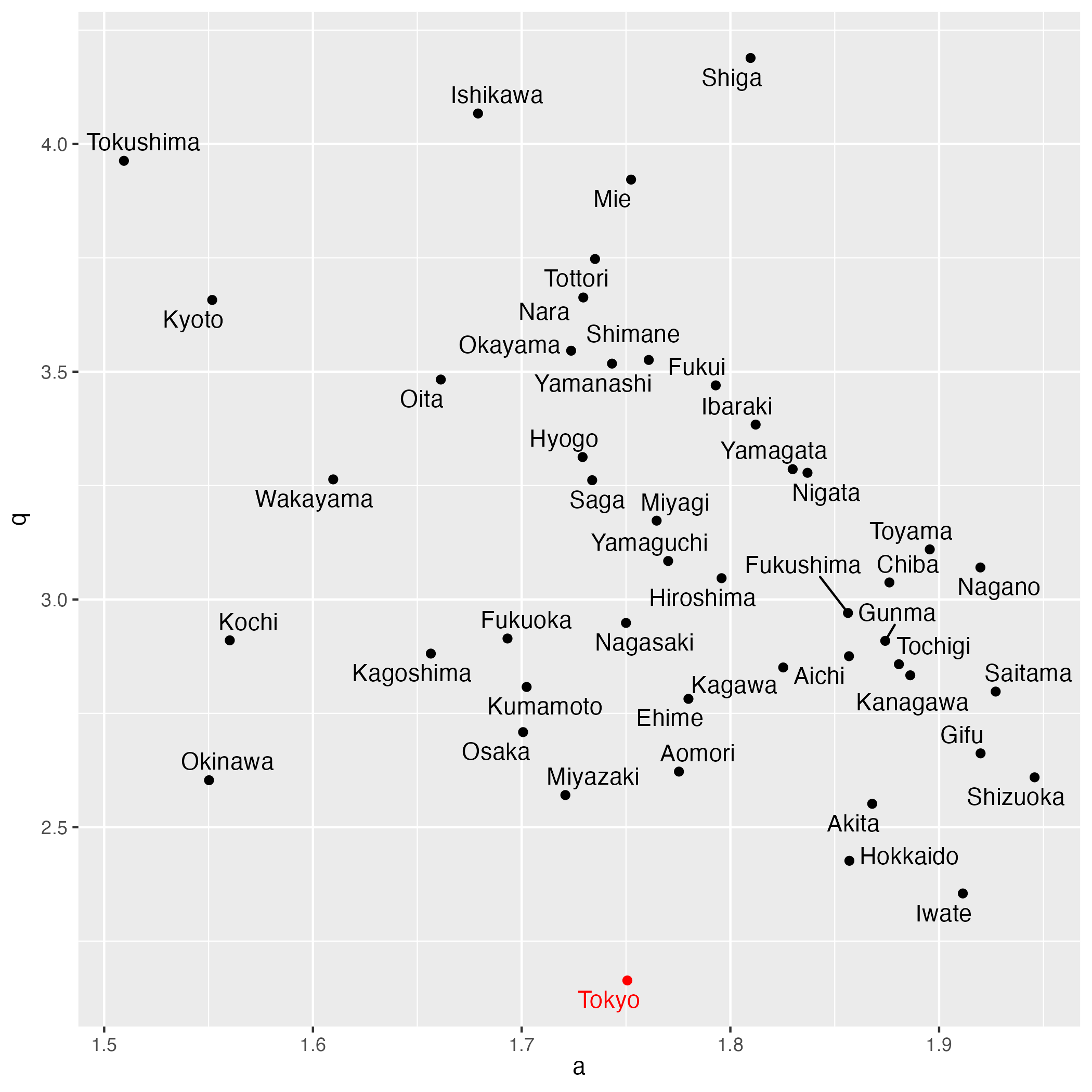}
    \includegraphics[width=5.2cm]{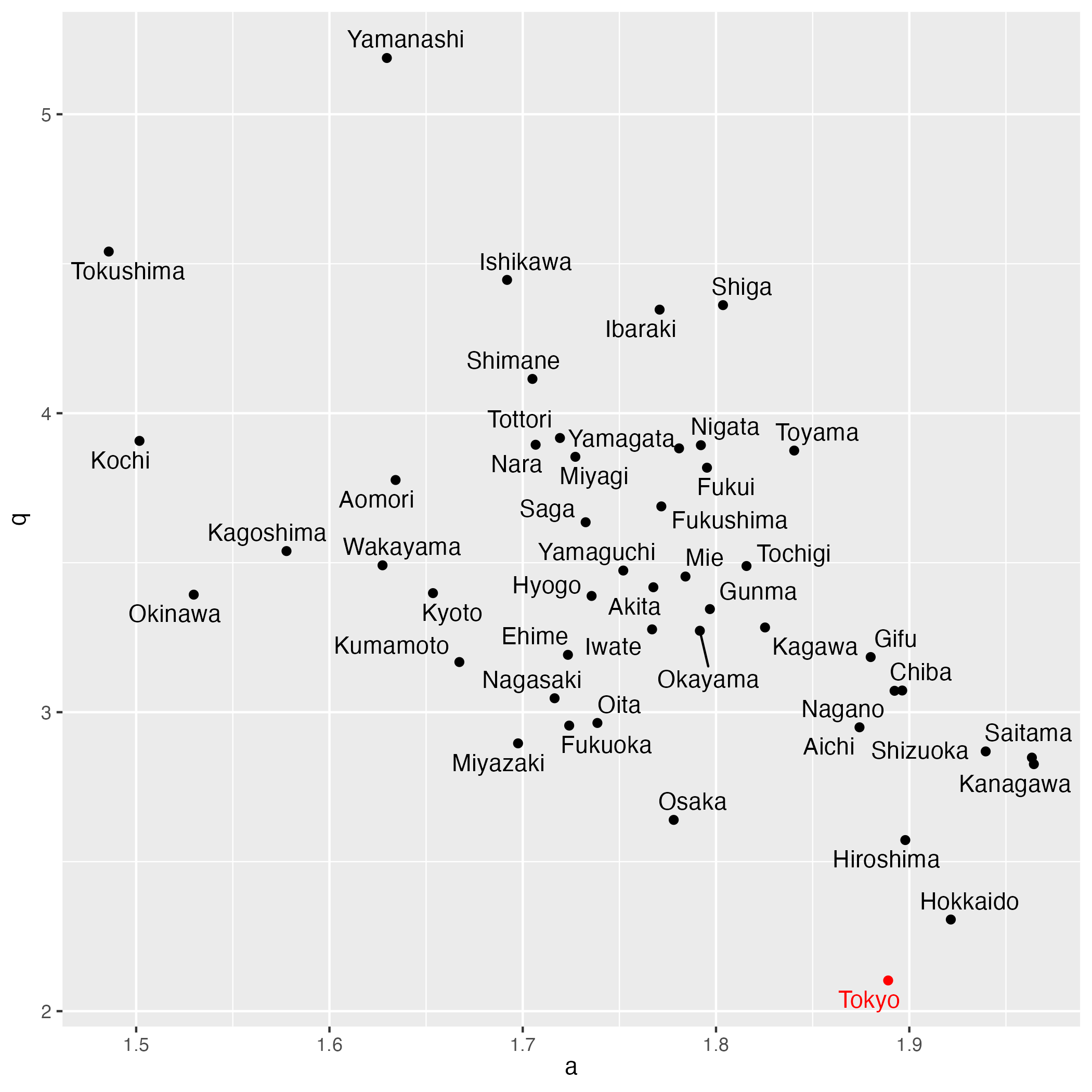}
    \includegraphics[width=5.2cm]{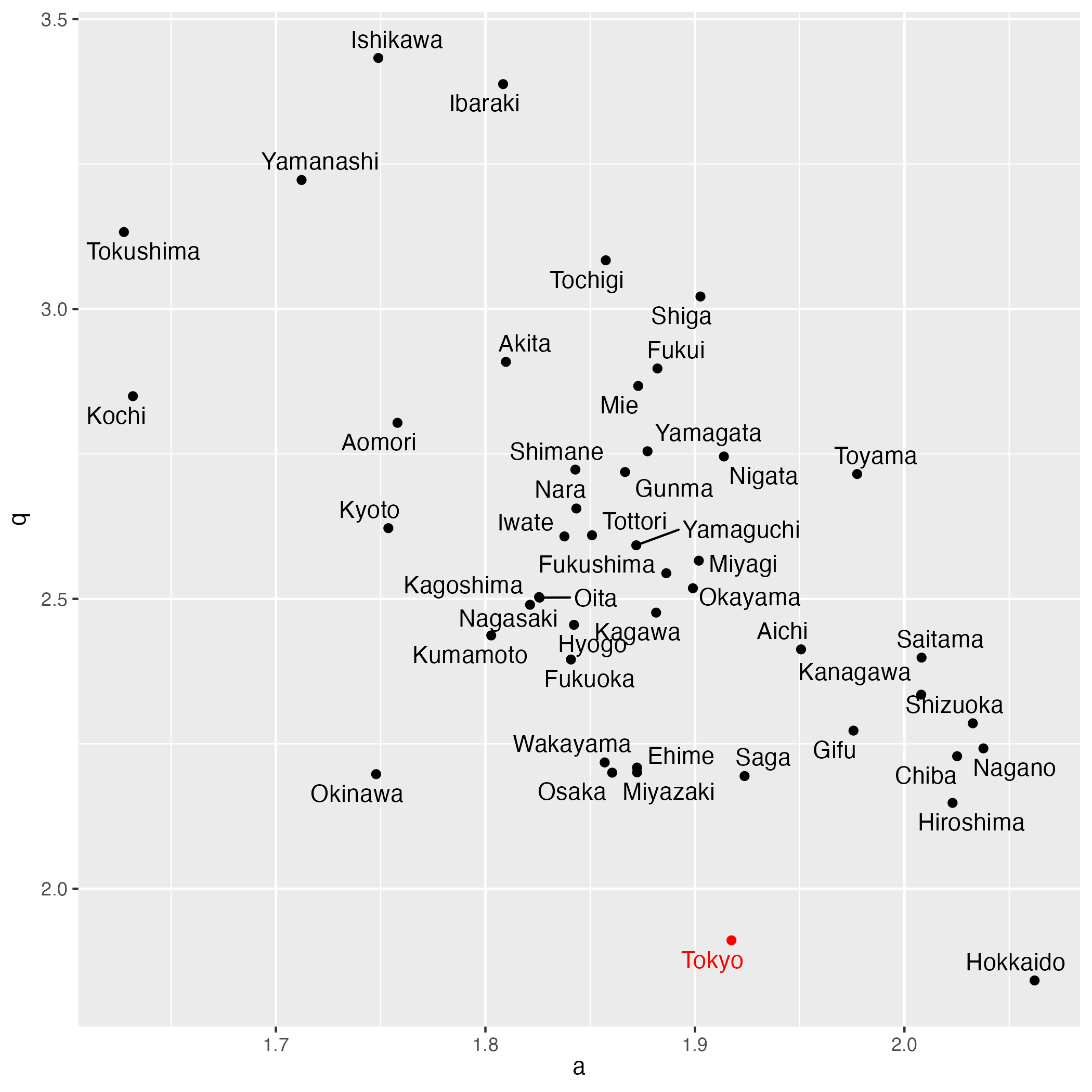}
    \caption{Scatter plots of posterior means of parameters $a$ vs $q$ for the Singh--Maddala distribution of each prefecture in 2003 (left), 2008 (middle) and 2013 (right).}
    \label{fig:aq}
\end{figure}

To examine differences in income inequality between prefectures, we analyze variations in parameters, as shown in Figure \ref{fig:aq}.
Although the Singh--Maddala distribution is more restrictive than the GB2 distribution due to the fixed value of $p=1$, it still includes two parameters, $a$ and $q$, both of which influence changes in income inequality.
Figure \ref{fig:aq} presents scatter plots of the posterior means of $a$ and $q$.
\footnote{For the relationship between the magnitudes of $a$ and $q$ and GE, refer to the appendix \ref{sec:parameters_GE}.}
It reveals a general tendency for a negative correlation between $a$ and $q$: regions with smaller values of $a$ tend to exhibit relatively larger values of $q$, and vice versa. Given this pattern, we investigate whether there are distinct characteristics between prefectures where a high degree of inequality is primarily attributable to a small $a$, and those where it is driven by a small $q$. In prefectures where inequality is heightened due to a small $a$ (e.g., Okinawa and Kochi), the GE index tends to be larger when $\theta$ is small. This indicates that inequality among the lower-income population is particularly pronounced in these regions. Conversely, in prefectures where inequality is primarily due to a small $q$ (e.g., Tokyo and Hokkaido, particularly in 2013), GE is larger at higher values of $\theta$, suggesting that inequality among the higher-income population is of greater concern. These findings align with the theoretical properties of the Singh--Maddala distribution, in which the parameters $a$ and $q$ respectively influence the lower and upper tails of the income distribution. Specifically, when both $a$ and $q$ are small, the distribution becomes more spread out at both ends, leading to heightened inequality across the entire income level. This behavior is illustrated in the GE contour plots with $a$ on the horizontal axis and $q$ on the vertical axis (see Appendix \ref{sec:parameters_GE}). In particular, among the prefectures characterized by small values of $q$, Tokyo is notable for having especially small $a$. Given the sensitivity of GE to changes in $a$, this suggests that inequality is pronounced even among the high-income group in Tokyo. This observation raises the question of how income inequality varies within Tokyo itself, prompting a closer look at the municipal-level distribution within the prefecture.

\begin{figure}
    \centering
    \includegraphics[width=5cm]{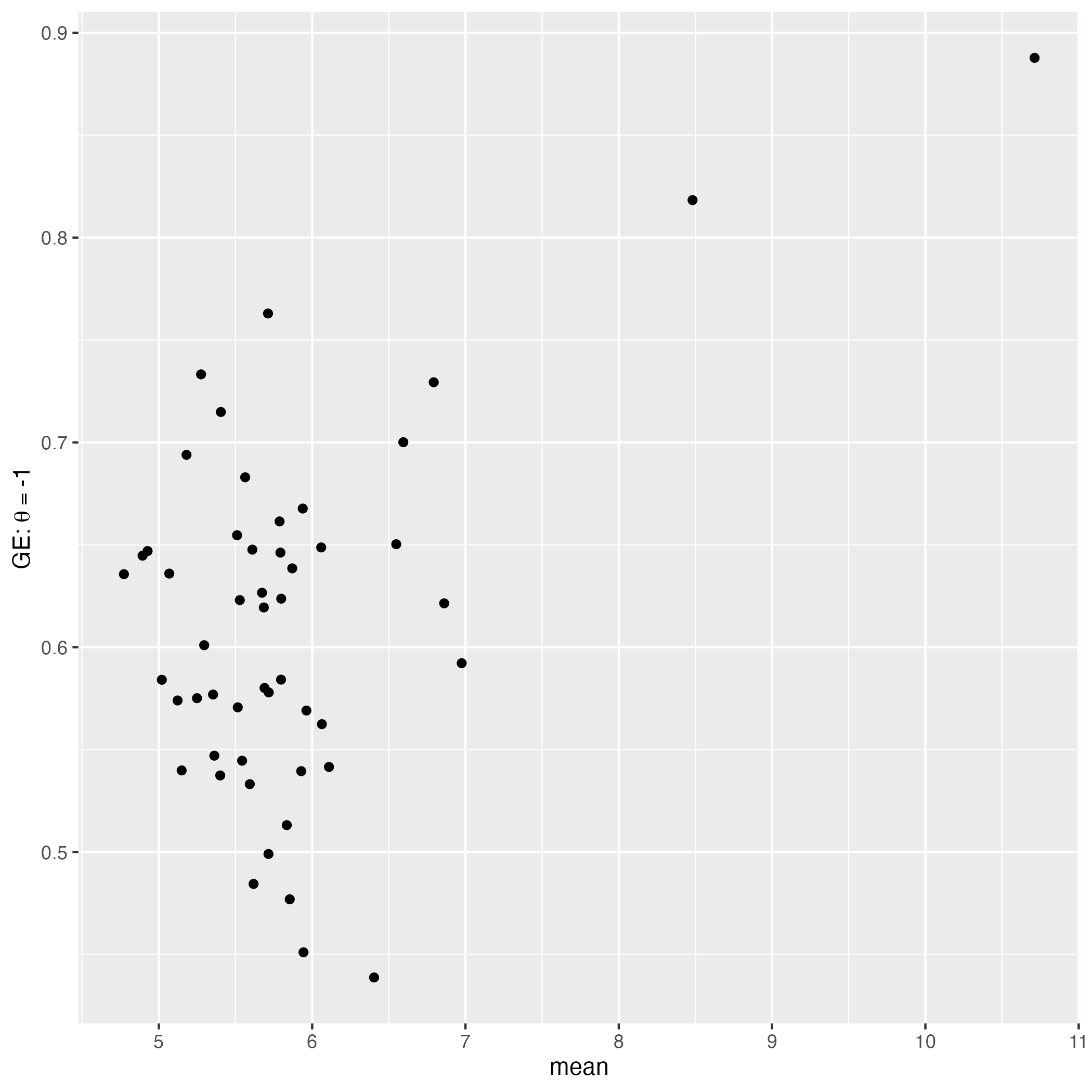}
    \includegraphics[width=5cm]{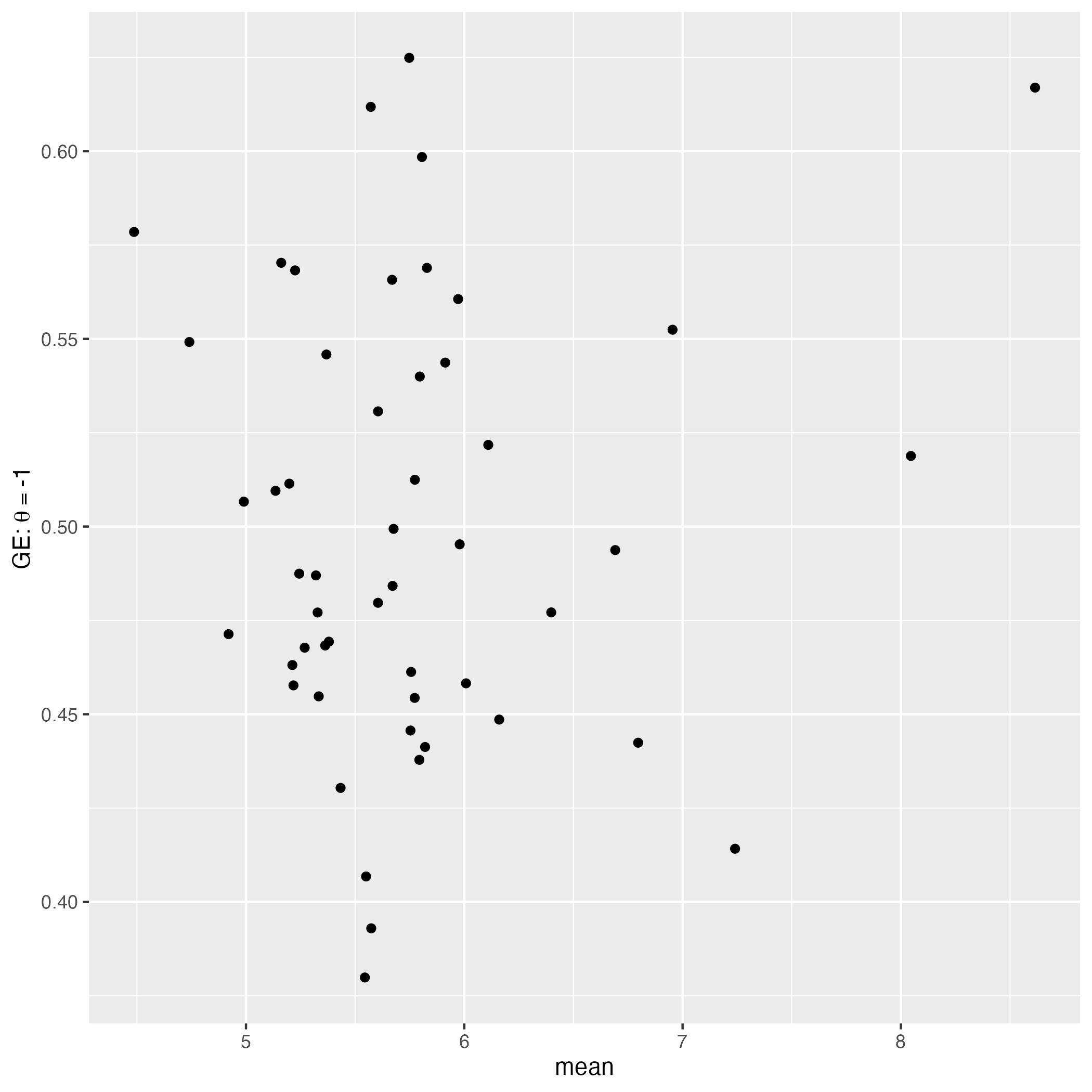}
    \includegraphics[width=5cm]{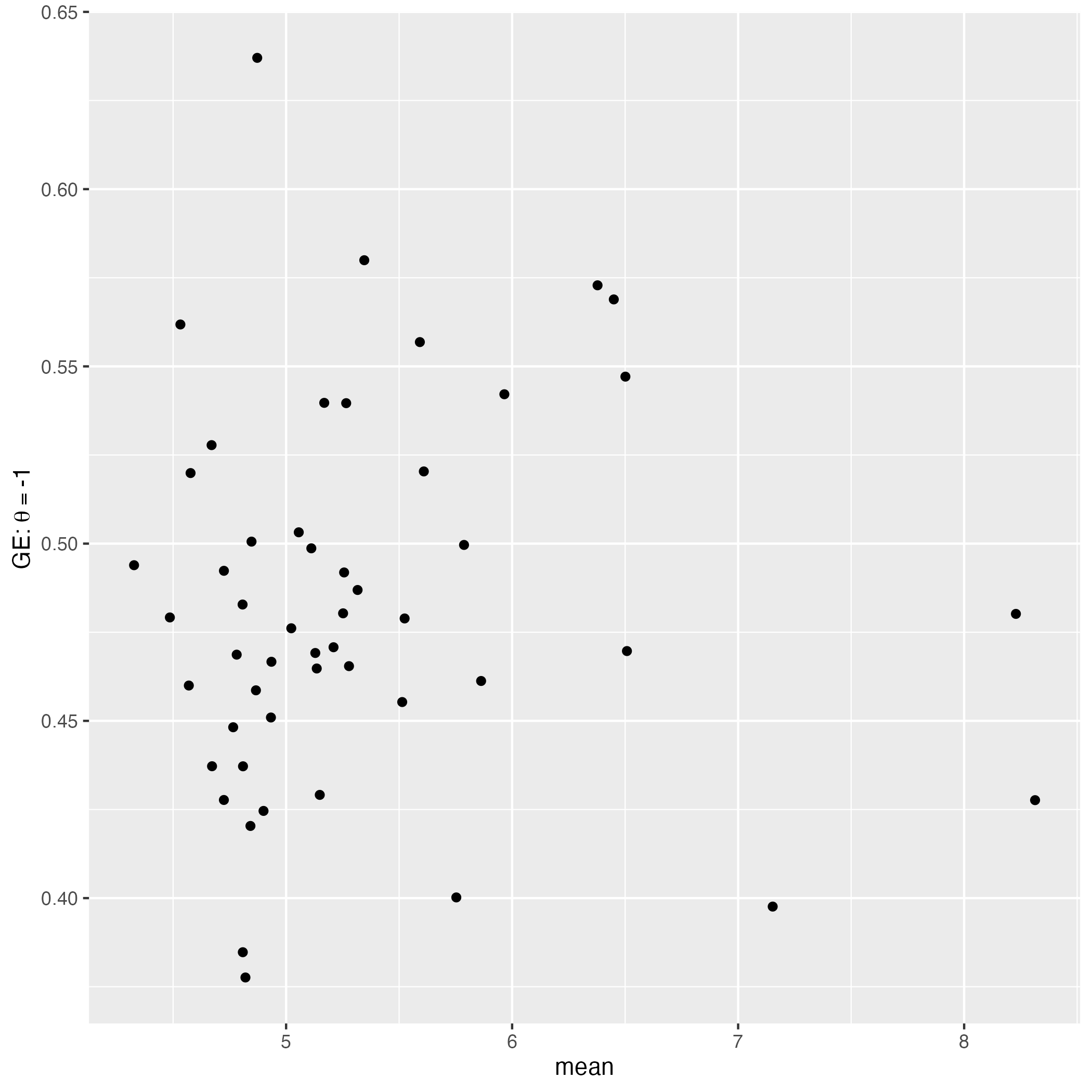}

    \includegraphics[width=5cm]{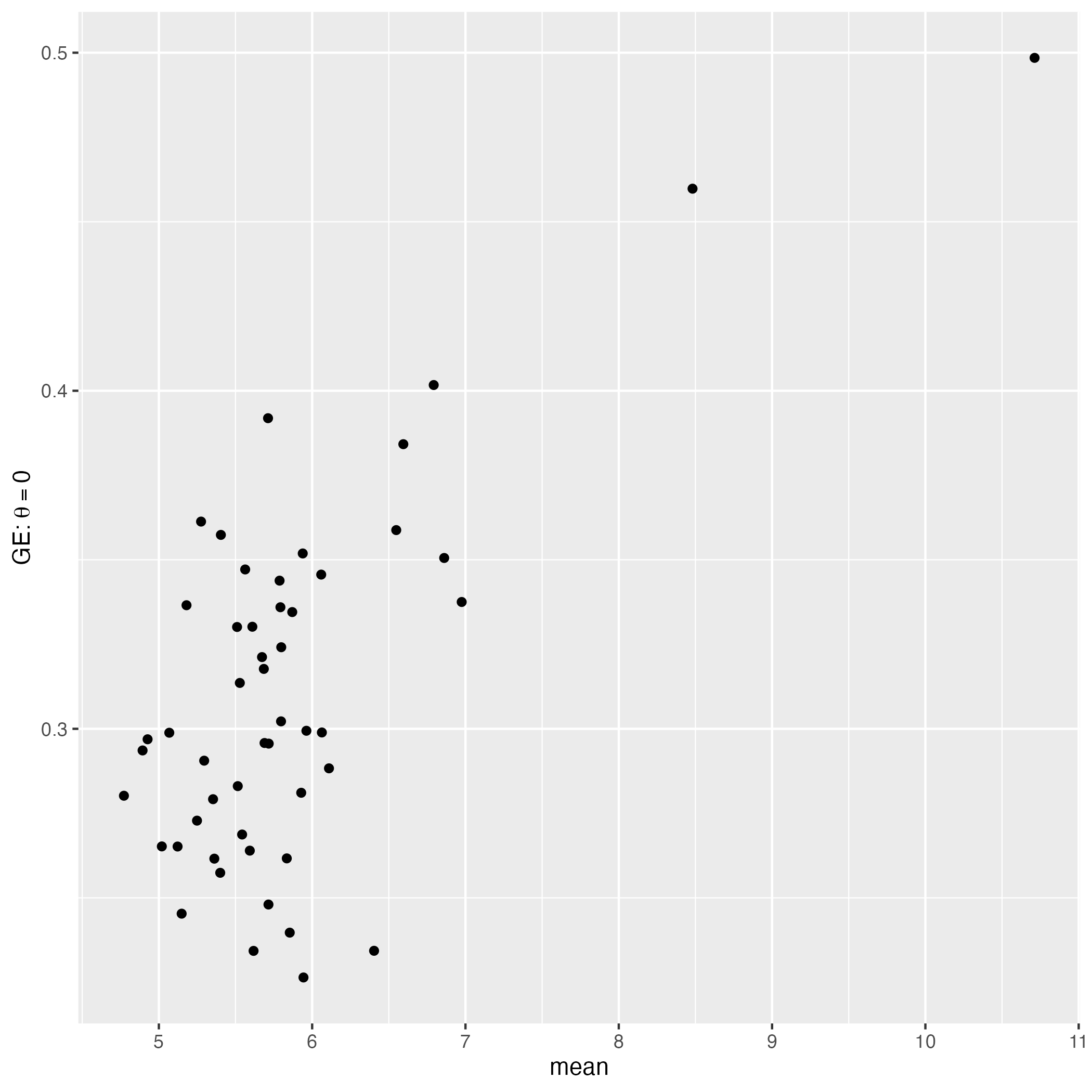}
    \includegraphics[width=5cm]{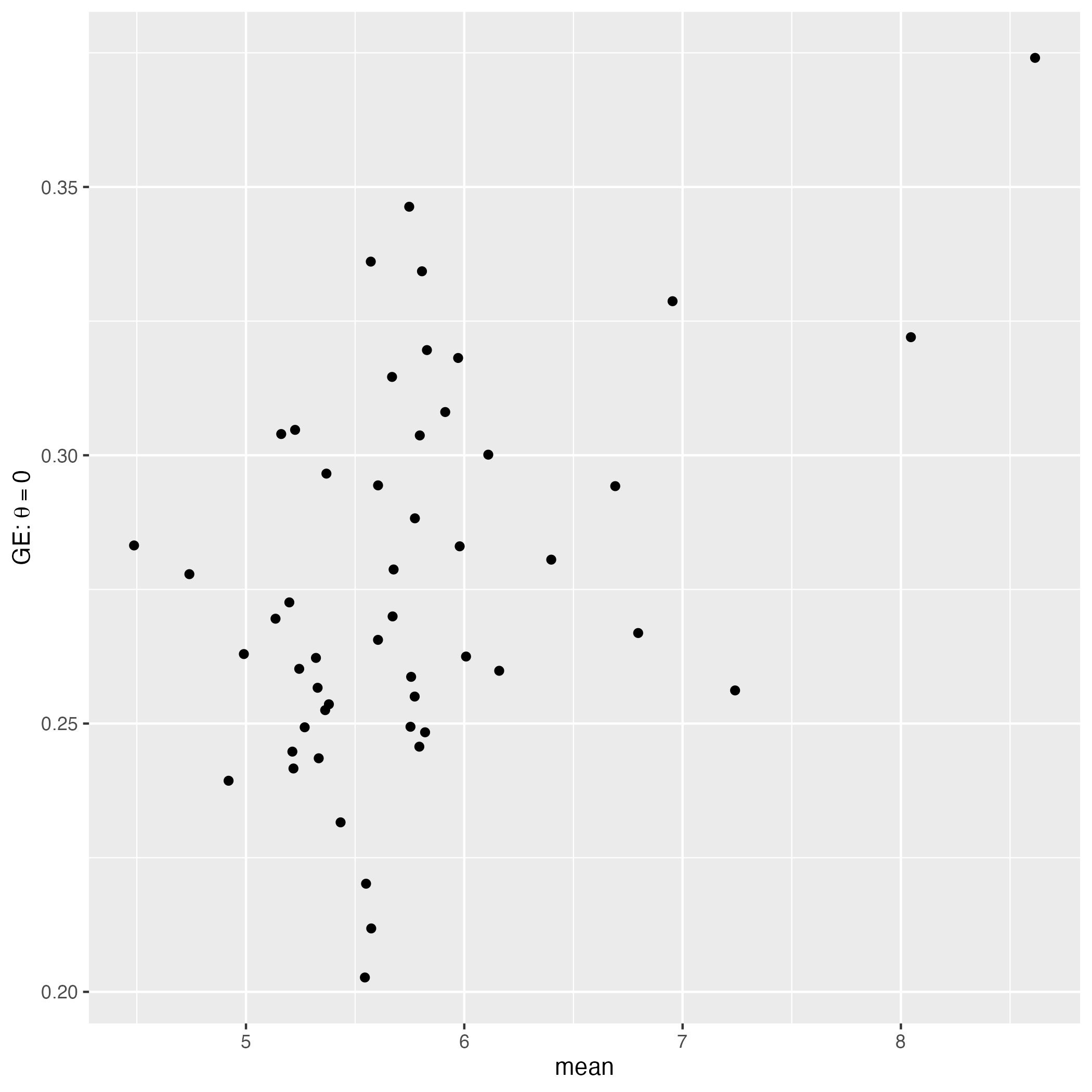}
    \includegraphics[width=5cm]{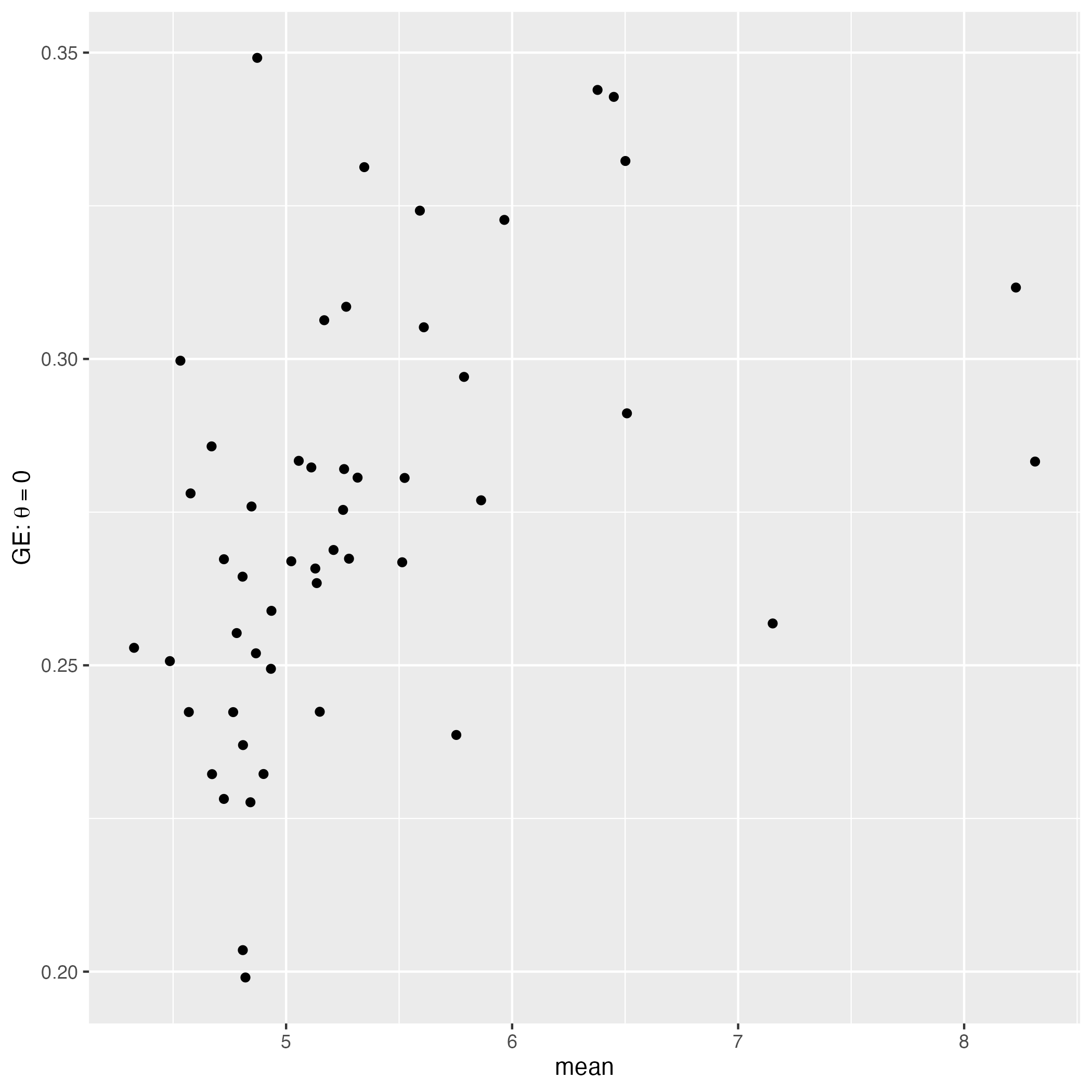}

    \includegraphics[width=5cm]{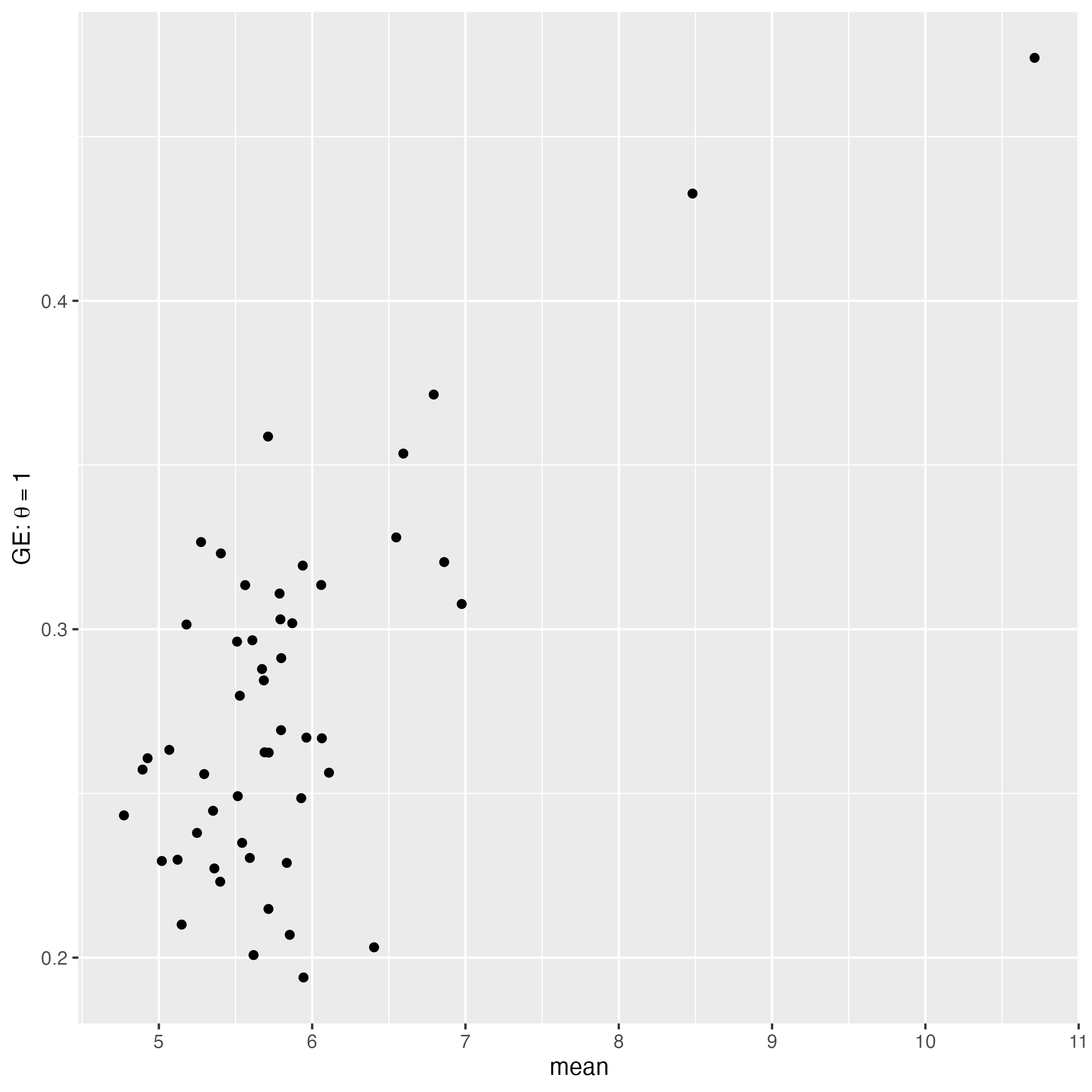}
    \includegraphics[width=5cm]{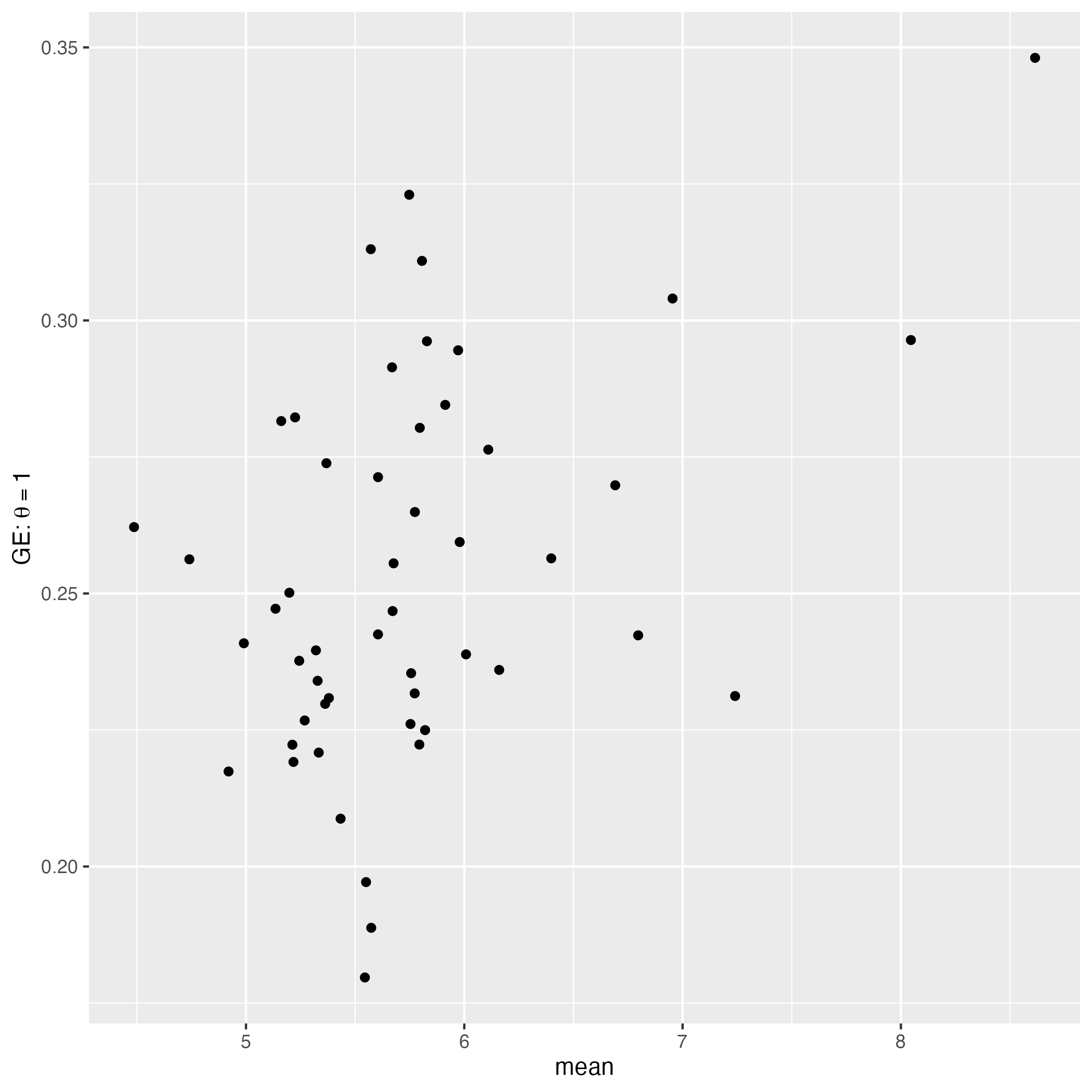}
    \includegraphics[width=5cm]{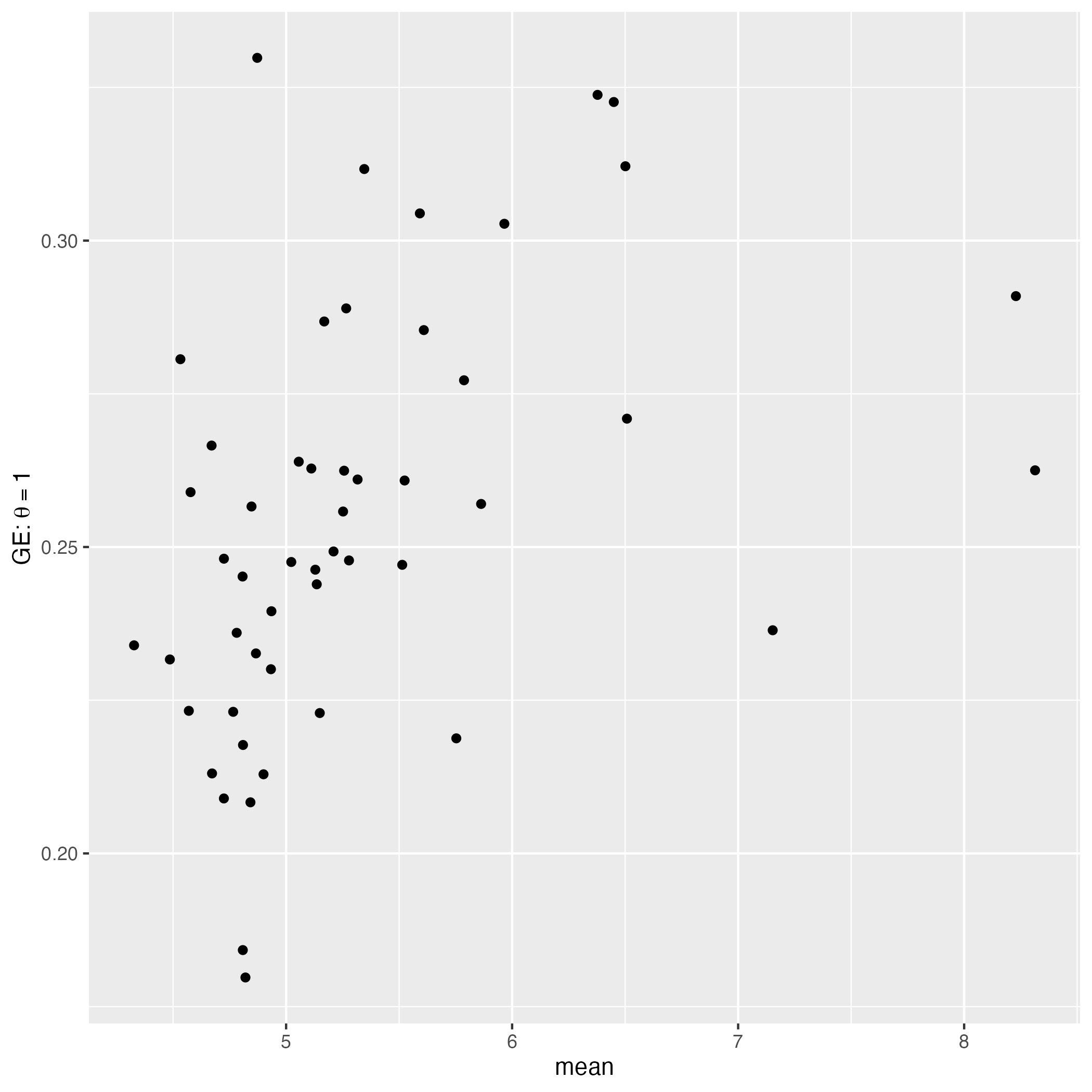}

    \includegraphics[width=5cm]{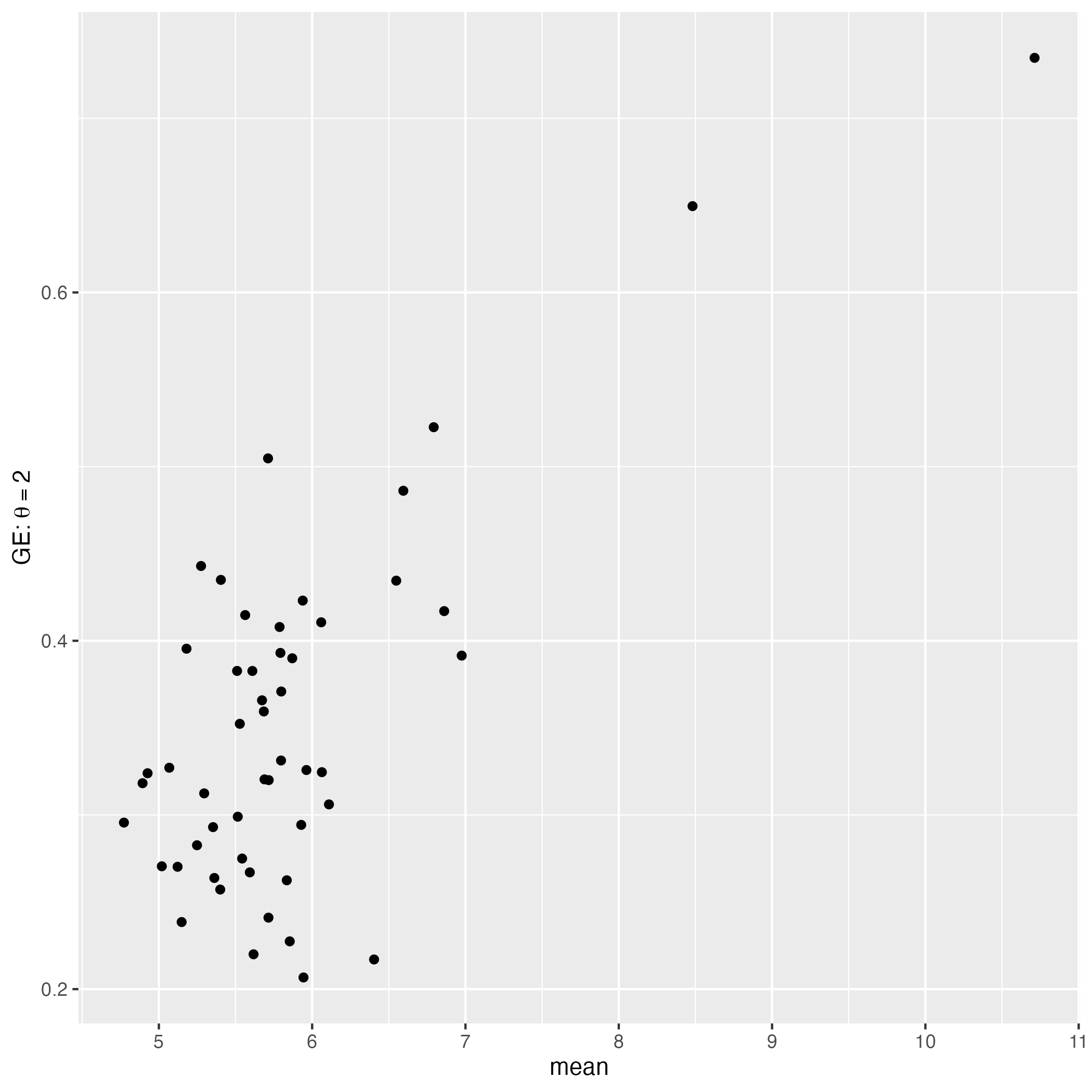}
    \includegraphics[width=5cm]{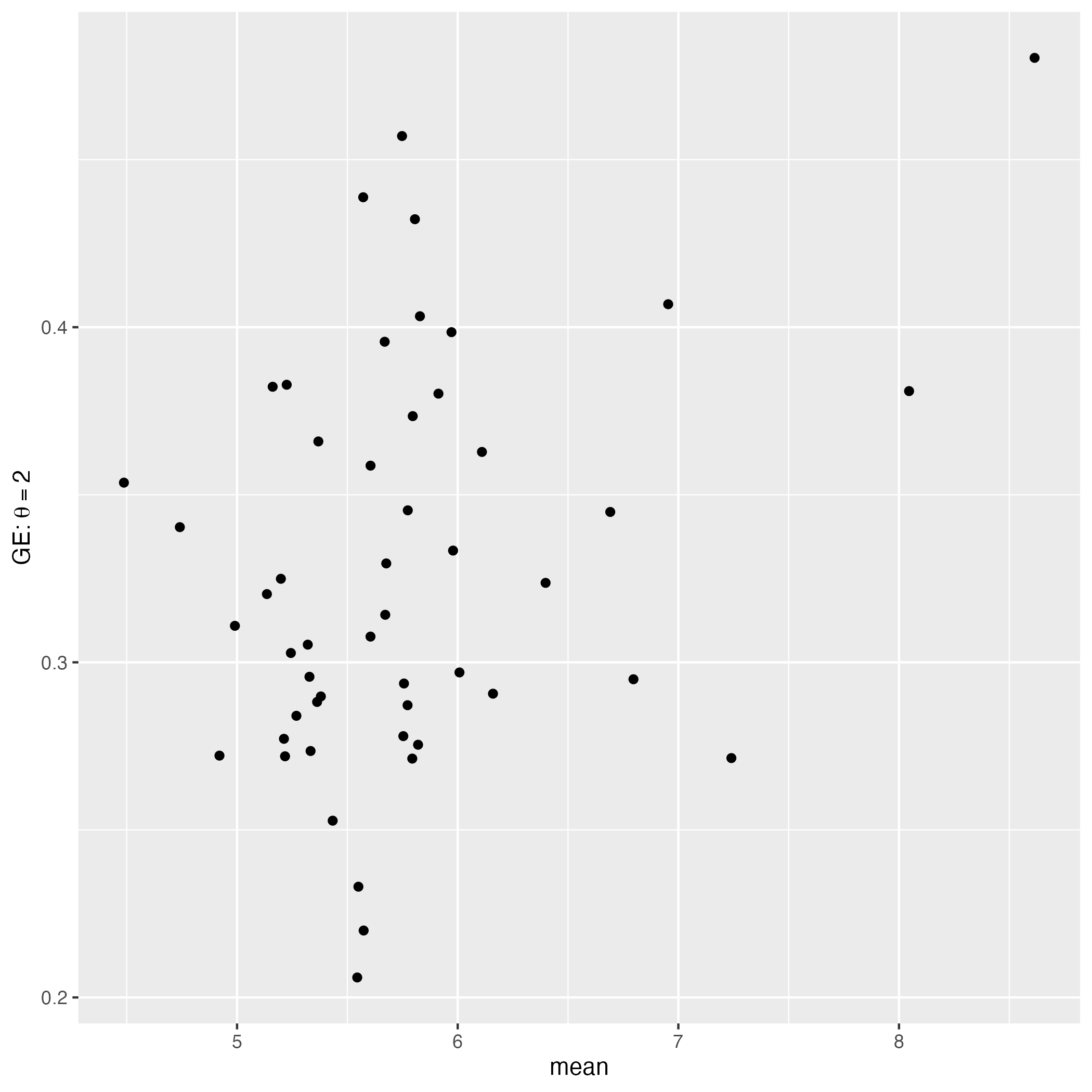}
    \includegraphics[width=5cm]{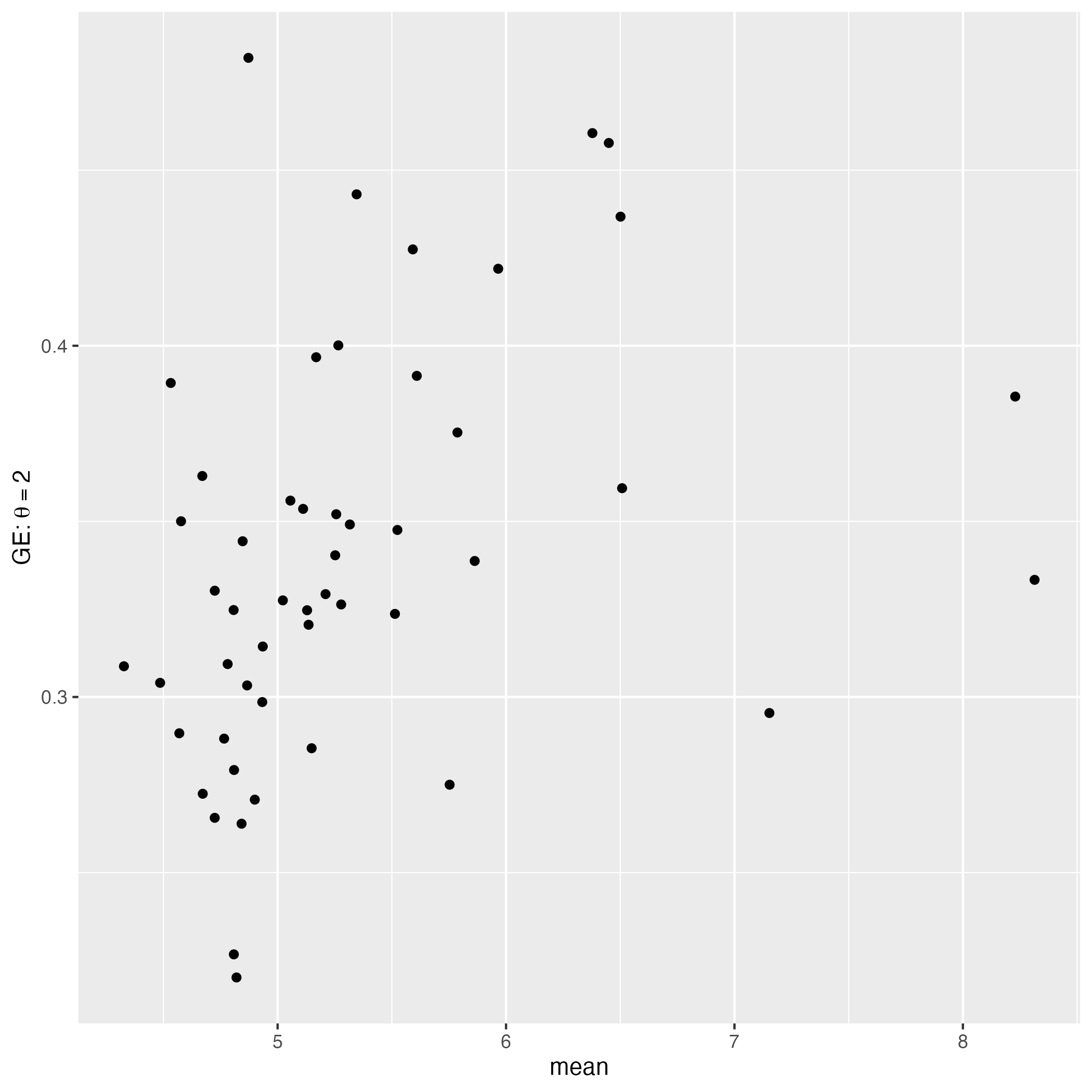}
    \caption{Scatter plots of mean income vs GE for different $\theta$ values for municipalities in Tokyo. The leftmost column shows the results for 2003, the middle column shows the results for 2008, and the rightmost column shows the results for 2013. $\theta$ values for GE are -1, 0, 1, and 2 from the top row, respectively.}
    \label{fig:MIvsGETokyo}
\end{figure}

\begin{figure}
    \centering
    \includegraphics[width=5cm]{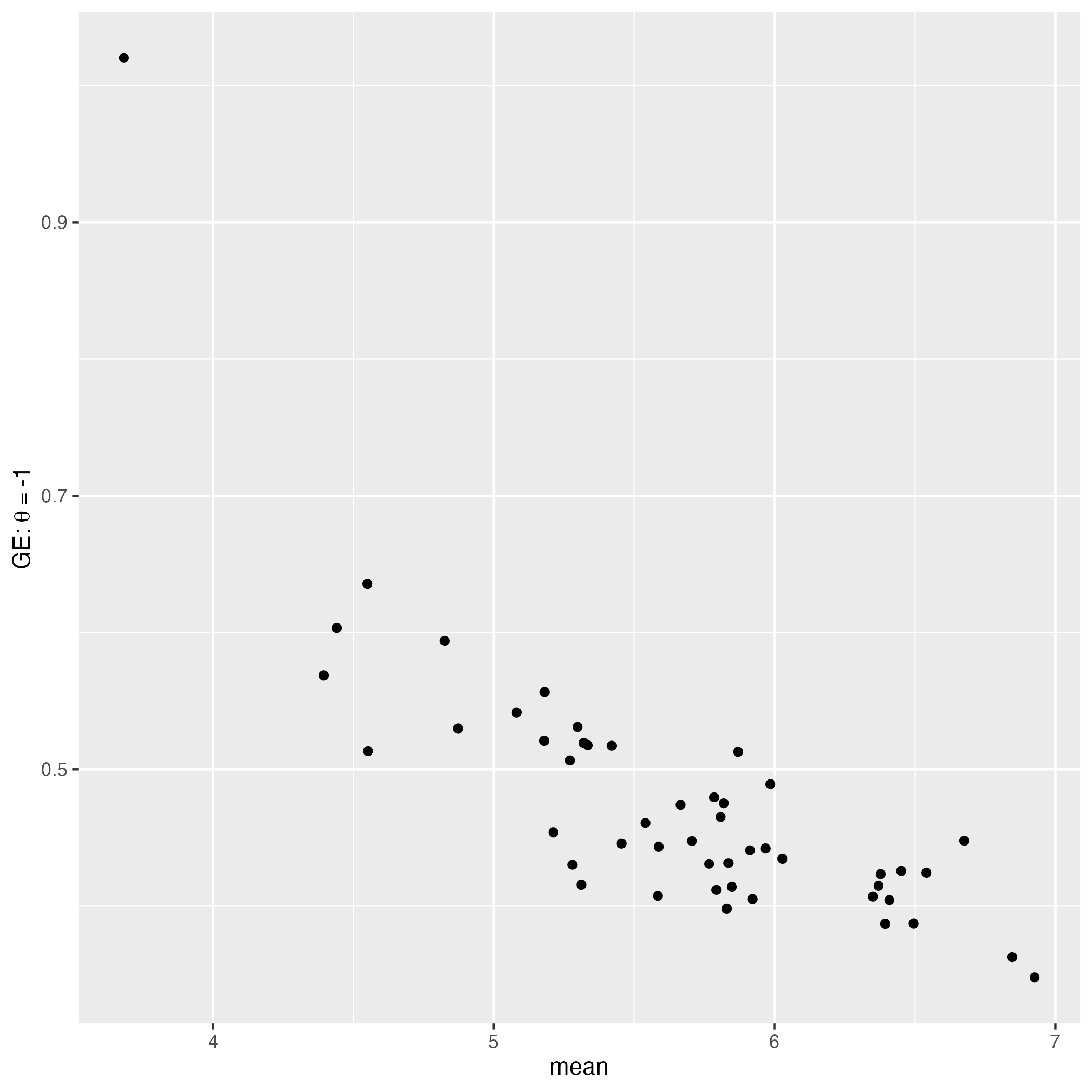}
    \includegraphics[width=5cm]{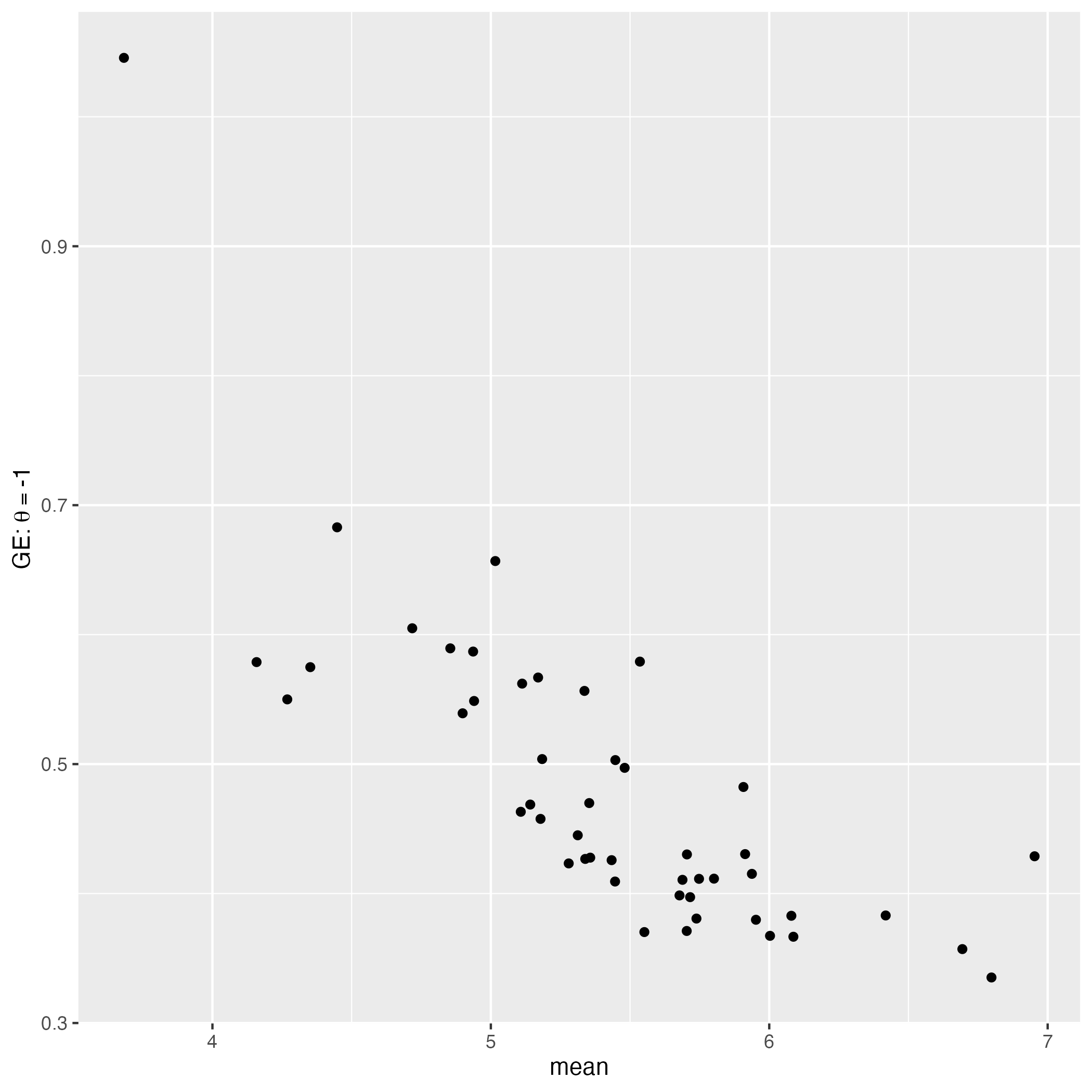}
    \includegraphics[width=5cm]{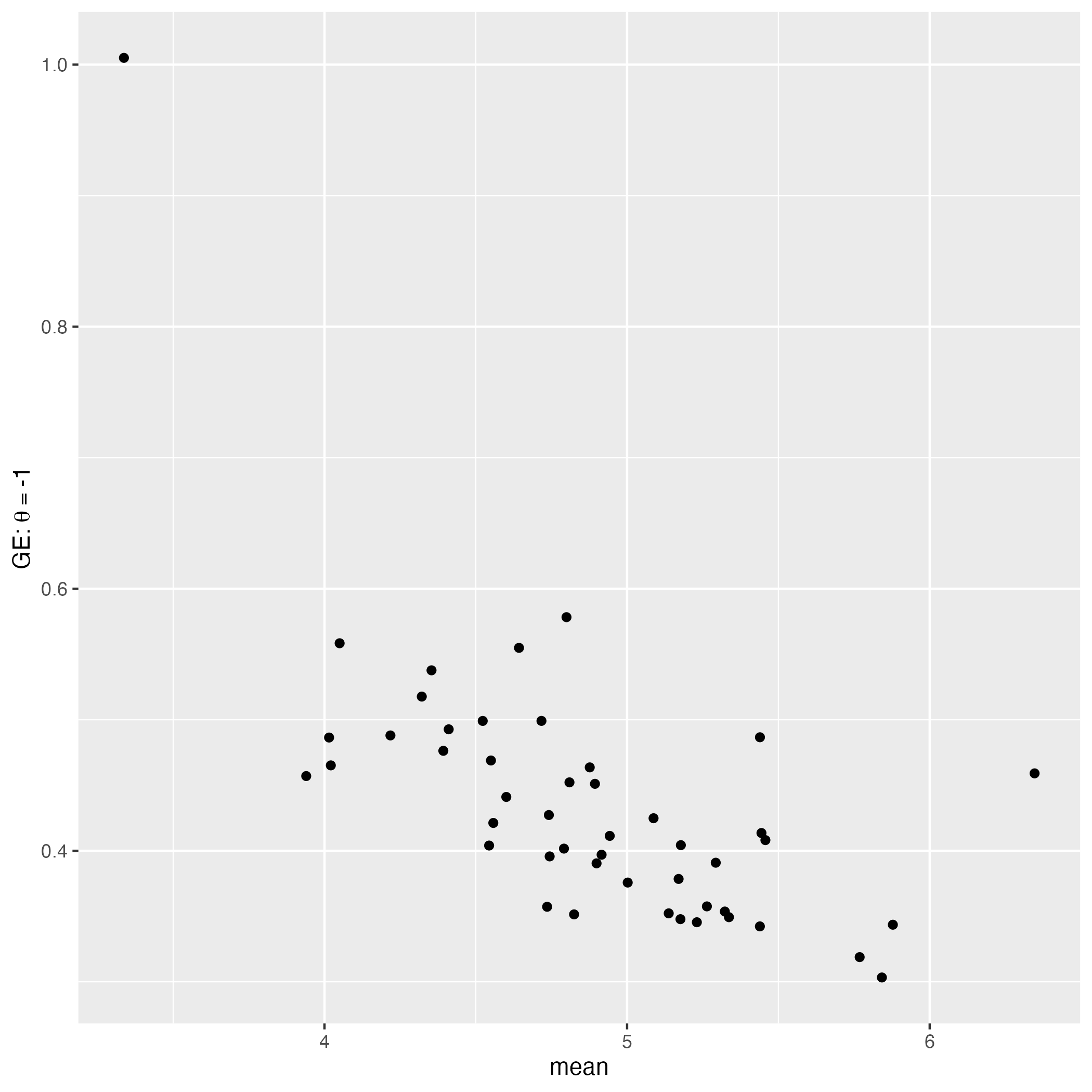}

    \includegraphics[width=5cm]{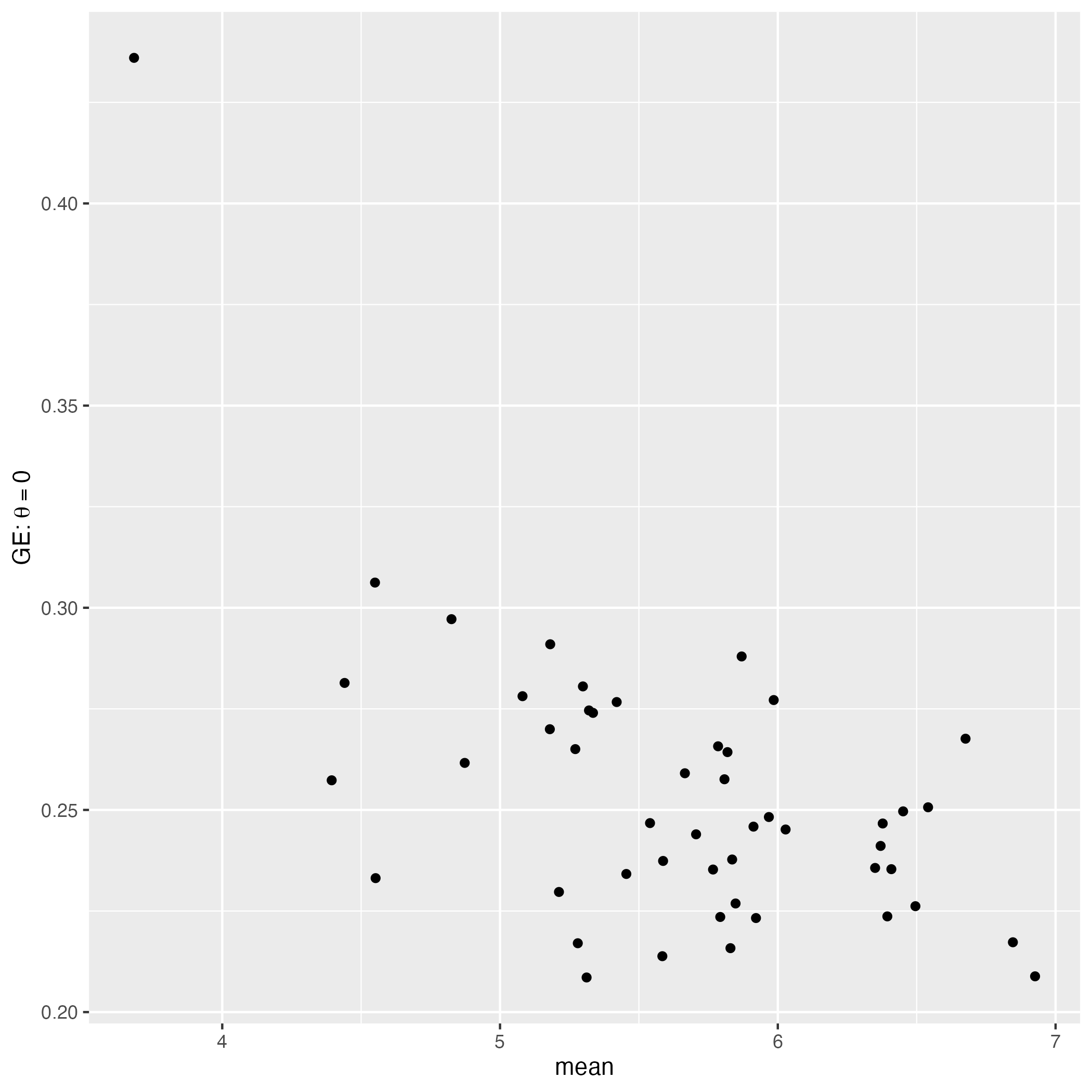}
    \includegraphics[width=5cm]{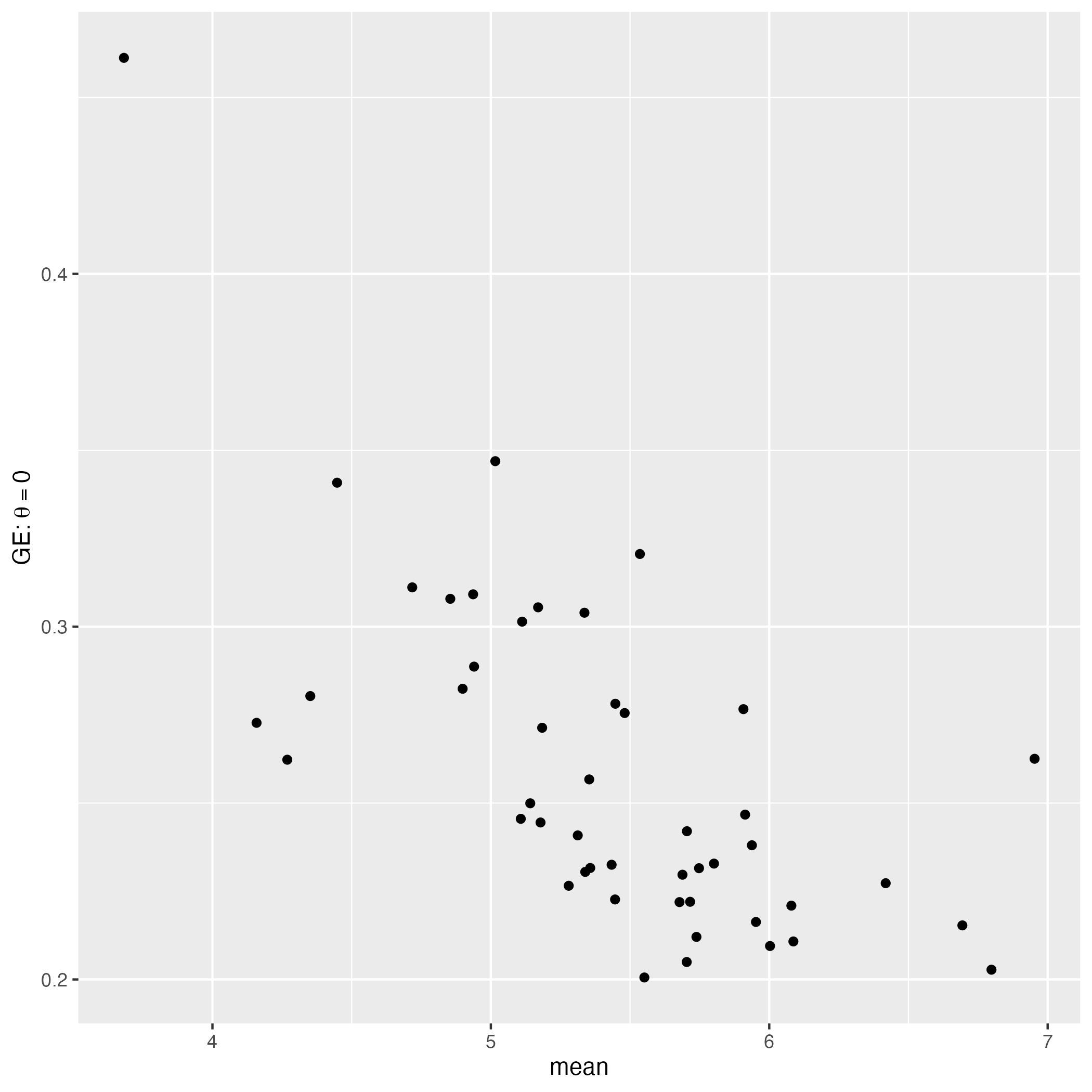}
    \includegraphics[width=5cm]{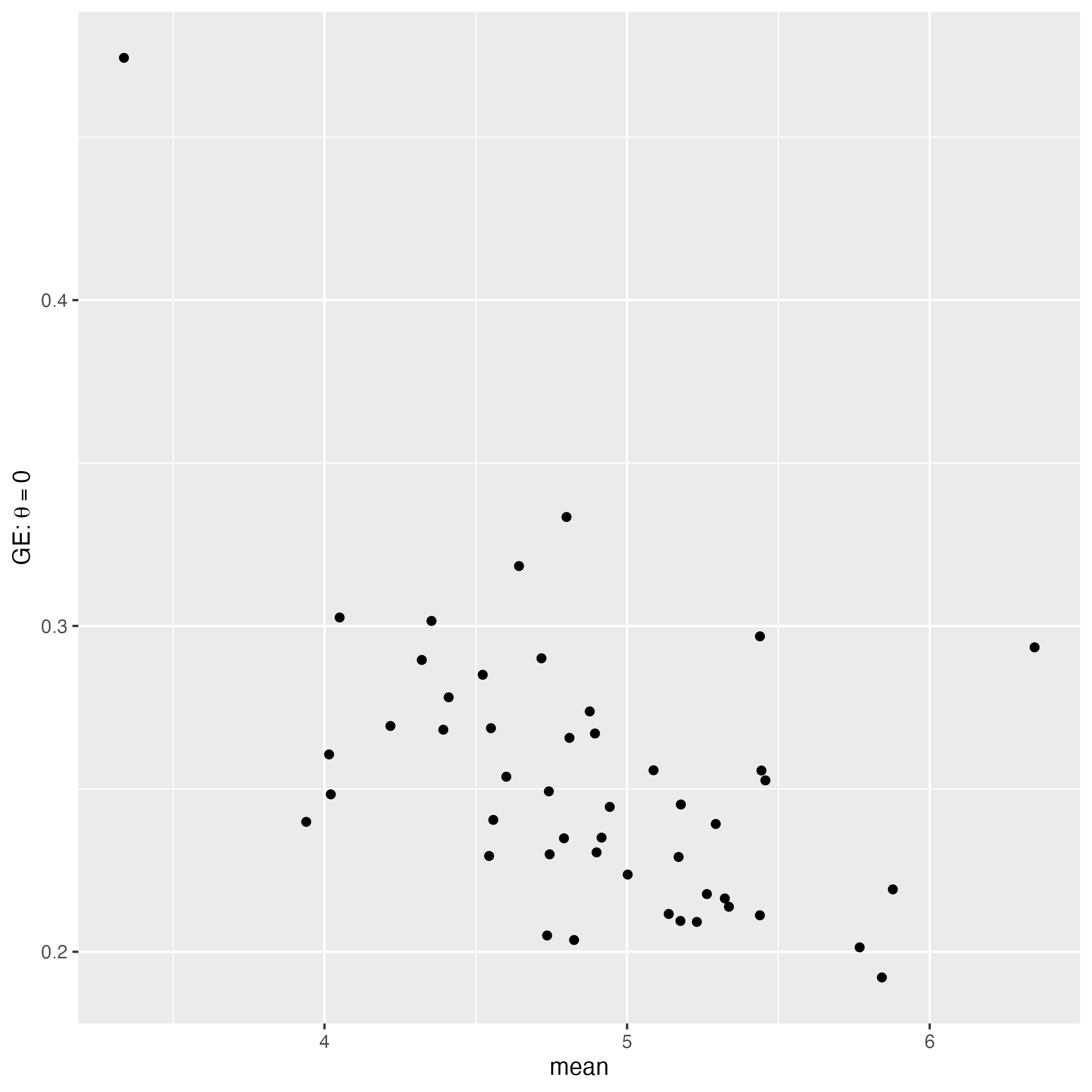}

    \includegraphics[width=5cm]{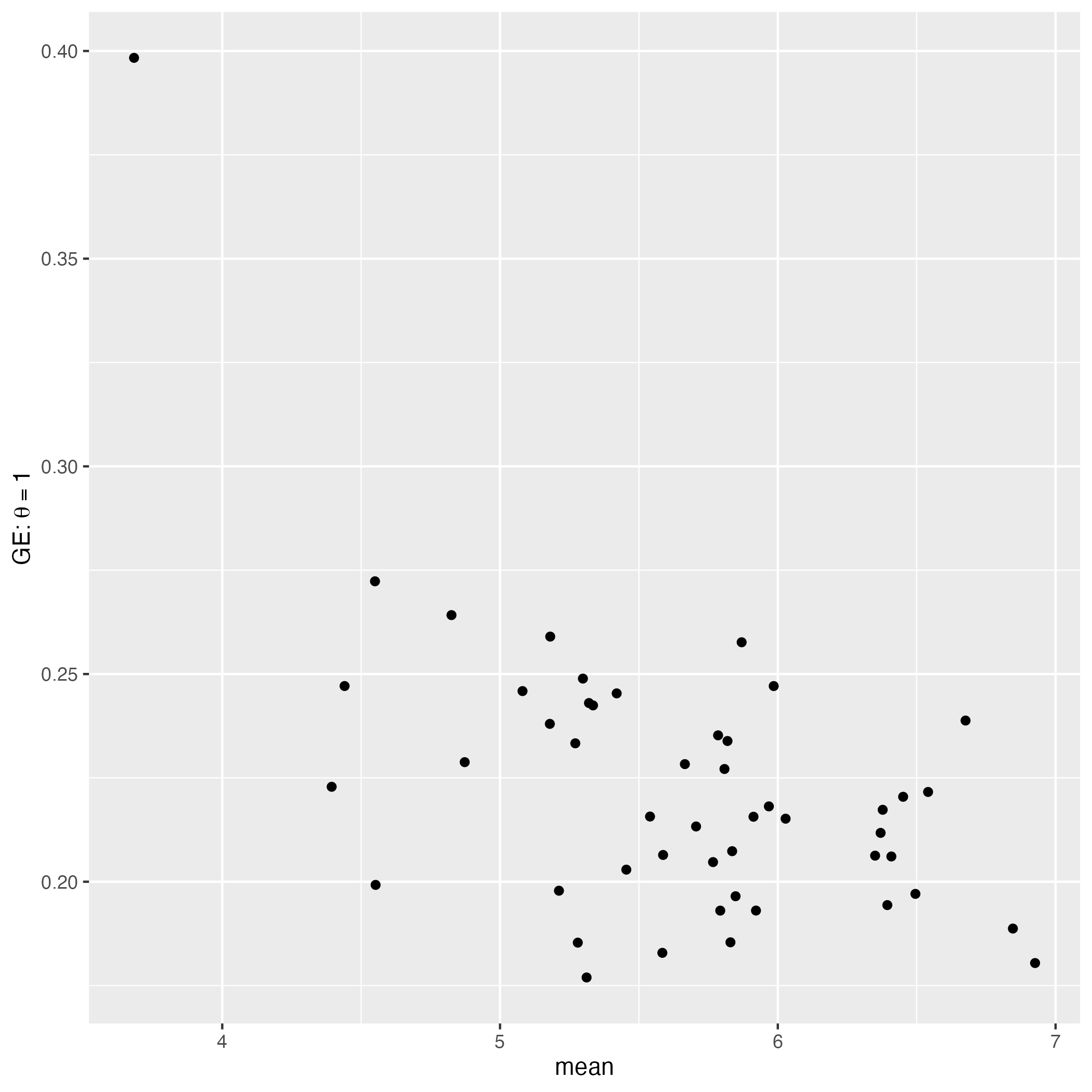}
    \includegraphics[width=5cm]{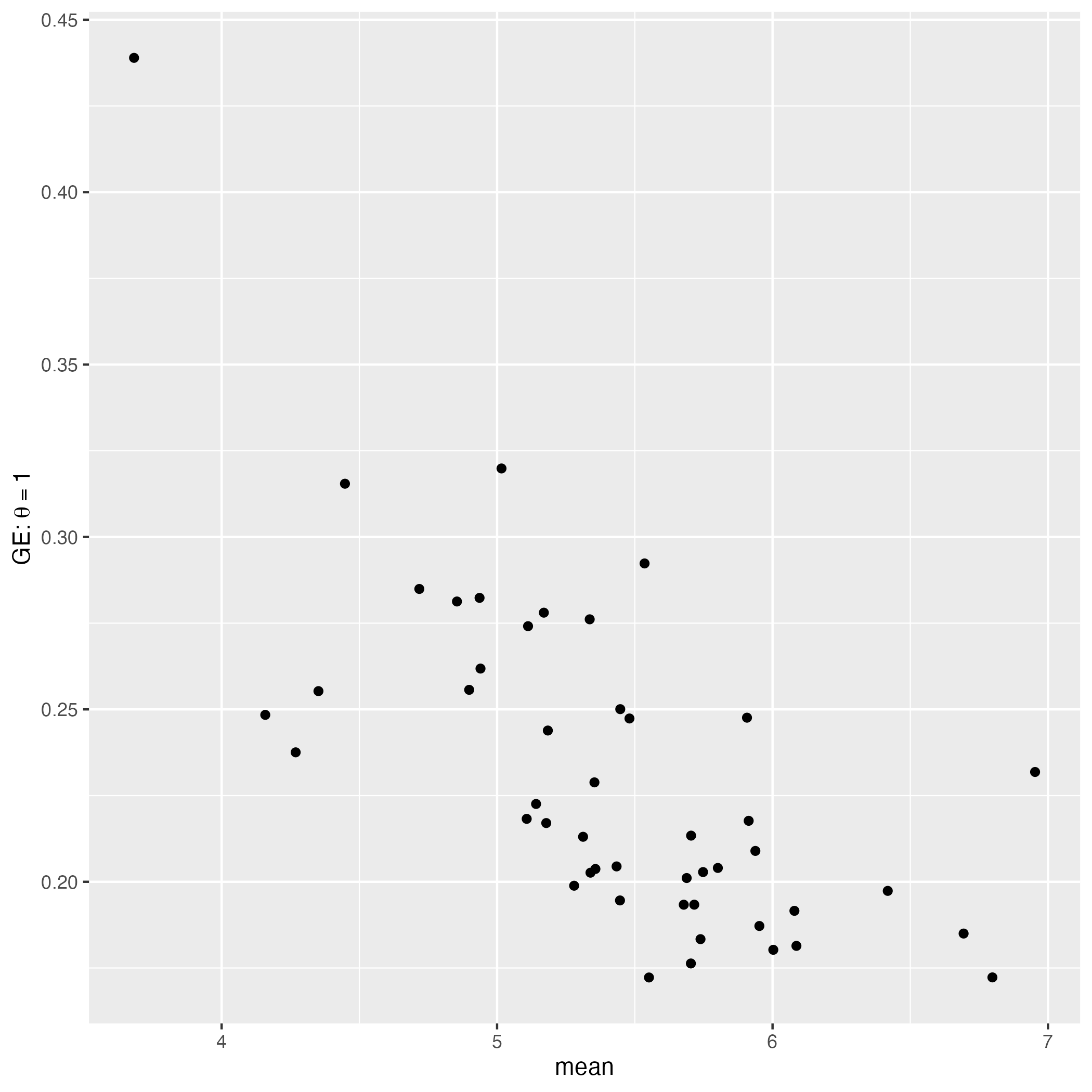}
    \includegraphics[width=5cm]{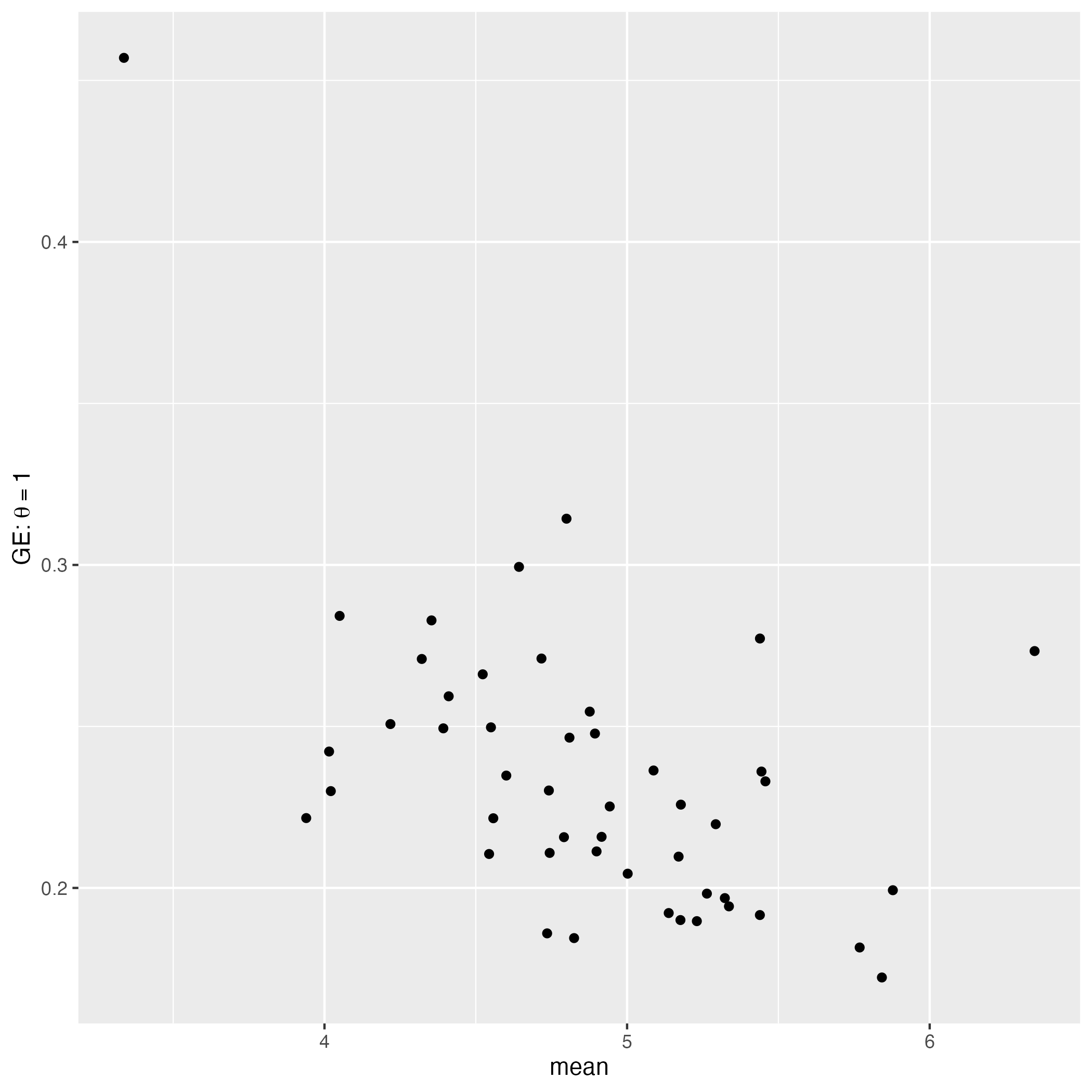}

    \includegraphics[width=5cm]{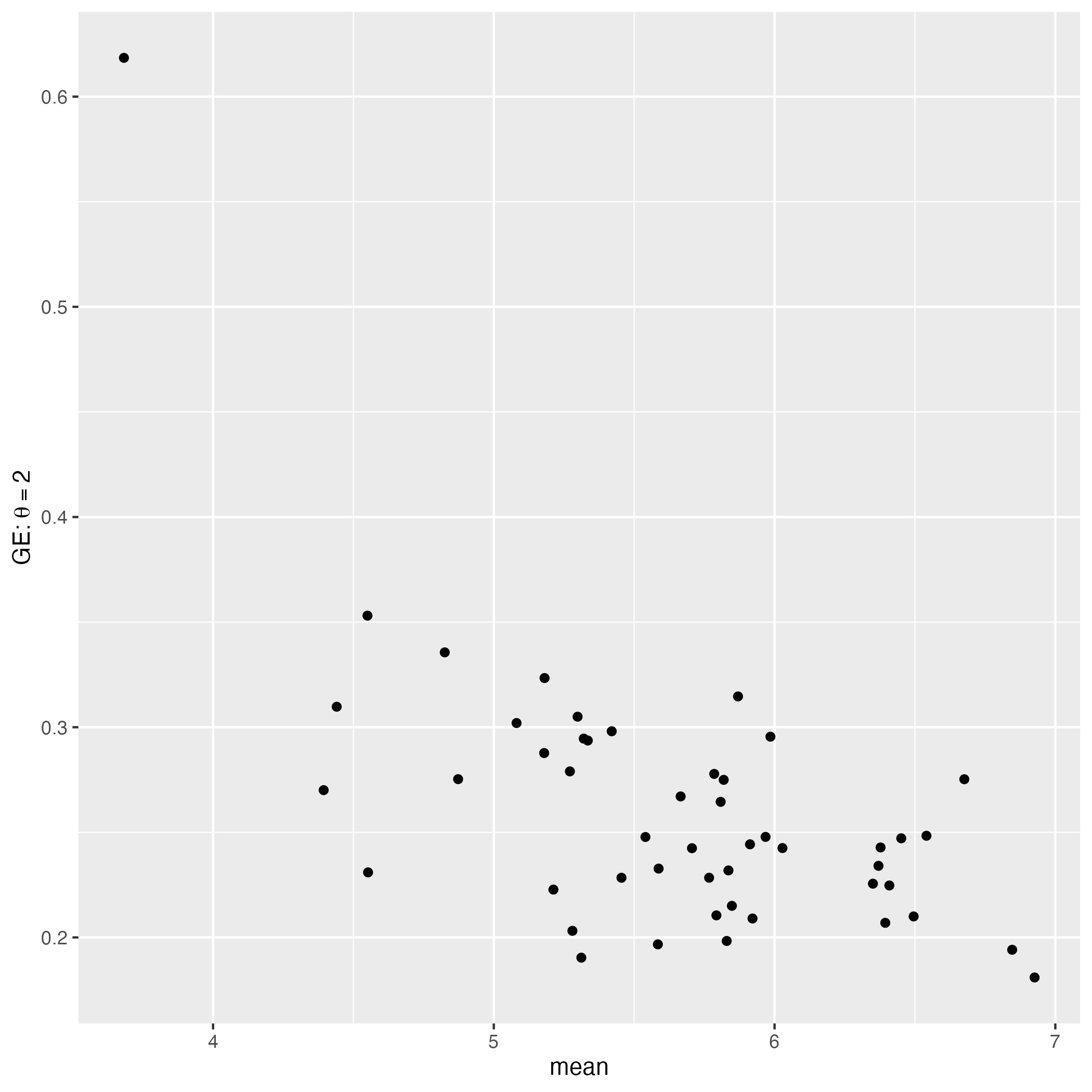}
    \includegraphics[width=5cm]{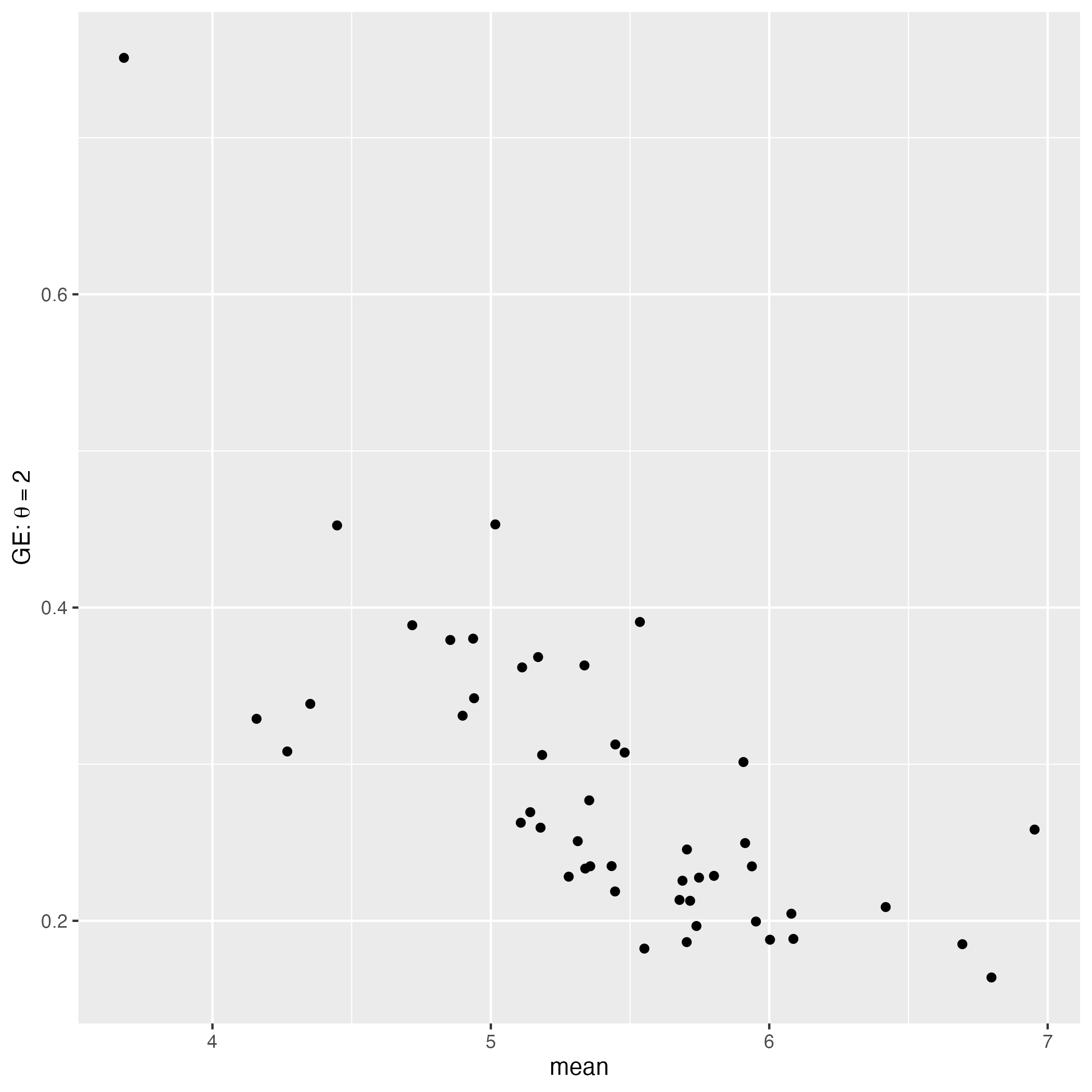}
    \includegraphics[width=5cm]{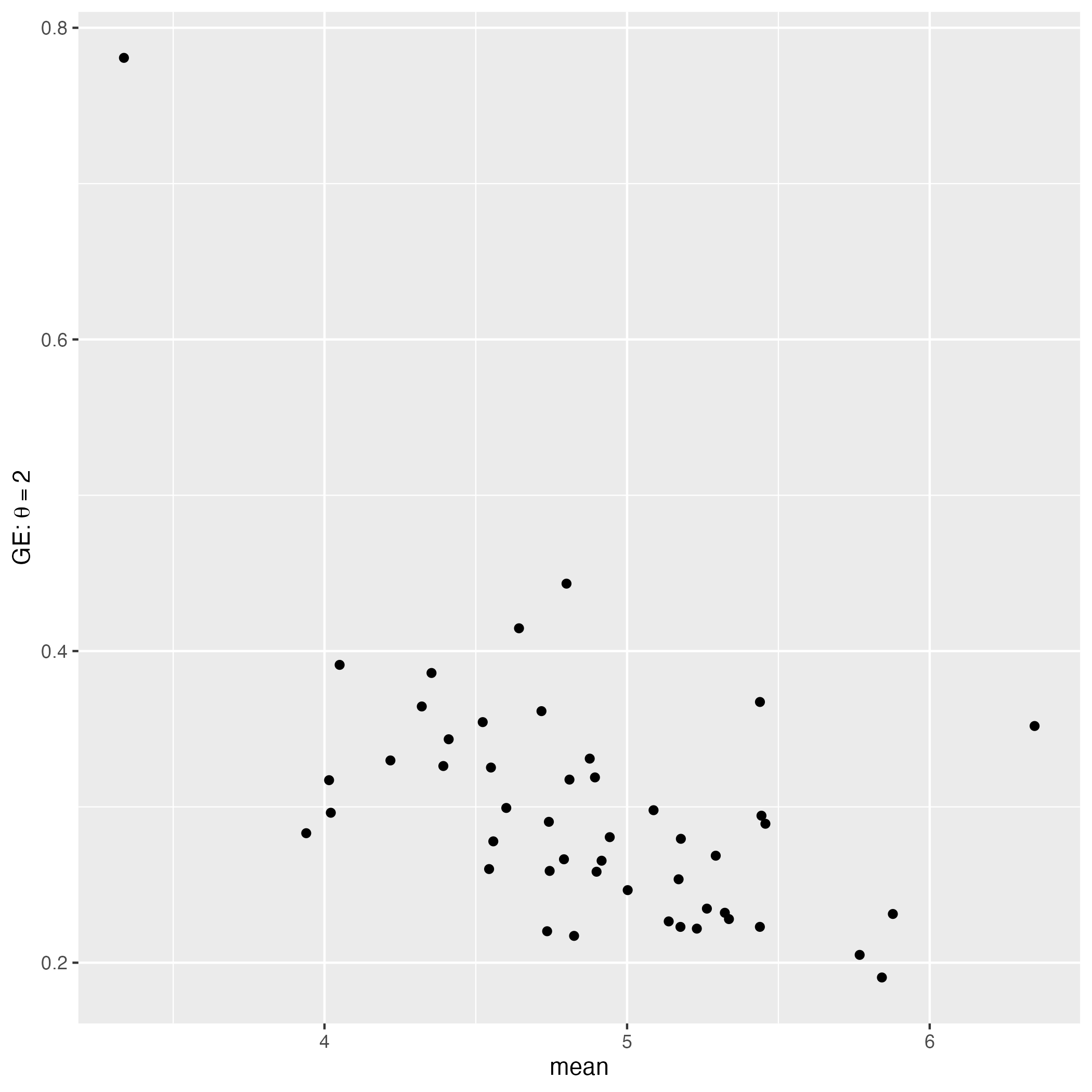}
    \caption{Scatter plots of mean income vs GE for different $\theta$ values for municipalities in Chiba prefecture. The leftmost column shows the results for 2003, the middle column shows the results for 2008, and the rightmost column shows the results for 2013. $\theta$ values for GE are -1, 0, 1, and 2 from the top row, respectively.}
    \label{fig:MIvsGEChiba}
\end{figure}

To further examine the peculiarity of Tokyo, we analyze the decomposition of GE measures for each prefecture into within-municipality and between-municipality inequality, as given by \eqref{eqn:decomp_sub}.
Although we assume a lognormal distribution for each municipality due to sample size constraints, it remains possible to estimate mean incomes.
To explore income distribution patterns within municipalities, we plot scatter diagrams of mean income against GE for different values of $\theta$.
Although similar trends are observed in all 47 prefectures, Tokyo exhibits a distinct pattern.
Therefore, we present the results for Tokyo and Chiba (which borders Tokyo) in Figures \ref{fig:MIvsGETokyo} and \ref{fig:MIvsGEChiba}, respectively.
From these figures, we observe a positive correlation in Tokyo, whereas a negative correlation appears in Chiba.
This finding suggests that in Tokyo, higher-income municipalities tend to have greater inequality, potentially contributing to increased inequality in the upper tail of the income distribution.

In summary, we confirm that our proposed method effectively examines income inequality in Japan through decomposition analysis.
First, our approach allows for a detailed examination of income inequality in Japan by analyzing parameter variations and different values of $\theta$.
Second, by decomposing inequality into between-prefecture and within-prefecture components, we identify a negative correlation between the parameters $a$ and $q$ in the Singh--Maddala distribution.
Furthermore, we find that income inequality in the upper tail becomes more pronounced when both $a$ and $q$ are small, as observed in Tokyo.
Finally, by applying a multilevel decomposition, we reveal that greater inequality in municipalities with higher average incomes may contribute to increased inequality in the upper tail.
Although the insights obtained depend on the assumed income distribution, our method, which enables estimations compatible with decomposition analysis, offers a significant advantage over conventional approaches that estimate income distributions separately, as it provides a more comprehensive understanding of income inequality.

\section{Discussion}
\label{sec:discussion}

In this paper, we propose a method to estimate regional GEs using a constrained Bayes estimator to be compatible with the decomposition of the national GE into within-region and between-region inequality, while assuming different parametric distributions for the entire country and regions, respectively.
Furthermore, we estimated the multilevel decomposition of the GE of Japan, taking into account the situation in which the whole of Japan is divided into 47 prefectures and each prefecture is further divided into municipalities.
For the income distribution in Japan, we assumed a GB2 distribution for the entire country and a Singh--Maddala distribution for the prefectures, but different parametric distributions may be more appropriate for other countries.
However, the proposed method is applicable even if different parametric distributions are assumed and can be widely applied.

A promising direction for future research is to enhance the modeling of income distributions at the municipal level.
In our current framework, we adopted the lognormal distribution for municipalities primarily due to data limitations; however, it is well known that the lognormal distribution often provides a poor fit to income data.
To improve the accuracy and robustness of the estimates---especially for small or missing areas---it would be beneficial to adopt a more flexible distribution such as the Singh--Maddala, which is known to better capture the tail behavior of income distributions.

In such cases, directly modeling the parameters of the parametric distribution as a function of auxiliary variables, such as demographic or socioeconomic indicators, could help borrow strength across areas.
Moreover, imposing spatial or hierarchical correlation structures on the parameters may further stabilize estimation in small areas and enable prediction for regions with missing data.
This approach is in line with recent advances in model-based small area estimation, and extending it to the context of multilevel inequality decomposition using the constrained Bayes estimator proposed in this study would represent a significant methodological contribution.

Additionally, modeling the relationship between the parameters and covariates opens the door to exploratory analysis of the determinants of income inequality.
Such an extension would not only improve predictive accuracy but also yield valuable policy-relevant insights into the structural factors underlying regional disparities.

\medskip
\bigskip
\noindent
{\bf Acknowledgments.}
This work was supported by JSPS KAKENHI Grant Number 20H00080 and 22K01421.

\small
\appendix
\section{Appendix}
\subsection{Derivation of the constrained Bayes estimator \eqref{eqn:CB}}
\label{sec:App_proof}
We take a Lagrange multiplier approach.
Define a Lagrange function $L(\bm{d}, \lambda)$ as
$$
L(\bm{d},\lambda) = \sum_{j=1}^J \phi_j E \left[ \left\{ d_j - \GE_{\theta,j}(\bm{\omega}_j) \right\}^2 \mid \bm{y}_{\mathrm{R}} \right] - \lambda \left( \widehat{\GE}_\theta - \widehat{B} - \sum_{j=1}^J w_jd_j \right),
$$
where $\bm{d} = (d_1,\dots,d_J)^\top$ and $\lambda$ is a Lagrange multiplier.
The equation $\partial L / \partial d_j = 0$ is written as
\begin{equation*}
2\phi_j \left( d_j - \widehat{\GE}_{\theta,j}^\mathrm{B} \right) + \lambda w_j = 0,
\end{equation*}
which implies that
\begin{equation}
\label{eqn:dj}
d_j = \widehat{\GE}_{\theta,j}^\mathrm{B} - \frac{\lambda r_j}{2}.
\end{equation}
The equation $\partial L / \partial \lambda = 0$ is written as
\begin{align*}
&\widehat{\GE}_\theta - \widehat{B} - \sum_{j=1}^J w_jd_j \\
=& \ \widehat{\GE}_\theta - \widehat{B} - \sum_{j=1}^J w_j \left( \widehat{\GE}_{\theta,j}^\mathrm{B} - \frac{\lambda r_j}{2} \right) \quad (\because \eqref{eqn:dj}) \\
=& \ \widehat{\GE}_\theta - \widehat{B} - \overline{\GE}_\theta^w + \frac{\lambda q}{2} = 0.
\end{align*}
Then, we obtain
\begin{equation}
\label{eqn:lambda}
\lambda = -\frac{2}{q} \left( \widehat{\GE}_\theta - \widehat{B} - \overline{\GE}_\theta^w \right).
\end{equation}
Pluggin \eqref{eqn:lambda} into \eqref{eqn:dj}, we obtain
$$
d_j = \widehat{\GE}_{\theta,j}^\mathrm{B} + \frac{r_j}{q} \left( \widehat{\GE}_\theta - \widehat{B} - \overline{\GE}_\theta^w \right).
$$
\hfill$\Box$

\subsection{Additional data analysis for evaluating the performance of constrained Bayes estimator}
\label{sec:App_DataAnalysis}
In this section, we conduct an additional data analysis to evaluate the performance of our proposed constrained Bayes estimator.
In this analysis, we first estimate the GE for the entire country assuming the GB2 distribution as in Section \ref{sec:realdata}.
Next, we consider the following three estimators of the GEs for the prefectures: the constrained Bayes estimator based on the Singh--Maddala distribution (SM\_CB), constrained Bayes estimator based on lognormal distribution (LN\_CB) and Bayes estimator without benchmarking based on the lognormal distribution (LN).
The first estimator SM\_CB is the one we propose in Section \ref{sec:realdata}.
From Tables \ref{tab:multi1} and \ref{tab:multi2}, we can see that the Singh--Maddala distribution provides a good fit to the prefectural income distributions, given the small residuals under the separate method.
Thus, we treat SM\_CB estimates as the pseudo-true and evaluate the relative difference of the two estimates, LN\_CB and LN, from SM\_CB, that is,
$$
\mathrm{RD}_i = \frac{\widehat{\mathrm{GE}}_i - \widetilde{\mathrm{GE}}_i}{\widetilde{\mathrm{GE}}_i},
$$
where $\widetilde{\mathrm{GE}}_i$ is SM\_CB estimate of the GE (pseudo-true) and $\widehat{\mathrm{GE}}_i$ is LN\_CB or LN estimate in the prefecture $i \ (=1,\dots,47)$.

\begin{figure}[h]
    \centering
    \includegraphics[width=7.5cm]{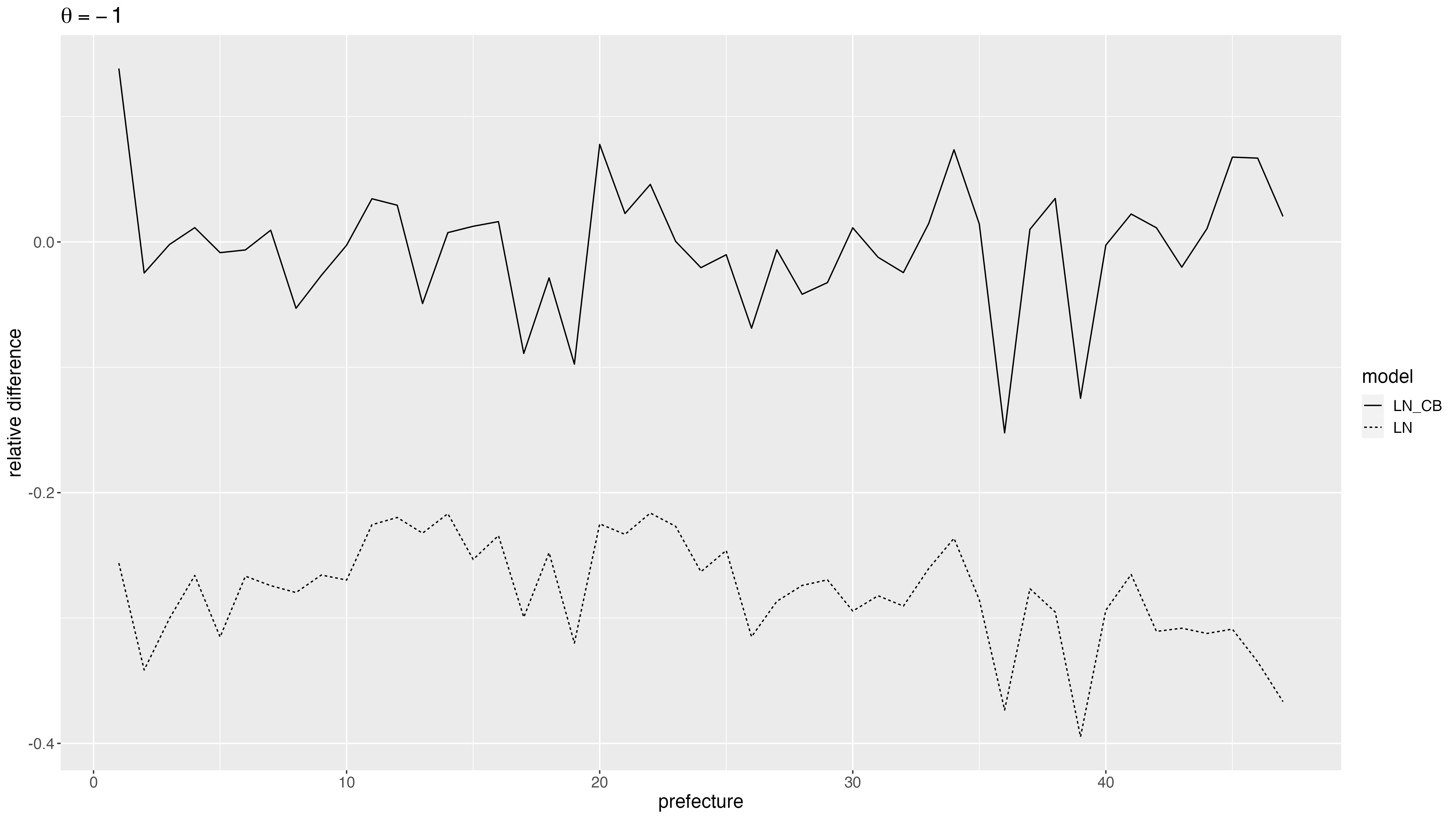}
    \includegraphics[width=7.5cm]{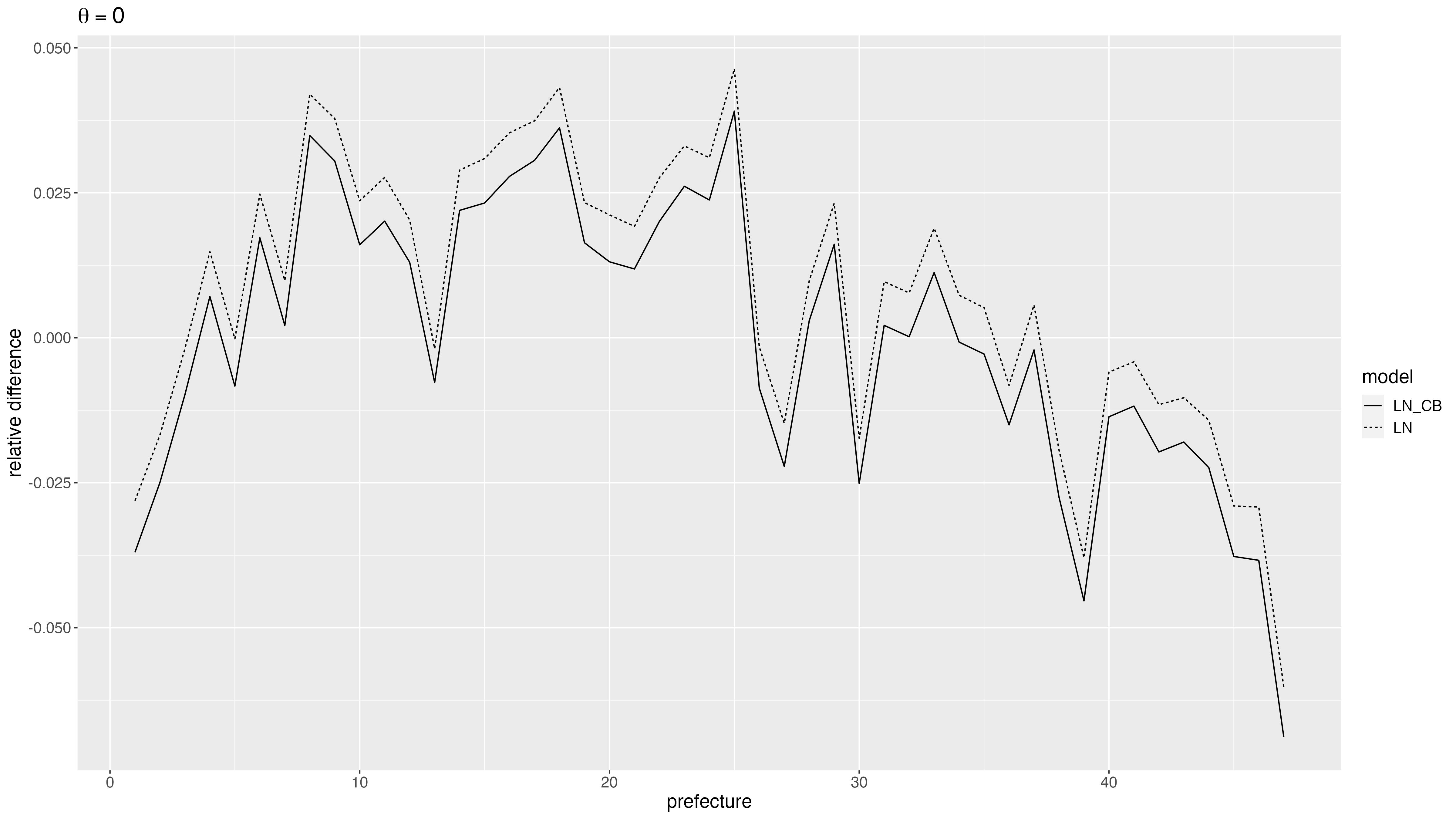}

    \includegraphics[width=7.5cm]{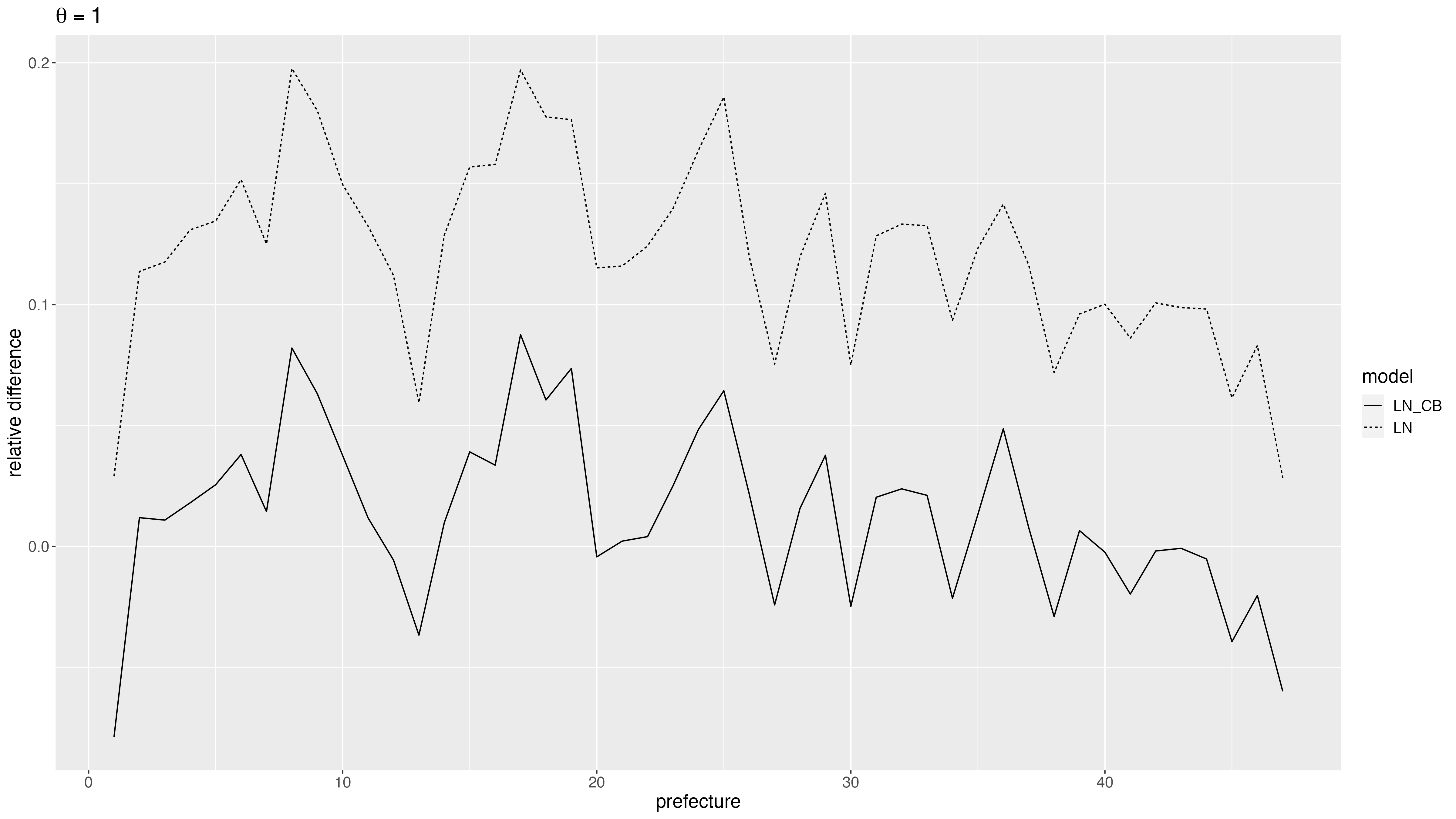}
    \includegraphics[width=7.5cm]{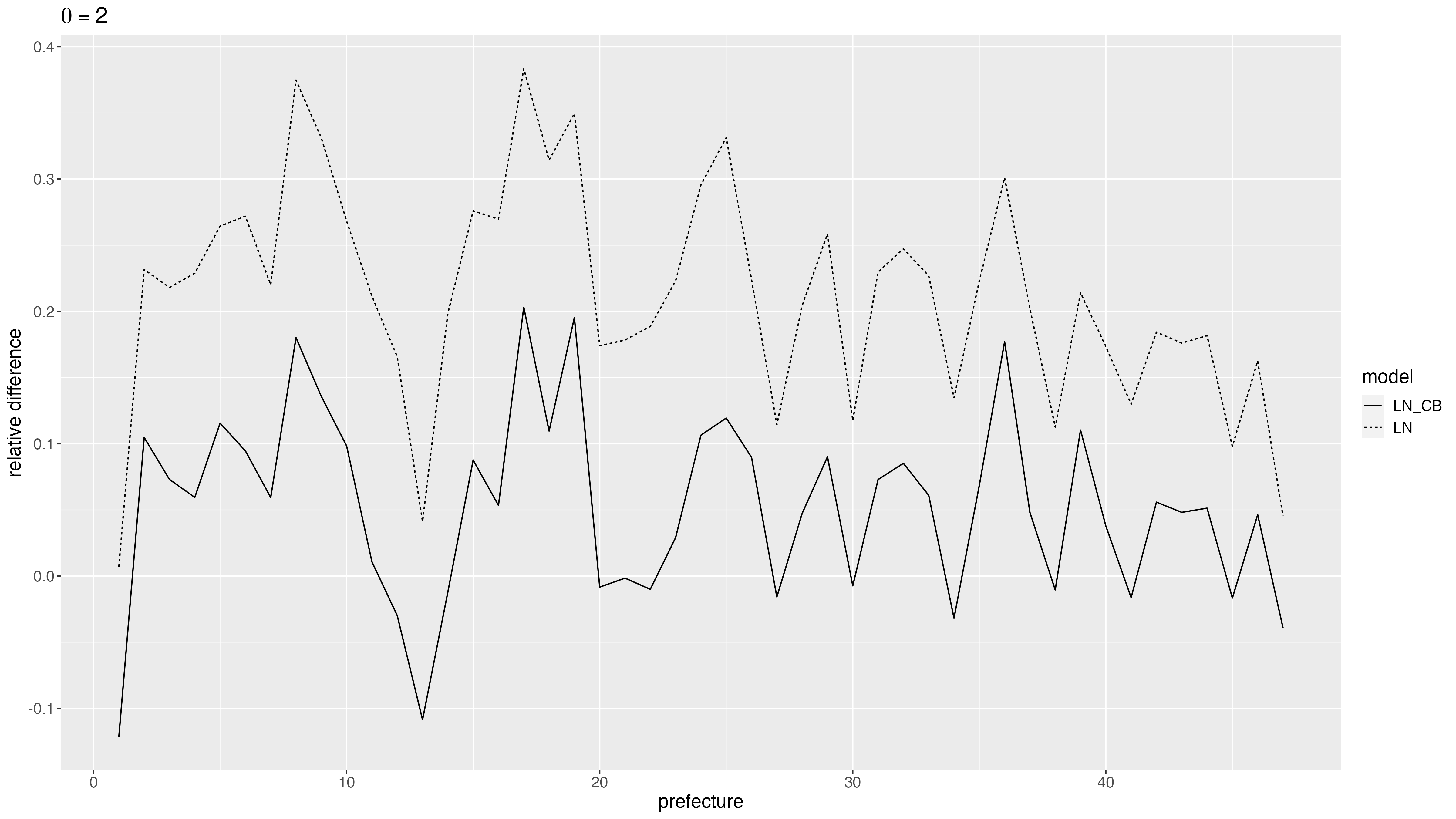}
    \caption{Relative differences of LN\_CB and LN estimates from SM\_CB (pseudo-true) estimates of GEs for the 47 prefectures for different $\theta$ values, $\theta = -1$ (top left), $\theta = 0$ (top right), $\theta = 1$ (bottom left) and $\theta = 2$ (bottom right).}
    \label{fig:RD}
\end{figure}

The results are shown in Figure \ref{fig:RD}.
First, we can see from the figure that the LN estimator has negative bias for $\theta = -1$ and has positive bias for $\theta = 1$ and $\theta = 2$.
This result is consistent with Section \ref{sec:realdata}.
On the other hand, for all $\theta$ values, LN\_CB estimator reduces the bias, which implies that our proposed constrained Bayes estimator performs well even if the assumed parametric distribution does not fit well.
When the sample size in the subpopulation is small, it is necessary to assume a distribution with a small number of parameters, such as a lognormal distribution, but even in such cases our proposed method is expected to have a good GE estimation performance for the subpopulation.

\subsection{The relationship between parameters and GE measures}
\label{sec:parameters_GE}

\begin{figure}[th]
    \centering
    \includegraphics[width=\linewidth]{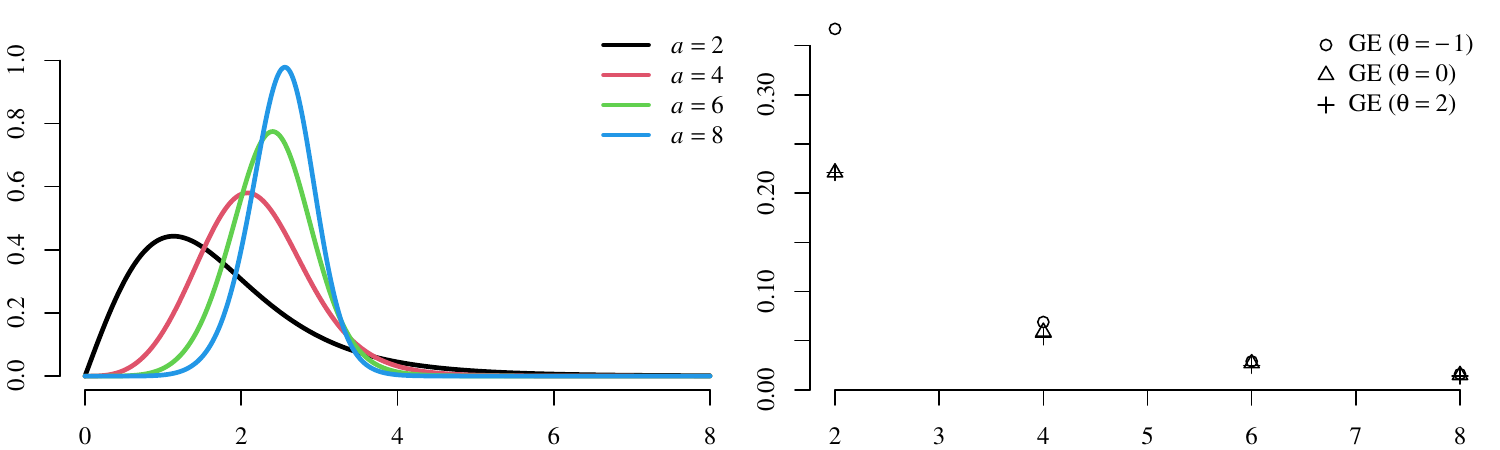}
    
    \includegraphics[width=\linewidth]{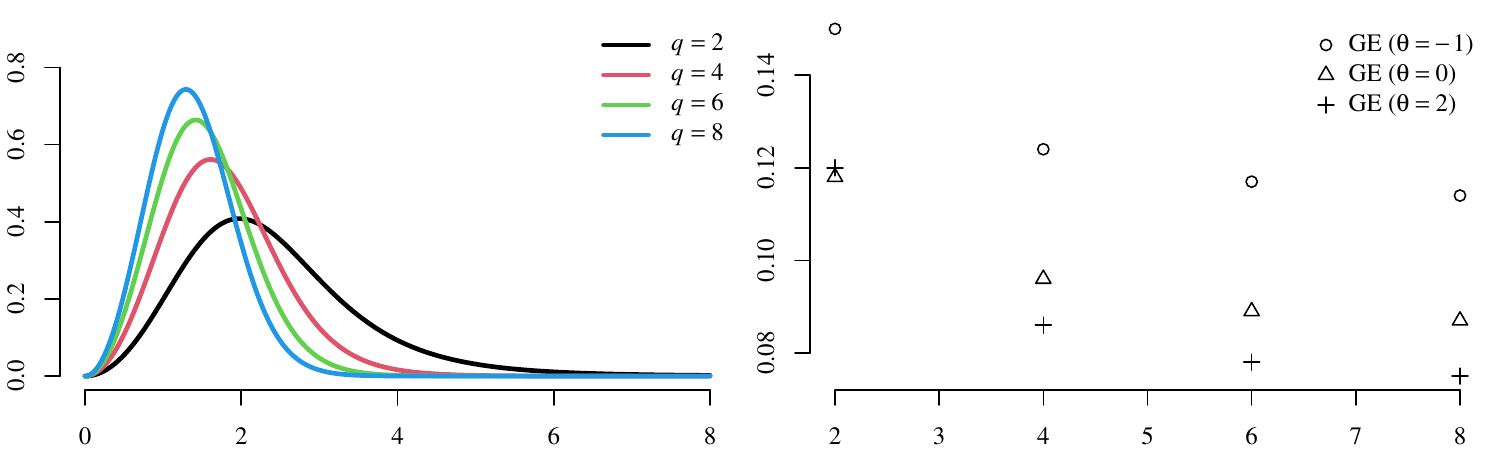}
    \caption{The SM distributions and corresponding GE measures}
    \label{fig:SM}
\end{figure}

In this section, we explain the relationship between the parameters of the SM distribution and the GE measures.
The top panel of Figure \ref{fig:SM} illustrates the SM distributions (left) and their corresponding GE measures (right) when the parameters are fixed at $b=3$ and $q=3$.
From the SM distributions, we observe that as $a$ increases, the lower tail becomes thinner, while the mode shifts to the right and becomes more pronounced.
These effects are reflected in the changes in the GE measures, particularly when $\theta=-1$.
On the other hand, the bottom panel of Figure \ref{fig:SM} presents the SM distributions (left) and their corresponding GE measures (right) when the parameters are fixed at $a=3$ and $b=3$.
Here, we observe that as $q$ increases, the upper tail becomes thinner and the mode shifts to the left and becomes more pronounced.
However, the impact of variations in $a$ and $q$ on the GE measures differs: changes in $a$ have a greater effect on lower-income groups, whereas changes in $q$ have a greater effect on higher-income groups.
Therefore, under certain values of $\theta$, even if the GE measure remains the same, different combinations of $a$ and $q$ can lead to the same GE measure.
This highlights the importance of considering the values of $a$ and $q$ in addition to the GE measure itself. 

\begin{figure}[th]
    \centering
    \includegraphics[width=5.2cm]{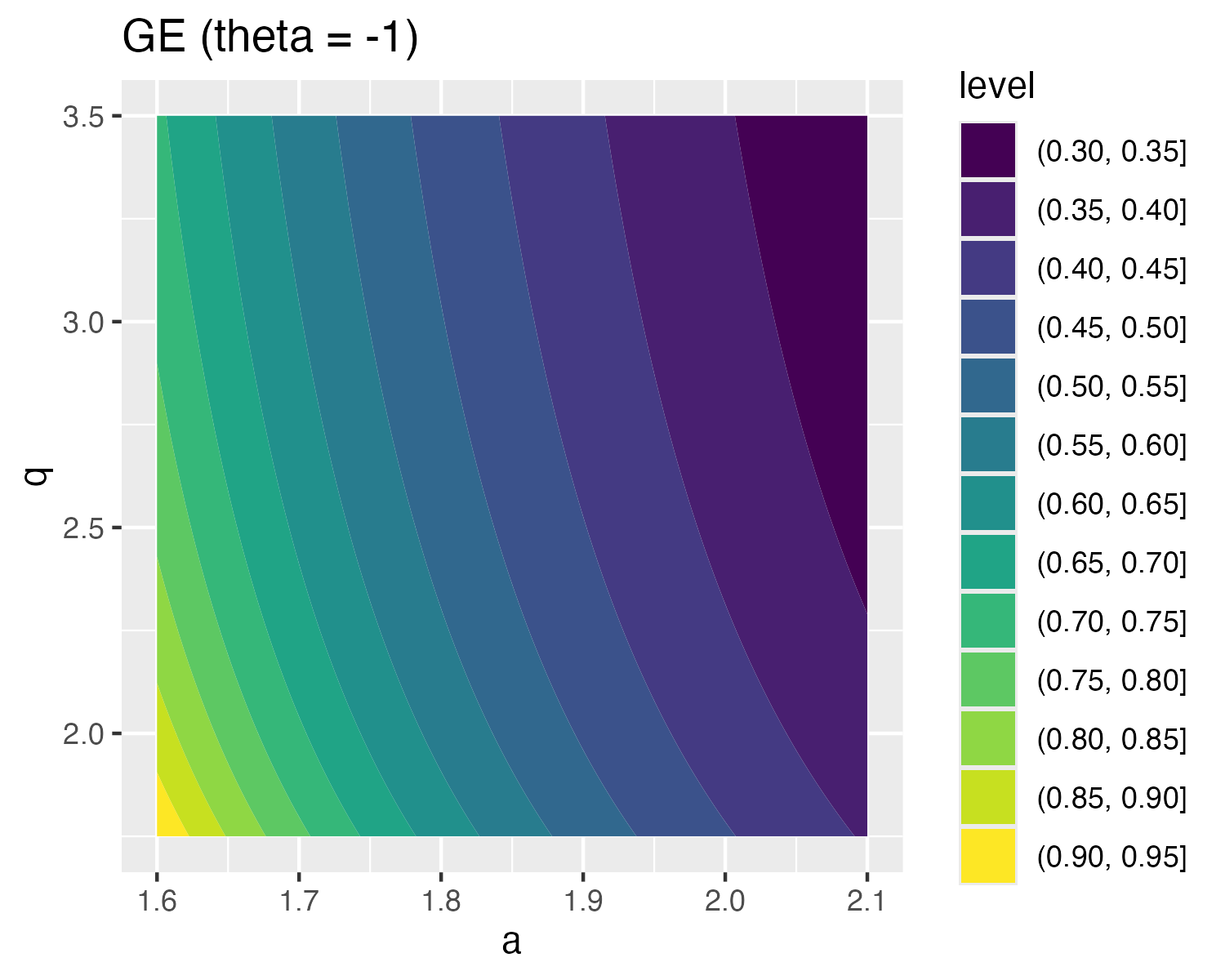}    
    \includegraphics[width=5.2cm]{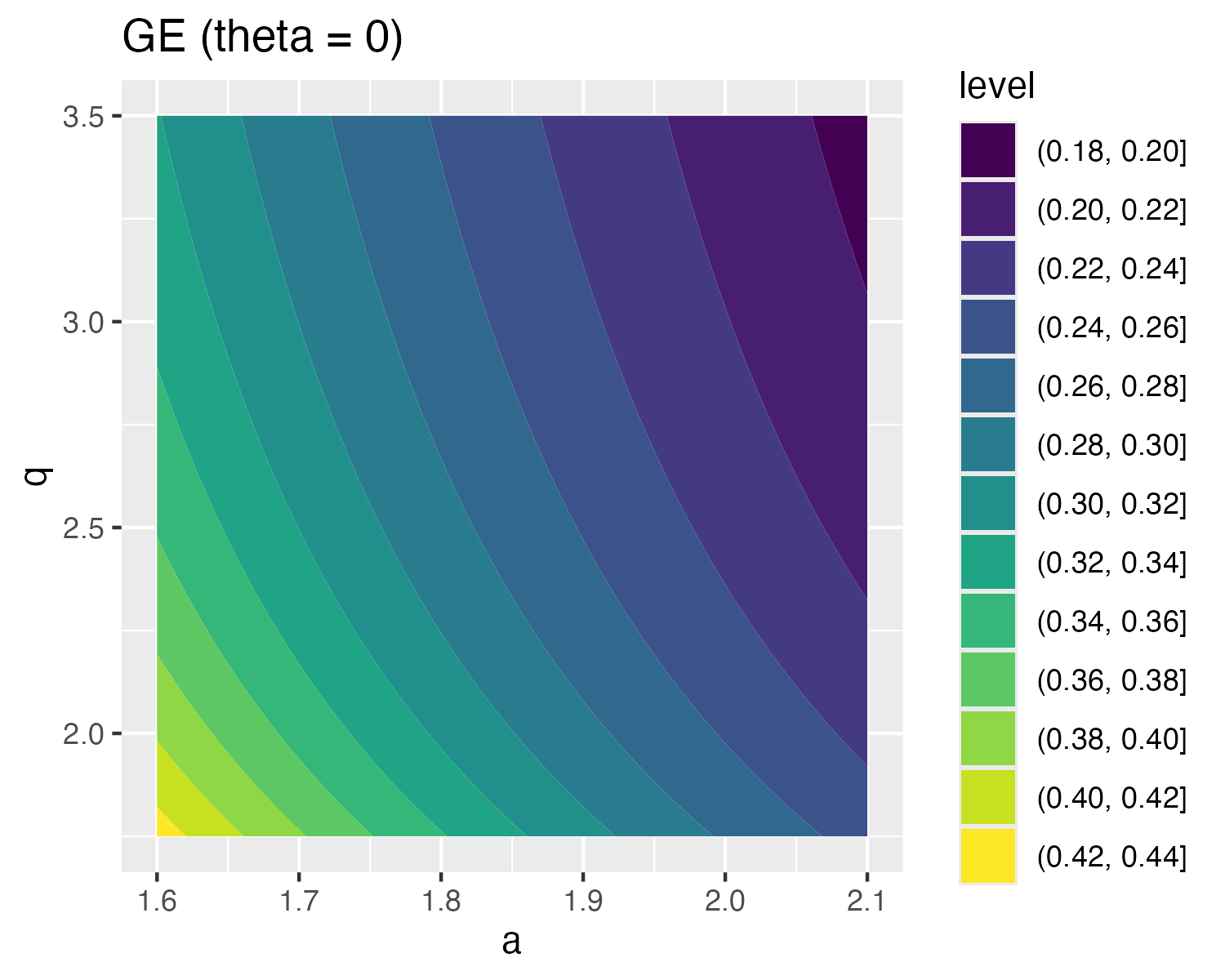}
    \includegraphics[width=5.2cm]{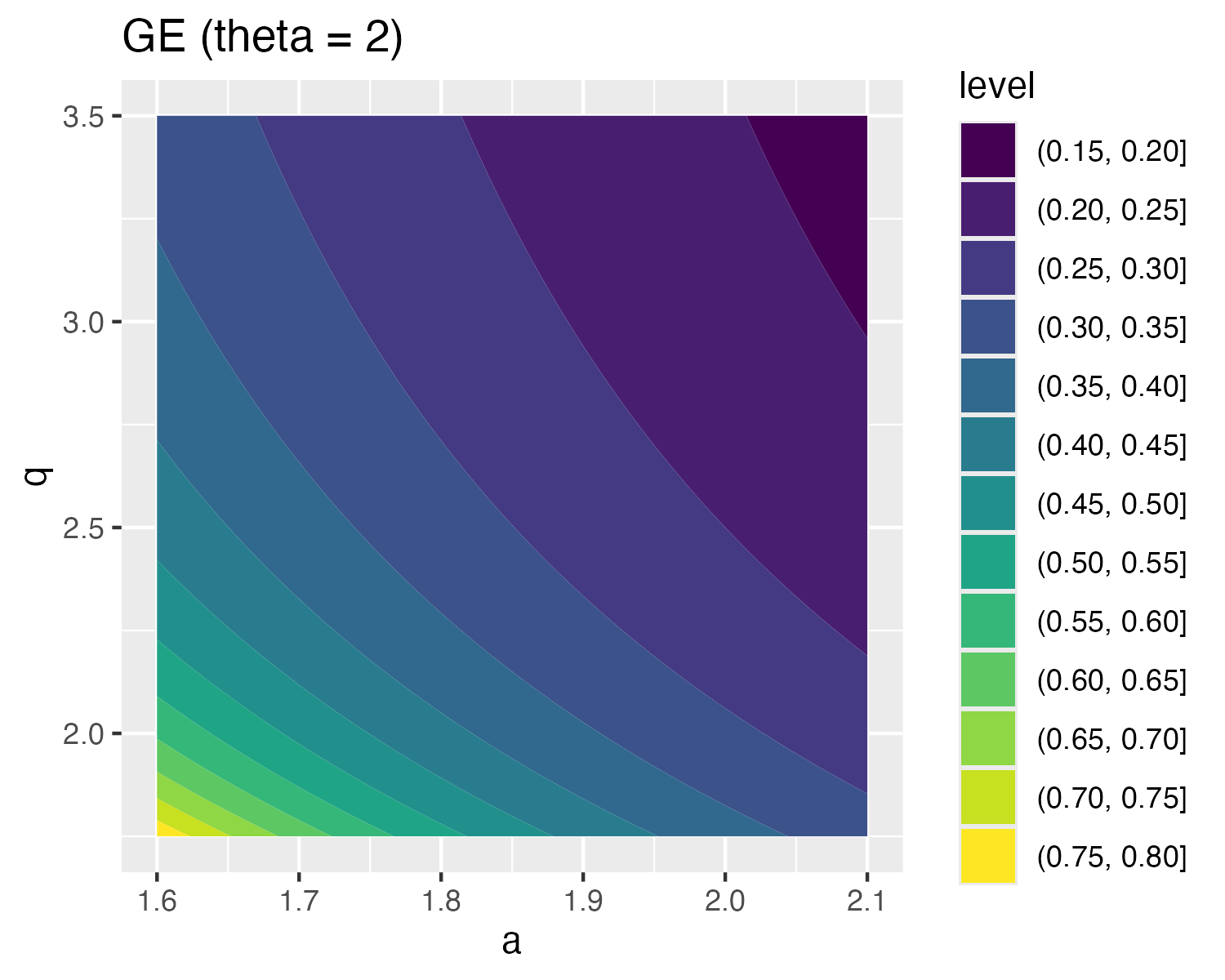}
    \caption{The contour plots of GE measures}
    \label{fig:SM-GE}
\end{figure}

Figure \ref{fig:SM-GE} provides contour plots that depict how parameters $a$ and $q$ are related to the GE under different $\theta$.
From the figure , it can be seen that as $a$ and/or $q$ decrease, GE increases.
Moreover, even with the same values of $a$ and $q$, the relative magnitude of GE varies depending on the value of $\theta$, and as $\theta$ decreases, GE becomes more sensitive to changes in $a$. This can also be observed in Figure \ref{fig:SM}.


\end{document}